\documentclass[conference]{IEEEtran}

\usepackage{xspace}
\usepackage[hyphens]{url}
\usepackage[
bookmarksopen,
bookmarksdepth=2,
colorlinks=false,
urlcolor=blue]{hyperref}
\usepackage{color}
\usepackage{soul}
\usepackage[margin=4pt,caption=false]{subfig}
\usepackage{diagbox}
\usepackage{multirow}
\usepackage{algorithm}
\usepackage{graphicx, tabularx}
\newcolumntype{C}{>{\centering\arraybackslash}X}

\usepackage{amsmath}
\usepackage{algpseudocode}
\usepackage{tikz}

\algblockdefx[Foreach]{Foreach}{EndForeach}[1]{\textbf{foreach} #1 \textbf{do}}{\textbf{end foreach}}
\algnewcommand\algorithmiccase{\textbf{case}}
\algdef{SE}[CASE]{Case}{EndCase}[1]{\algorithmiccase\ #1}{\algorithmicend\ \algorithmiccase}%
\algtext*{EndCase}%
\makeatletter
\def\algbackskip{\hskip-\ALG@thistlm}
\makeatother

\usepackage{amssymb}
\usepackage{alltt}
\usepackage{bm}
\usepackage{comment}
\usepackage{makecell}

\hypersetup{draft}
\definecolor{lavender}{rgb}{0.75, 0.58, 0.89}

\newcommand{\FM}{$\mathcal{M_F}$\xspace}
\newcommand{\TM}{$\mathcal{M_T}$\xspace}
\newcommand{\CM}{$\mathcal{M_C}$\xspace}
\newcommand{\F}{$\mathcal{F}$\xspace}
\newcommand{\A}{$\mathcal{A}$\xspace}
\newcommand{\B}{$\mathcal{B}$\xspace}
\newcommand{\AM}{$\mathcal{M_A}$\xspace}
\newcommand{\BM}{$\mathcal{M_B}$\xspace}
\newcommand{\1}{$Zone_1$\xspace}
\newcommand{\2}{$Zone_2$\xspace}
\newcommand{\3}{$Zone_3$\xspace}
\allowdisplaybreaks
\usepackage{xparse}
\NewDocumentCommand{\ceil}{s O{} m}{%
  \IfBooleanTF{#1} 
{$\left\lceil#3\right\rceil$} 
    {#2\lceil#3#2\rceil} 
}
\usepackage{mathtools}
\DeclareMathOperator*{\argmax}{arg\,max}
\DeclarePairedDelimiter{\norm}{\lVert}{\rVert}

\newcolumntype{R}{>{\raggedleft\arraybackslash}X}
\usepackage[utf8]{inputenc}
\usepackage[english]{babel}

\usepackage{enumitem}
\usepackage{amsthm}
\usepackage{xcolor}
\theoremstyle{plain}
\makeatletter
\def\thm@space@setup{%
  \thm@preskip=1pt plus 1pt minus 1pt
  \thm@postskip=\thm@preskip 
}
\makeatother
\newtheorem{theorem}{Theorem}[section]

\newtheorem{lemma}[theorem]{Lemma}
\newtheorem{definition}{Definition}[section]
\newtheorem{assumption}{Assumption}
\newcommand{\blue}[1]{{#1}}

\author{\IEEEauthorblockN{Yujin Kwon\IEEEauthorrefmark{1},
Hyoungshick Kim\IEEEauthorrefmark{2}, Jinwoo Shin\IEEEauthorrefmark{1},
Yongdae Kim\IEEEauthorrefmark{1}}
\IEEEauthorblockA{\IEEEauthorrefmark{1}KAIST\\
\{dbwls8724, jinwoos, yongdaek\}@kaist.ac.kr\\
\IEEEauthorrefmark{2}Sungkyunkwan University\\
hyoung@skku.edu}}

\begin{document}

\title{Bitcoin vs. Bitcoin Cash: \\
    Coexistence or Downfall of Bitcoin Cash?}

\maketitle

\begin{abstract}
Bitcoin has become the most popular cryptocurrency based on a peer-to-peer network. 
In Aug. 2017, Bitcoin was split into the original Bitcoin (BTC) and Bitcoin Cash (BCH). 
Since then, miners have had a choice between BTC and BCH mining because they have compatible proof-of-work algorithms. 
Therefore, they can freely choose which coin to mine for higher profit, where the profitability depends on both the coin price and mining difficulty. 
Some miners can immediately switch the coin to mine only when mining difficulty changes because the difficulty changes are more predictable than that for the coin price, and we call this behavior \emph{fickle mining}. 

In this paper, we study the effects of fickle mining  
by modeling a game between two coins.  
To do this, we consider both fickle miners and some \textit{factions} (e.g., BITMAIN for BCH mining) that stick to mining one coin to maintain that chain.
In this model, we show that fickle mining leads to a Nash equilibrium in which only a faction sticking to its coin mining remains as a loyal miner to the less valued coin (e.g., BCH), where loyal miners refer to those who conduct mining even after coin mining difficulty increases. 
This situation would cause severe centralization, weakening the security of the coin system. 

To determine which equilibrium the competing coin systems (e.g., BTC vs. BCH) are moving toward, we traced the historical changes of mining power for BTC and BCH and found that BCH often lacked loyal miners until Nov. 13, 2017, when the difficulty adjustment algorithm of BCH mining was changed. 
However, the change in difficulty adjustment algorithm of BCH mining led to a state close to the stable coexistence of BTC and BCH. 
We also demonstrate that the lack of BCH loyal miners may still be reached when a fraction of miners automatically and repeatedly switches to the most profitable coin to mine (i.e., automatic mining). 
According to our analysis, as of Dec. 2018, loyal miners to BCH would leave if more than about 5\% of the total mining capacity for BTC and BCH has engaged in the automatic mining. 
In addition, we analyze the recent ``hash war'' between Bitcoin ABC and SV, which confirms our theoretical analysis. 
Finally, we note that our results can be 
applied to any competing cryptocurrency systems in which the same hardware (e.g., ASICs or GPUs) can be used for mining.  
Therefore, our study brings new and important angles in competitive coin markets: a coin can \emph{intentionally} weaken the security and decentralization level of the other rival coin when mining hardware is shared between them, allowing for automatic mining.  
\end{abstract}


\IEEEpeerreviewmaketitle

\section{Introduction}
\label{sec:intro}

Bitcoin~\cite{nakamoto2008bitcoin} is the most popular cryptocurrency based on a distributed and public digital ledger called \emph{blockchain}. 
Nodes in the Bitcoin network store the blockchain, where transactions are recorded in a unit of a \textit{block}, and the blockchain is extended by generating new blocks. 
The process of generating new blocks is referred to as \textit{mining}, and nodes conducting mining activities are referred to as \textit{miners}. 
To successfully mine, miners should find a solution called the \textit{proof-of-work} (PoW)~\cite{pow}. 
In Bitcoin, miners are required to solve a cryptographic puzzle finding a hash value to satisfy specific conditions such as a certain number of leading zeroes. 
To solve a puzzle, miners spend their computational power, and the miner who finds the solution obtains 12.5 coins and the transaction fees in the new block as a reward.
In addition, Bitcoin has an average block interval of 10 minutes by adjusting the mining difficulty (i.e., the difficulty of the puzzles).

As Bitcoin has gained popularity, the transaction scalability issue has risen, and several solutions have been proposed to address the issue. 
However, there were also several conflicts over these solutions. 
As a result, in Aug. 2017, the Bitcoin system was split into the original Bitcoin (BTC) and Bitcoin Cash (BCH)~\cite{bch, split}. 
The key idea of BCH is to increase a maximum block size to process more transactions than BTC. However, even with different block size limits, they have compatible proof-of-work mechanisms with each other. 
Therefore, miners can freely alternate between BTC and BCH mining to boost their profits~\cite{stability}.
The mining profitability changes when the mining difficulty and coin price change, but some miners may be concerned only with the change in former because it is relatively easier to predict the former than the latter. 
More precisely, rational miners can decide which cryptocurrency is better to mine depending on the coin mining difficulty --- BCH mining would be conducted by the miner only if the BCH mining difficulty is low compared to the BTC mining difficulty; 
otherwise, the miner does BTC mining rather than BCH mining. 
We call this miner's behavior ``\textbf{fickle mining}'' in this paper.
Note that the fickle miner may change the coin to mine at a specific time period whenever the coin mining difficulty changes.
Thus, fickle mining leads to instability of mining power, which may eventually cause unstable coin prices~\cite{stability}.

\smallskip
\noindent \textbf{Game model and analysis. }
In this study, we aim to analyze the economics of fickle mining rigorously, which can later be extended to show how one coin can lead to a lack of loyal miners for other less valued coins. 
Here, a loyal miner represents one who conducts mining the less valued coin even after the coin mining difficulty increases. 
To study the economics of fickle mining, we propose a game theoretical framework of players who can conduct fickle mining between two coins (e.g., BTC and BCH). 
Moreover, our game model reflects \textit{coin factions} that stick to mining their own coins, as they are interested in only the maintenance of their systems rather than the payoffs. 
Then we analyze Nash equilibria and dynamics in the game; two types of equilibria exist: the stable coexistence of two coins and the lack of loyal miners for the less valued coin. 
More specifically, in the latter case, only some factions (e.g., BITMAIN for BCH mining) remain as loyal miners for the less valued coin, and this fact can eventually make the coin system severely centralized, weakening its security. 
We describe the game model in Section~\ref{sec:model} and analyze the game in Section~\ref{sec:analysis}. 

\smallskip
\noindent\textbf{Data analysis for BTC vs. BCH.}
Next, as a case study, we analyzed the mining power changes in BTC and BCH to see if our theoretical analysis matches with actual mining power changes. 
In this paper, we refer to the \textit{Bitcoin system} as a coin system consisting of BTC and BCH. 
We examine the mining power history in the Bitcoin system from the release date of BCH until Dec. 2018 to 1) analyze which equilibrium its state has been moving to and 2) evaluate our theoretical analysis empirically. 
Our analysis results show that until the BCH mining difficulty adjustment algorithm changed (on Nov. 13, 2017), the Bitcoin state reached a lack of loyal miners for BCH. 
Therefore, BCH periodically became severely centralized before the update of the BCH protocol. 
For example, we observe a period when only five miners exist, of which two miners possess about 70 \% power. 
However, since Nov. 13, 2017, the Bitcoin state has been close to coexistence because the change in the BCH mining difficulty adjustment algorithm with a shorter difficulty adjustment time interval (i.e., every block) has affected the game as an external factor. 

Nevertheless, we explain that 
the state would still get closer to a lack of BCH loyal miners 
if \emph{automatic mining}, in which miners automatically choose the most profitable coin to mine, is popularly used. 
Note that the main difference between fickle mining and automatic mining is that fickle miners \textit{immediately} change their coin only when the mining difficulty changes while automatic miners can \textit{immediately} change their coin when not only the mining difficulty but also the coin price changes. 
As a result, at the time of writing (Dec. 2018), if 5\% of the total mining power of the Bitcoin system involves automatic mining, the current loyal miners for BCH would leave, weakening its security.

\smallskip
\noindent\textbf{Data analysis for Bitcoin ABC vs. SV.}
As another case study in our game model, we also analyze the changes in the hash rate distributions of Bitcoin ABC and Bitcoin SV, before and after the recent ``hash war"  between those two coins. 
The analysis results of these case studies are presented in Section~\ref{sec:application} and \ref{sec:data}.

\smallskip
\noindent\textbf{Generalization.}
Moreover, we remark that our analysis can be generalized to any circumstance wherein two coins have compatible PoW mechanisms with each other. 
\blue{We believe that the generalized results bring new important angles in competitive coin markets; a coin can attempt to steal loyal miners from other rivalry coins that have compatible PoW mechanisms.}
In Section~\ref{sec:attack}, a risk of automatic mining and the way to intentionally reduce the number of loyal miners for other coins are described.
Then, in Section~\ref{sec:discuss}, we discuss countermeasures and environmental factors that may make the actual coin states deviate from our game analysis.

In summary, our main contributions are as follows: 

\begin{enumerate}
\item To analyze the economics of fickle mining, we first model a game between two coins, considering some coin factions that stick to mining their own coin.

\item We analyze Nash equilibria and dynamics in the game and find two types of equilibria: 1) stable coexistence of two coins and 2) a lack of loyal miners to the less valued coin.
Then, we apply this game to the Bitcoin system.

\item To determine if real-world miners' behaviors follow our model, 
we investigate the mining power history in the Bitcoin system. 
Then we show that the state reached the lack of BCH loyal miners until Nov.\ 13, 2017, and we confirm that this fact periodically led the BCH system to be centralized and insecure. 
Moreover, for generalization, we also analyze the recent ``hash war'' situation between Bitcoin ABC and Bitcoin SV according to our game model.

\item We introduce a risk of automatic mining and predict that the current BCH loyal miners would leave when 5\% of the total mining power in BTC and BCH involves automatic mining.

\item Finally, our game is generalized to any mining-compatible coins (e.g. Ethereum vs. Ethereum Classic).
Therefore, our study brings a threat that one coin can intentionally steal loyal miners from other less valued coin.
\end{enumerate}

\section{Preliminary}
\label{sec:preliminary}

\subsection{Cryptocurrency}

Many cryptocurrencies such as Bitcoin, Ethereum, and Litecoin adopt the PoW mechanism as a consensus algorithm. 
In the PoW mechanism, when a node solves a cryptographic puzzle, the node can generate and propagate a valid block. 
Then other nodes append the generated block to the existing blockchain. 
The puzzle is to find an inverse image of a hash function satisfying the certain condition, and thus the node should spend computational power to solve the cryptographic puzzle.
The process of generating a block is called \textit{mining}, and nodes participating in mining are called \textit{miners}. 
In systems, the mining difficulty is adjusted to maintain the average time of generating one block.
In particular, Bitcoin mining difficulty is adjusted to keep the average period of generating one block at 10 minutes.
In addition, to incentivize mining, whenever a miner finds a valid block, the miner earns the reward for one block in compensation for the computational power spent. 
For example, currently, miners earn the block reward of 12.5 coins in the Bitcoin system when they find one block.

Many people have become involved in mining because of the incentive for mining, and specialized hardware for efficient mining such as application-specific integrated circuits (ASICs) has appeared.
Based on the above reasons, the vast computational power is used for mining, and mining difficulty has increased significantly. 
Therefore, it should take a \textit{solo miner}, who mines alone, a significantly long time to find a valid block, and this causes solo miners to wait for a long time to earn block rewards.
To reduce not only node costs and but also the variance of their rewards, \textit{mining pools} where miners gather together for mining have been organized.
Most pools are composed of workers and a manager.
The manager gives puzzles to workers, and they solve the puzzles.
If a worker solves a given puzzle, the block reward is distributed to the workers in the pool.

In the past years, there have been many attacks on and problems with cryptocurrency systems, and these attacks or problems have even caused cryptocurrency systems to split.
For example, because Bitcoin has become a popular cryptocurrency, the system needs to provide high transaction throughput.
To address the scalability issue, several solutions such as Segregated Witness~\cite{segwit} and unlimited block size have been proposed. 
Because of the debate on the proposed solutions, Bitcoin was eventually split into BTC and BCH in early Aug. 2017. 
Even though BCH chose to increase the block size limit in order to allow more transactions per block, the mining protocol of BCH was designed to be compatible with that of BTC. 
Therefore, miners can conduct both BTC and BCH mining with one hardware device. 

\subsection{Fickle mining}

Before Nov. 13, 2017, BCH adjusted the mining difficulty every 2016 block to ensure that the average time period for generating a block is 10 minutes, like in the case of BTC.
In doing so, if the time required for generating past 2016 blocks is longer than two weeks, 
the mining difficulty decreases, and miners can generate subsequent blocks more easily.
In addition, BCH added a new difficulty adjustment algorithm called 
emergency difficulty adjustment (EDA)~\cite{eda} to decrease the mining difficulty without waiting for 2016 blocks to be generated when it is significantly difficult to find a valid block. 

Because BTC and BCH have a PoW mechanism compatible with each other, miners can freely switch between them depending on the mining difficulty and the coin price. 
However, because the change in coin price is hard to predict, some miners immediately change their coin only when mining difficulty changes, where we call this behavior \emph{fickle mining}. 
Concretely, 
the fickle miners first conduct BTC mining, observing the changes in the mining difficulties of BTC and BCH. 
Then, if the BCH mining difficulty is low, they immediately shift to BCH mining. 
When the BCH mining difficulty increases again thanks to its difficulty adjustment algorithm, fickle miners immediately shift to BTC mining. 
Fickle mining can boost profits of miners; however, this behavior might cause instability of both BTC and BCH. 

\begin{figure}[ht]
\centering
\includegraphics[width=\columnwidth]{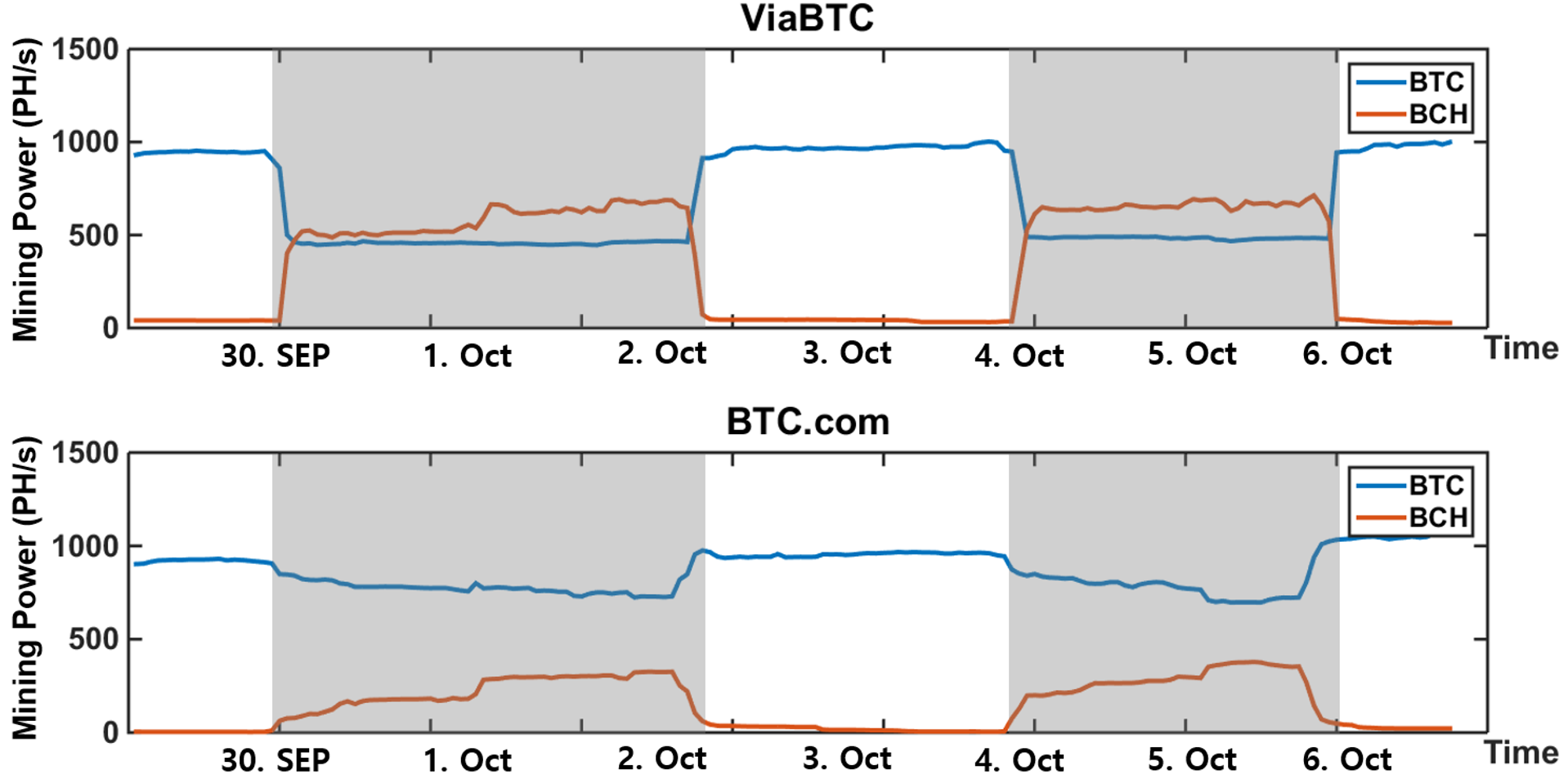}
\caption{
Mining power history of ViaBTC and BTC.com (Sep. 29, 2017 $\sim$ Oct. 6, 2017). 
The grey regions represent movements of mining power from BTC to BCH.}
\label{fig:mining pattern}
\end{figure}

This mining behavior was easily observed in Bitcoin when we monitored the mining power in pools. 
We collected mining power history data over the course of a week from two popular pools: ViaBTC~\cite{viabtc} and BTC.com~\cite{btc.com}. 
These two pools support both BTC and BCH mining; miners in the pools can choose either BTC or BCH mining by just clicking one button.
Figure~\ref{fig:mining pattern} represents the mining power data of ViaBTC and BTC.com for a week. In the figure, the grey regions show movements of mining power from BTC to BCH mining.

\begin{figure}[ht]
\centering
\includegraphics[width=\columnwidth]{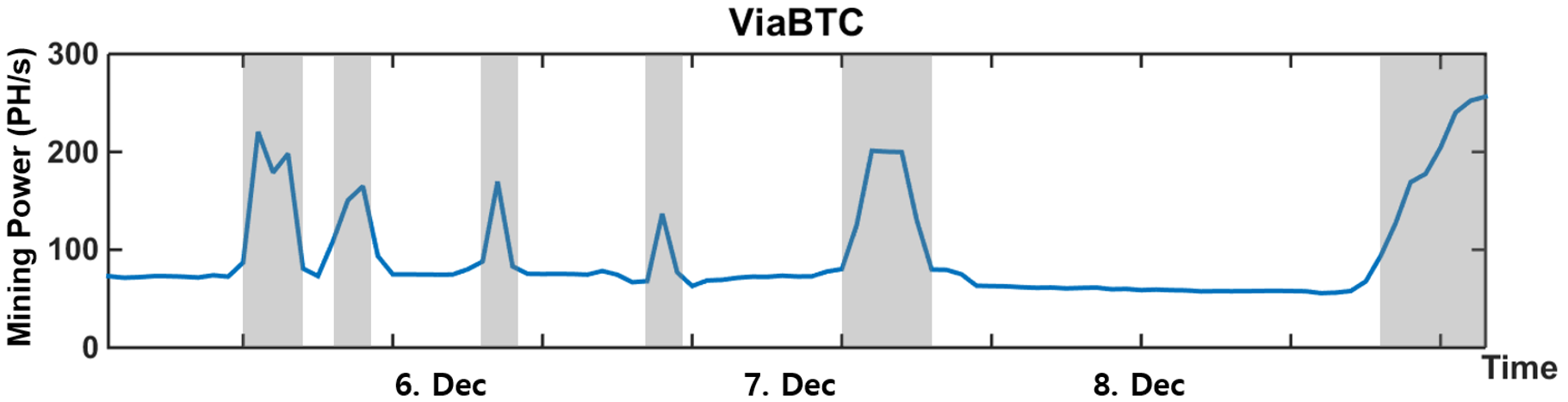}
\caption{
Mining power history of ViaBTC (Dec. 5, 2017 $\sim$ Dec. 8, 2017).
Grey regions represent movements of mining power from BTC to BCH. Note that we only displayed the mining power history of ViaBTC because BTC.com did not evidently execute fickle mining for this period.}
\label{fig:mining pattern2}
\end{figure}

As fickle mining causes a sudden increase in mining power as shown in the grey zones of Figure~\ref{fig:mining pattern}, many blocks were generated quite quickly in the BCH system. 
For example, in the BCH system, 2016 blocks were generated within only three days in each grey zone. 
This caused the blockchain of BCH to be thousands of blocks ahead of BTC, and the halving time of the block reward in BCH was brought forward.
To address this issue, BCH performed another hard fork on Nov. 13, 2017~\cite{hardfork}. 
Currently, BCH adjusts the difficulty for \textit{each} block based on the previous 144 blocks as a moving window~\cite{newdaa}.
To determine if it is possible that miners conduct fickle mining even after the hard fork of Nov. 13, 2017, 
we investigated the BCH mining power data of ViaBTC for four days (Dec. 5, 2017 $\sim$ Dec. 8, 2017). 
Figure~\ref{fig:mining pattern2} represents the BCH mining power data of ViaBTC during this time period; as is evident from the figure, some miners still conduct fickle mining. 
Because the BCH mining difficulty is more quickly adjusted than before the hard fork of BCH, fickle miners should switch their mining power more quickly than before the hard fork. 
Indeed, fickle mining can occur in any mining difficulty adjustment algorithm.
\section{Related work}
\label{sec:related}

In this section, we review previous studies related to mining in PoW systems. Kroll et al. considered the Bitcoin mining process as a game among multiple 
players~\cite{kroll2013economics} and showed that a miner possessing 51\% mining power can be motivated to disrupt the Bitcoin system.
Several works~\cite{johnson2014game, laszka2015bitcoin} modeled and analyzed a game between two pools that can launch denial of service attacks against each other.
Eyal and Sirer introduced the selfish mining strategy, where a malicious miner successfully mines blocks but does not immediately broadcast the blocks; instead, the attacker temporarily withholds the block~\cite{eyal2014majority}. 
Many researchers have intensively studied ways to optimize and extend selfish mining~\cite{sapirshtein2016optimal,nayak2016stubborn, gervais2016security, zhang2017necessity}.
Bonneau introduced bribery attacks as a way for an attacker to increase her mining power~\cite{B16a}.
Lewenberg et al. considered a mechanism of sharing rewards among pool miners as a cooperative game~\cite{lewenberg2015bitcoin}.
In 2015, Eyal modeled a game between two pools that execute block withholding (BWH) attacks~\cite{eyal2015miner}.
As a concurrent work, Luu et al.~\cite{luu2015power} modeled a power splitting game to find an optimized strategy for a BWH attacker.
Kwon et al.~\cite{kwon2017selfish} proposed a new attack called a fork after withholding (FAW) attack against pools~\cite{kwon2017selfish}. 
Also, several works~\cite{carlsten2016instability, tsabary2018gap} analyzed a transaction-fee regime in PoW systems, where miners receive incentives for mining as transaction fees.
Moreover, because many cryptocurrencies are competing with each other, there can be another incentive to execute 51\% attacks. 
Considering this fact, Bonneau revisited the 51\% attack with some basic analysis~\cite{bonneau2018hostile}.

Recently, Ma et al.~\cite{ma2018market} considered a mining game of multiple miners and concluded that openness of the Bitcoin system causes the need for vast mining power.
Another study~\cite{prat2018equilibrium} examined the relation between the Bitcoin/USD exchange rate and Bitcoin mining power.
They first proposed an industry equilibrium model to forecast the mining power depending on the Bitcoin/USD exchange rate.
Then, they showed that the real mining power data and simulated mining power according to their model are similar.
Our study focuses on the relation between two coins that have compatible PoW mechanisms with each other and the miners' behavior between two coins.
Furthermore, our model can be used to forecast the ratio of mining power between two coins. 
To the best of our knowledge, this is the first to study the effects of fickle mining.
\section{Model}
\label{sec:model}

In this section, we formally model a game to represent fickle mining between two coins. 

\subsection{Notation and assumptions}
\label{subsec:game model}

We consider two coins, $coin_{\tt A}$ and $coin_{\tt B}$, which have compatible PoW mechanisms with each other. 
In this case, a miner with a hardware device can alternately conduct mining of $coin_{\tt A}$ and $coin_{\tt B}$; 
that is, he can conduct fickle mining between them. 
Meanwhile, a $coin_{\tt B}$-faction can stick to $coin_{\tt B}$-mining rather than fickle mining or $coin_{\tt A}$-mining to maintain its own coin, and the set of $coin_{\tt B}$-factions sticking to $coin_{\tt B}$-mining is denoted by $\Omega_{\tt stick}$. 
For example, in the case where BCH is $coin_{\tt B}$, BITMAIN~\cite{bitmain}, one of the main supporters of BCH, may belong to $\Omega_{\tt stick}$. 
We aim to formalize a game considering the fickle mining and $\Omega_{\tt stick}$. 

The proposed game consists of many players (i.e., miners), where the set of all players is denoted by $\Omega.$
Player $i\in \Omega$ chooses one of three strategies, $s_i \in \{\mathcal{F},\mathcal{A},\mathcal{B}\}$: 
Fickle mining (\F), $coin_{\tt A}$-only mining (\A), and $coin_{\tt B}$-only mining (\B). 
The payoff function of player $i$ is denoted by $U_i : \{\mathcal{F},\mathcal{A},\mathcal{B}\}^n \rightarrow \mathbb{R},$
which we will formally define later as well as fickle mining. 
We also define three sets \FM$=\{i\in\Omega\,|s_i=\mathcal{F}\}$, 
\AM$=\{i\in\Omega\,|s_i=\mathcal{A}\}$, and \BM$=\{i\in\Omega\,|s_i=\mathcal{B}\}$, indicating a set of players who conduct fickle mining, $coin_{\tt A}$-only mining, and $coin_{\tt B}$-only mining, respectively. 
Note that $\Omega_{\tt stick}$ is a subset of \BM because players in $\Omega_{\tt stick}$ always choose strategy \B.
The sum of mining powers in $coin_{\tt A}$ and $coin_{\tt B}$ is regarded as 1; 
mining power of a coin is expressed as a ratio to the total mining power.
The mining power possessed by player $i$ is denoted by $c_i,$ and the total computational power possessed by $\Omega_{\tt stick}$ is denoted by $c_{\tt stick}.$ 
We also define $c_{\tt max}$ as the maximum of $\{c_i\,|\,i\in \Omega\backslash \Omega_{\tt stick}\}.$ 
Moreover, because our game analysis result would depend on the computational power possessed by players, we use the notation $\mathcal{G}(\bm{c}, c_{\tt stick})$ to refer to the game, where $\bm{c}$ indicates a vector of computational power possessed by players except for $\Omega_{\tt stick}$ 
(i.e., $\bm{c}=(c_i)_{i\in \Omega\backslash \Omega_{\tt stick}}$). 
Lastly, we denote the total mining power of \FM, \AM, and \BM as $r_{\mathcal F}$ 
(i.e., $\sum_{i\in\mathcal{M}_{\mathcal F}}{c_i}$), 
$r_{\mathcal A}$ (i.e., $\sum_{i\in\mathcal{M}_{\mathcal A}}{c_i}$), 
and $r_{\mathcal B}$ (i.e., $\sum_{i\in\mathcal{M}_{\mathcal B}}{c_i}$), 
respectively.
Observe that $r_{\mathcal A}=1-r_{\mathcal F}-r_{\mathcal B}$ and $c_{\tt stick} \leq r_{\mathcal B}.$
Namely, $(r_{\mathcal F},r_{\mathcal B})$ represents the full status of mining powers where $r_{\mathcal{B}}$ is not less than $c_{\tt stick}$. 

For the analysis of the game, we assume the following:

\begin{assumption}
A miner conducts either only $coin_{\tt A}$ or $coin_{\tt B}$-mining (not both) at each time instance; for example,
an ASIC miner cannot execute both BTC and BCH mining simultaneously.
However, their choices can be time-varying; that is, miners can change their coin to mine.
\end{assumption}

\smallskip
\begin{assumption}\label{ass:k}
The price of 1 $coin_{\tt B}$ is equal to that of $k$ $coin_{\tt A}$. 
We assume that $0<k\leq 1$ without loss of generality.
In addition, rewards for mining a block in both coins are 1 $coin_{\tt A}$ and 1 $coin_{\tt B}$, respectively.
\end{assumption}

\smallskip
\begin{assumption}\label{ass:c}
In both $coin_{\tt A}$ and $coin_{\tt B}$ systems, mining difficulties are adjusted to maintain the average period of generating a block as the same specific time period, which we denote by 1 $P_{\tt{ag}}$ time and regard as \textit{a time unit}; 
for example, 1 $P_{\tt{ag}}$ = 10 minutes in the Bitcoin system. 
Furthermore, we consider a generalized model in which mining difficulties of $coin_{\tt A}$ and $coin_{\tt B}$ are adjusted in proportion to the mining power for the previous time window, and we consider a normalized difficulty.
Thus, if $x$ mining power has been engaged in coin mining, the mining difficulty would be $x.$
More precisely, in our model, the coin mining difficulty 
decreases and increases again, considering the generation time of a specific number of blocks since the last update of coin mining difficulty. 
In particular, for the mining difficulty of $coin_{\tt B},$ we denote the number of considered blocks when the $coin_{\tt B}$-mining difficulty decreases and increases as $N_{\tt de}$ and $N_{\tt in}$, respectively.\footnote{In Section~\ref{sec:application}, we will show that our results can be applied to the coin system regardless of the mining difficulty adjustment algorithm of $coin_{\tt B}$.}
Note that $N_{\tt de}$ and $N_{\tt in}$ cannot be zero.
In the case of BTC and Litecoin, $N_{\tt de}$ and $N_{\tt in}$ are 2016. 
\end{assumption}

\smallskip
As described previously, a fickle miner may change the preferred coin when the coin mining difficulty changes.
Here we define fickle mining formally. 

\smallskip
\begin{definition}[Fickle mining]
Let $D_{\tt A}$ and $D_{\tt B}$ denote the $coin_{\tt A}$ and $coin_{\tt B}$-mining difficulties, respectively.
If $D_{\tt B}< \min\{r_{\mathcal F}+r_{\mathcal B}, k\cdot D_{\tt A}\}$ or $D_{\tt B}\leq r_{\mathcal B}$ when $D_{\tt A}$ or $D_{\tt B}$ is updated, 
fickle miners (\FM) decide to conduct $coin_{\tt B}$-mining until $D_{\tt A}$ or $D_{\tt B}$ is adjusted again. Otherwise, they conduct $coin_{\tt A}$-mining.
\label{def:fickle}
\end{definition}

\smallskip\noindent 
We also emphasize that 
if $r_{\mathcal F}$ is 0, no miner engages in fickle mining, and mining powers of $coin_{\tt A}$ and $coin_{\tt B}$ are stably maintained. 
\blue{
On the other hand, if $r_{\mathcal B}$ is $c_{\tt stick}$, only $coin_{\tt B}$-factions $\Omega_{\tt stick}$ would conduct $coin_{\tt B}$-mining after an increase in the mining difficulty of $coin_{\tt B}.$
In other words, in this case, only the factions remain as \textit{loyal miners} for $coin_{\tt B}.$ 
Therefore, if the number of such factions ($|\Omega_{\tt stick}|$) is small, the state would be a lack of loyal miners. 
Note that loyal miners refer to players who continue to conduct $coin_{\tt B}$-mining even after an increase in $coin_{\tt B}$-mining. 
In particular, if all $coin_{\tt B}$-factions stop $coin_{\tt B}$-mining for higher payoff (i.e., $|\Omega_{\tt stick}|=0$), $r_{\mathcal B}$ is 0, and no player conducts $coin_{\tt B}$-mining after an increase in the mining difficulty of $coin_{\tt B}.$} 
Note that the $coin_{\tt B}$-mining difficulty cannot decrease in this case because $N_{\tt de}$ cannot be zero.
Therefore, the case $r_{\mathcal B}=0$ indicates the complete downfall of $coin_{\tt B}$ while only $coin_{\tt A}$ survives.  

Parameters used in this paper are summarized in Table~\ref{tab:par}. 
The last parameter in the table will be introduced later. 

\begin{figure}[t]
\centering
\includegraphics[width=\columnwidth]{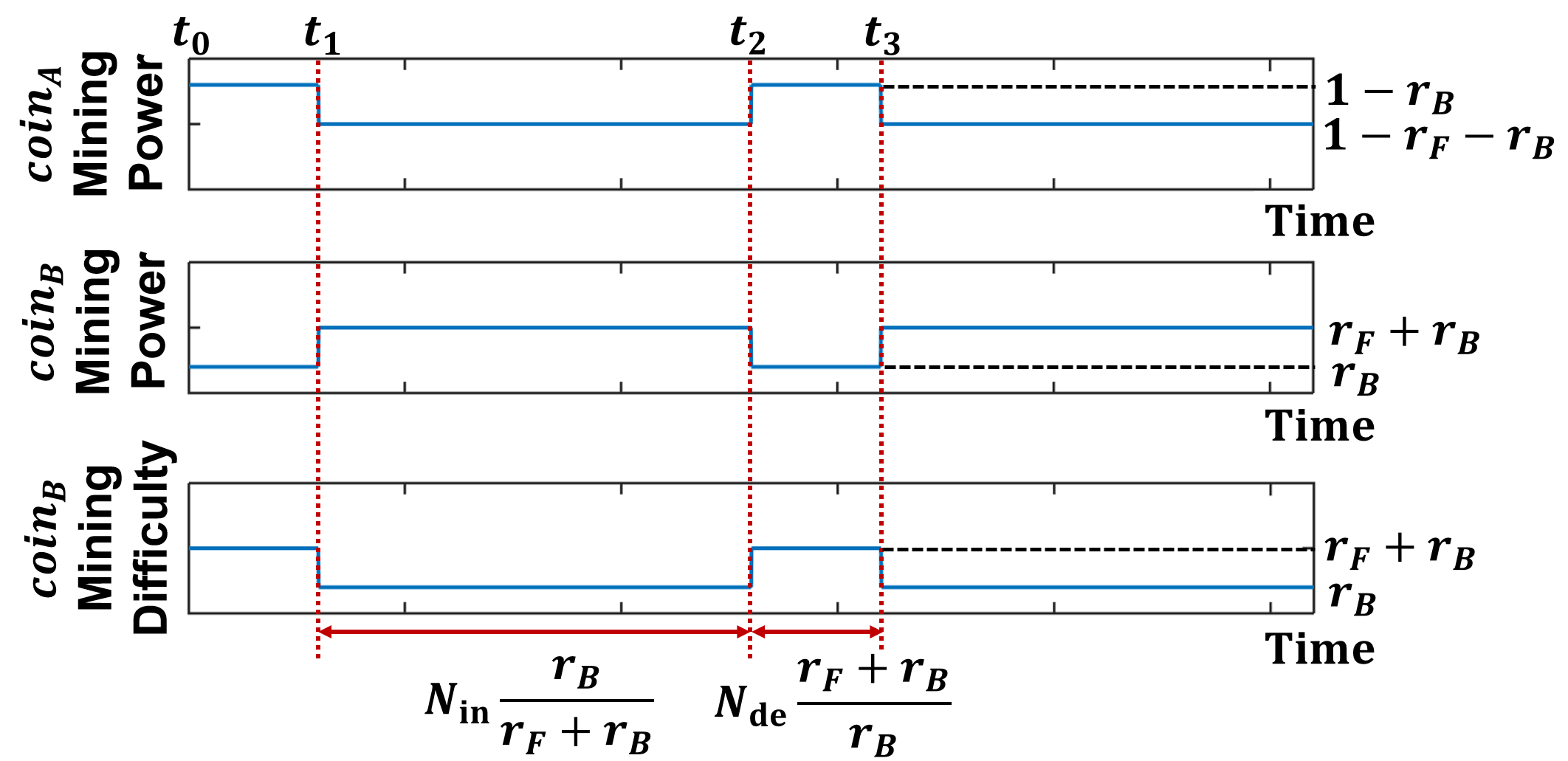}
\caption{Changes in the mining power of $coin_{\tt A}$ and $coin_{\tt B}$, and mining difficulty of $coin_{\tt B}$.}
\label{fig:analysis}
\end{figure}

\begin{table}[t] 
\renewcommand{\tabcolsep}{1.5pt}
\centering
\caption{List of parameters.}
\label{tab:par}
\begin{tabular}{|l||l|l|l|l|}
\hline
\parbox[c][0.35cm][c]{2.1cm}{\centering \blue{$\Omega_{\tt stick}$}} & \multicolumn{4}{l|}{\makecell{The set of $coin_{\tt B}$-factions sticking to $coin_B$ \\ mining to maintain their own coin}} \\ \hline
\parbox[c][0.35cm][c]{2.1cm}{\centering \blue{$\Omega$}} & \multicolumn{4}{l|}{\makecell{The set of all players}} \\ \hline
\parbox[c][0.35cm][c]{2.1cm}{\centering $s_i$} & \multicolumn{4}{l|}{\makecell{Player $i$'s strategy}} \\ \hline
\parbox[c][0.35cm][c]{2.1cm}{\centering $U_i$} & \multicolumn{4}{l|}{\makecell{Player $i$'s payoff}} \\ \hline
\parbox[c][0.35cm][c]{2.1cm}{\centering \F, \A, \B} & \multicolumn{4}{l|}{\makecell{Fickle, $coin_{\tt A}$-only, $coin_{\tt B}$-only mining}} \\ \hline
\parbox[c][0.35cm][c]{2.1cm}{\centering \FM, \AM, \BM} & \multicolumn{4}{l|}{\makecell{The set of players with \F, \A, \B}} \\ \hline
\parbox[c][0.35cm][c]{2.1cm}{\centering \blue{$c_i$}} & \multicolumn{4}{l|}{\makecell{Computational power of player $i$}} \\ \hline
\parbox[c][0.35cm][c]{2.1cm}{\centering \blue{$c_{\tt stick}$}} & \multicolumn{4}{l|}{\makecell{Computational power possessed by $\Omega_{\tt stick}$}} \\ \hline
\parbox[c][0.35cm][c]{2.1cm}{\centering \blue{$c_{\tt max}$}} & \multicolumn{4}{l|}{\makecell{The maximum of $\{c_i\,|\,i\in \Omega\backslash \Omega_{\tt stick}\}$}} \\ \hline
\parbox[c][0.35cm][c]{2.1cm}{\centering $\bm{c}$} & \multicolumn{4}{l|}{\makecell{The vector of computational power \\ possessed  by players in $\Omega\backslash \Omega_{\tt stick}$}} \\ \hline
\parbox[c][0.35cm][c]{2.1cm}{\centering \blue{$\mathcal{G}(\bm{c}, c_{\tt stick})$}} & \multicolumn{4}{l|}{\makecell{The game of players and $\Omega_{\tt stick}$ with \\ computational power $\bm{c}$ and $c_{\tt stick}$}} \\ \hline
\parbox[c][0.35cm][c]{2.1cm}{\centering $r_{\mathcal{F}}, r_{\mathcal{A}}, r_{\mathcal{B}}$} & \multicolumn{4}{l|}{\makecell{The total computational power \\ fraction of \FM, \AM, \BM}} \\ \hline
\parbox[c][0.35cm][c]{2.1cm}{\centering $k$} & \multicolumn{4}{l|}{\makecell{The relative price of $coin_{\tt B}$ to $coin_{\tt A}$}} \\ \hline
\parbox[c][0.7cm][c]{2.1cm}{\centering $P_{\tt ag}$} & \multicolumn{4}{l|}{\makecell{The time unit representing the average \\period of generating one block}} \\ \hline
\parbox[c][0.7cm][c]{2.1cm}{\centering $N_{\tt de}, N_{\tt in}$} & \multicolumn{4}{l|}{\makecell{The number of considered past blocks when the \\mining difficulty of $coin_{\tt B}$ decreases or increases}} \\ \hline
\parbox[c][0.35cm][c]{2.1cm}{\centering $D_{\tt A}, D_{\tt B}$} & \multicolumn{4}{l|}{\makecell{The mining difficulty of $coin_{\tt A}$, $coin_{\tt B}$}} \\ \hline
\parbox[c][0.35cm][c]{2.1cm}{\centering \blue{$\mathcal{E}(\bm{c}, c_{\tt stick})$}} & \multicolumn{4}{l|}{\makecell{The set of all Nash equilibrium in $\mathcal{G}(\bm{c}, c_{\tt stick})$}} \\ \hline
\end{tabular}
\end{table}

\smallskip
\noindent \textbf{Illustration of fickle mining.}
Figure~\ref{fig:analysis} illustrates
a stream of mining power in $coin_{\tt A}$ and $coin_{\tt B}$, as well as the mining difficulty of $coin_{\tt B}$ over time, caused by the strategies of players. 
\\ 
\noindent - Time $t_0~$: At the beginning, $1-r_{\mathcal{B}}$ and $r_{\mathcal B}$ mining powers are used for $coin_{\tt A}$ and $coin_{\tt B}$-mining, respectively. \\ 
\noindent - Time $t_1~$: The mining difficulty of $coin_{\tt B}$ decreases because it is relatively difficult to find PoWs with $r_{\mathcal B}$ mining power. 
At the moment, \FM shifts from $coin_{\tt A}$ to $coin_{\tt B}$, and each of $1-r_{\mathcal F}-r_{\mathcal B}$ and $r_{\mathcal F}+r_{\mathcal B}$ mining powers is used for $coin_{\tt A}$ and $coin_{\tt B}$-mining, respectively. \\
\noindent - Time $t_2~$: Because the mining difficulty of $coin_{\tt B}$ is again adjusted (increases) after $N_{\tt in}$ blocks are found in the $coin_{\tt B}$ system since the last adjustment of the mining difficulty of $coin_{\tt B}$,
the mining difficulty of $coin_{\tt B}$ would increase after $\frac{N_{\tt in}r_{\mathcal B}}{r_{\mathcal F}+r_{\mathcal B}}$ $P_{\tt{ag}}$ time since it takes $\frac{r_{\mathcal B}}{r_{\mathcal F}+r_{\mathcal B}}$ $P_{\tt{ag}}$ to find one valid block on average.
Then, \FM shifts again from $coin_{\tt B}$ to $coin_{\tt A}$ and conducts $coin_{\tt A}$-mining until the mining difficulty of $coin_{\tt B}$ decreases.  \\ 
\noindent - Time $t_3~$: Until when the mining difficulty of $coin_{\tt B}$ decreases after $N_{\tt de}$ blocks are found in the $coin_{\tt B}$ system, 
\FM would conduct $coin_{\tt A}$-mining (for $\frac{N_{\tt de}(r_{\mathcal F}+r_{\mathcal B})}{r_{\mathcal B}}$ $P_{\tt{ag}}$ time). \\
\noindent - This process is continually repeated.

\subsection{Payoff function}

Next, we describe payoff functions for our game model.
All payoffs are expressed as a unit of $coin_{\tt A}$ and are calculated as a profit density, which is defined as an average earned reward for $1~ P_{\tt{ag}}$ time divided by the player's mining power.
In other words, if player $i$ earns a reward $R$ for 1~$P_{\tt{ag}}$ time on average, the payoff would be $\frac{R}{c_i}.$ 
Player $i$'s payoff function $U_i(s_i, \mathbf{s_{-i}})$ is expressed as follows: 
\begin{equation}
U_i(s_i, \mathbf{s_{-i}})=
\begin{cases}
& U_{\mathcal{F}}(r_{\mathcal F},r_{\mathcal B})
\text{  if $s_i=\mathcal{F}$}\\
&U_{\mathcal{A}}(r_{\mathcal F},r_{\mathcal B})
\text{  if $s_i=\mathcal{A}$}\\
&U_{\mathcal{B}}(r_{\mathcal F},r_{\mathcal B}) \text{  if $s_i=\mathcal{B}$}
\end{cases}
\label{eq:payoff}
\end{equation}
where $\mathbf{s_{-i}}$ indicates other players' strategies.
Here, it suffices to define 
$U_\mathcal{F}, U_\mathcal{A}, U_\mathcal{B}$ 
in the range $0< r_{\mathcal F}\leq 1,$ $0< r_{\mathcal A} \leq 1$, and $0< r_{\mathcal B} \leq 1,$ respectively;
for example,
$U_\mathcal{F}$ would be defined when $s_i=\mathcal F$ (i.e, a fickle miner exists, and $0< r_{\mathcal F}$). 

First, we define the payoff $U_{\mathcal F}$ for a player in \FM.
As shown in  Figure~\ref{fig:analysis}, \FM conducts $coin_{\tt B}$-mining for $\frac{N_{\tt in}r_{\mathcal B}}{r_{\mathcal F}+r_{\mathcal B}}$~$P_{\tt{ag}}$ time.
\blue{Therefore, a player in \FM earns the profit $\frac{k\cdot c_i}{r_{\mathcal B}}$ per 1~$P_{\tt{ag}}$ time on average for $\frac{N_{\tt in}r_{\mathcal B}}{r_{\mathcal F}+r_{\mathcal B}}$~$P_{\tt{ag}}$ time.
After that, \FM conducts $coin_{\tt A}$-mining for $\frac{N_{\tt de}(r_{\mathcal F}+r_{\mathcal B})}{r_{\mathcal B}}$ $P_{\tt{ag}}$ time during which a player in \FM earns the following profit per 1 $P_{\tt{ag}}$ time on average:
\begin{equation}
\resizebox{.9\hsize}{!}{$
\mbox{AP}_{\mathcal F} :=
c_i \frac{\frac{N_{\tt in}r_{\mathcal B}}{r_{\mathcal F}+r_{\mathcal B}}+\frac{N_{\tt de}(r_{\mathcal F}+r_{\mathcal B})}{r_{\mathcal B}}}{(1-r_{\mathcal F}-r_{\mathcal B}) \frac{N_{\tt in}r_{\mathcal B}}{r_{\mathcal F}+r_{\mathcal B}}+(1-r_{\mathcal B}) \frac{N_{\tt de}(r_{\mathcal F}+r_{\mathcal B})}{r_{\mathcal B}}}.
$}
\label{eq:pd1}
\end{equation}}
\noindent 
The above formulation is due to the fact that
mining powers $1-r_{\mathcal F}-r_{\mathcal B}$ and $1-r_{\mathcal B}$ engage in $coin_{\tt A}$-mining for $\frac{N_{\tt in}r_{\mathcal B}}{r_{\mathcal F}+r_{\mathcal B}}$~$P_{\tt{ag}}$ and $\frac{N_{\tt de}(r_{\mathcal F}+r_{\mathcal B})}{r_{\mathcal B}}$ $P_{\tt{ag}}$ times, respectively, and thus, the second factor in the right-hand side of \eqref{eq:pd1} represents an inverse number of the mining difficulty of $coin_{\tt A}$. 
\blue{Consequently, the payoff of a player in \FM can be expressed as

\begin{equation*}
\resizebox{\hsize}{!}{$
U_{\mathcal{F}}(r_{\mathcal F},r_{\mathcal B}) =\left (\frac{k\cdot c_i}{r_{\mathcal B}}\cdot\frac{N_{\tt in}r_{\mathcal B}}{r_{\mathcal F}+r_{\mathcal B}}+\mbox{AP}_{\mathcal F} \times\frac{N_{\tt de}(r_{\mathcal F}+r_{\mathcal B})}{r_{\mathcal B}}\right)\times Z,$}
\end{equation*}
where 
\begin{equation*}
    Z=\frac{1}{c_i\left(\frac{N_{\tt in}r_{\mathcal B}}{r_{\mathcal F}+r_{\mathcal B}}+\frac{N_{\tt de}(r_{\mathcal F}+r_{\mathcal B})}{r_{\mathcal B}}\right )}.
\end{equation*}

Next, we provide payoffs $U_{\mathcal A}$ and $U_{\mathcal B}$ as follows:
\begin{align*}
U_{\mathcal{A}}(r_{\mathcal F},r_{\mathcal B}) =& \frac{\mbox{AP}_{\mathcal F}}{c_i},\\
U_{\mathcal{B}}(r_{\mathcal F},r_{\mathcal B}) =&
\left(\frac{kN_{\tt in}}{r_{\mathcal F}+r_{\mathcal B}}+\frac{kN_{\tt de}}{r_{\mathcal B}}\right)\times c_i\cdot Z,
\end{align*}
where we observe that 
a player in \BM earns the profit $\frac{k\cdot c_i}{r_{\mathcal B}}$ per 1 $P_{\tt{ag}}$ for $\frac{N_{\tt in}r_{\mathcal B}}{r_{\mathcal F}+r_{\mathcal B}}$ $P_{\tt{ag}}$ time and profit $\frac{k\cdot c_i}{r_{\mathcal F}+r_{\mathcal B}}$ per 1 $P_{\tt{ag}}$ for $\frac{N_{\tt de}(r_{\mathcal F}+r_{\mathcal B})}{r_{\mathcal B}}$ $P_{\tt{ag}}$ time, on average.}
\section{Game analysis}
\label{sec:analysis}

In this section, we analyze Nash equilibria and dynamics in game $\mathcal{G}(\bm{c}, c_{\tt stick}).$  

\subsection{Equilibrium in game $\mathcal{G}(\bm{c}, c_{\tt stick})$} \label{sec:finite}


\noindent \textbf{Characterization of equilibria.}
Before finding Nash equilibria of $\mathcal{G}(\bm{c}, c_{\tt stick}),$ 
we define a pure Nash equilibrium.

\smallskip
\begin{definition}[Pure Nash equilibrium]
A strategy vector $\mathbf{s}=(s_1, s_2, \cdots s_n)$ is a Nash equilibrium if
$$U_i(\mathbf{s})= \max_{s_i^\prime\in \{\mathcal{F}, \mathcal{A}, \mathcal{B}\}}
U_i(s_i^\prime, \mathbf{s_{-i}}), \qquad\mbox{for all $i$}.$$
\end{definition}

\smallskip
\noindent At an equilibrium, all rational players would not change their strategy, that is,
$r_{\mathcal F}$ and $r_{\mathcal B}$ are not updated.
We map a strategy vector $\mathbf{s}=(s_1, s_2, \cdots s_n)$ to state $(r_{\mathcal F},r_{\mathcal B})$ and denote by $\mathcal E(\bm{c}, c_{\tt stick})$ the set of all Nash equilibria in $\mathcal{G}(\bm{c}, c_{\tt stick}).$ 
We first determine the dynamics of player $i$ with small $c_i$ through Lemma~\ref{lem:dynamics} to establish the characterization of $\mathcal E(\bm{c}, c_{\tt stick}).$ 


\blue{\smallskip
\begin{lemma}
There is $\varepsilon>0$ such that, any player $i$ possessing $c_i<\varepsilon$ does not change its strategy at state $(r_{\mathcal F},r_{\mathcal B})$ if and only if 
\begin{equation*}
\resizebox{\hsize}{!}{$
    (r_{\mathcal F},r_{\mathcal B})=\begin{cases}
(f_{\varepsilon}(c_{\tt stick}), c_{\tt stick}) \quad\text{if} c_{\tt stick}>0,\\
(\frac{k}{2}+\frac{\sqrt{{N_{\tt de}}^2k^2+4N_{\tt de}N_{\tt in}(k\cdot c_i-c_i^2)}}{2N_{\tt de}}\leq r_{\mathcal F}\leq 1, 0) \,\,\text{otherwise,}
\end{cases}$}
\end{equation*}
where $f_{\varepsilon}$ is a decreasing function of which input is $c_{\tt stick}$ and output ranges between 0 and $1-c_{\tt stick}.$ 
Parameters $k, N_{\tt de},$ and $N_{\tt in}$ are defined in Assumption~\ref{ass:k} and \ref{ass:c}.
\label{lem:dynamics}
\end{lemma}}

\blue{
\smallskip
\noindent Note that $f_{\varepsilon}(c_{\tt stick})$ is $1-c_{\tt stick}$ for a small value of $c_{\tt stick}$ while $f_{\varepsilon}(c_{\tt stick})$ is 0 for a large value of $c_{\tt stick}.$
The above lemma implies that, considering miners with small computational power, if a Nash equilibrium exists, only $\Omega_{\tt stick}$ would remain as loyal miners to $coin_{\tt B}$ in the equilibrium. 
This is because $(r_{\mathcal F}, r_{\mathcal B})$ would continually change when $r_{\mathcal B}$ is greater than $c_{\tt stick}.$ 
From Lemma~\ref{lem:dynamics}, we can characterize the set $\mathcal E(\bm{c}, c_{\tt stick})$ as stated in Theorem~\ref{thm:eq}. 
We present the proof of Lemma~\ref{lem:dynamics} and Theorem~\ref{thm:eq} in Appendix~\ref{sec:pf_eq}.} 


\smallskip
\blue{\begin{theorem}
There is $\varepsilon>0$ such that, when $c_{\tt max}<\varepsilon,$ 
the set $\mathcal E(\bm{c}, c_{\tt stick})$ is as follows.  
\begin{equation*}
\resizebox{\hsize}{!}{$
\mathcal E(\bm{c}, c_{\tt stick})=
\begin{cases}
    \{(r_{\mathcal F}, r_{\mathcal B}): X \leq r_{\mathcal F}\leq 1, r_{\mathcal B}=0\} & \text{if} c_{\tt stick}=0, \\
    \{(1-c_{\tt stick}, c_{\tt stick})\} &\text{else if} c_{\tt stick}<x,\\
    \{(0, c_{\tt stick})\} &\text{else if} c_{\tt stick}>y, \\
\end{cases}$}
\end{equation*}
where $$X=\max_{i\in \Omega\backslash \Omega_{\tt stick}}\left\{\frac{k}{2}+\frac{\sqrt{{N_{\tt de}}^2k^2+4N_{\tt de}N_{\tt in}(k\cdot c_i-c_i^2)}}{2N_{\tt de}}\right\},$$
$x$ and $y$ $(>x)$ range between 0 and 1.
\label{thm:eq}
\end{theorem}}

\smallskip
\noindent 
\blue{
As described above, Theorem~\ref{thm:eq} shows that, in a game where players except for $\Omega_{\tt stick}$ possess small computational power, there exist only Nash equilibria where the $coin_{\tt B}$-factions sticking to $coin_{\tt B}$-mining are loyal miners for $coin_{\tt B}$. 
In the case where $c_{\tt stick}$ is small, we can certainly see that the overall health of the $coin_{\tt B}$ system would be weakened in terms of scalability, decentralization, and security, which will be discussed in more detail in Section~\ref{subsec:btc}. 
Indeed, even if $c_{\tt stick}$ is large, the case where $r_{\mathcal B}$ is equal to $c_{\tt stick}$ would make the $coin_{\tt B}$ system significantly centralized because only a few players possessing large power are loyal miners to $coin_{\tt B}$ (this example is presented in Section~\ref{subsec:stick}).  
In particular, if $\Omega_{\tt stick}$ is empty, no miner exists in the $coin_{\tt B}$ system in all Nash equilibria. 
Remark that this case indicates the complete downfall of $coin_B.$
As a result, Theorem~\ref{thm:eq} implies that fickle mining can be dangerous.} 

\blue{
\smallskip
\noindent\textbf{When players possess infinitesimal mining power.}
Under the game $\mathcal{G}(\bm{c}, c_{\tt stick})$, it is not easy to analyze movement of state $(r_{\mathcal F}, r_{\mathcal B})$ (this movement will be used for data analysis in Section~\ref{sec:data}) due to a large degree of freedom in $\bm{c}$. 
Thus, we further assume that players except for $\Omega_{\tt stick}$ (i.e., $\Omega\backslash \Omega_{\tt stick}$) possess infinitesimal computational power (i.e., $\norm{\bm{c}}_2\approx 0$).
We show that this assumption is reasonable by analyzing the real-world dataset in the Bitcoin system (see Section~\ref{sec:application}). 
We again study the equilibria of $\mathcal{G}(\bm{c}, c_{\tt stick})$ in this case.

\smallskip
\begin{theorem}\label{thm:inf_nash}
When players except for $\Omega_{\tt stick}$ possess infinitesimal mining power, the set $\mathcal{E}(\bm{c}, c_{\tt stick})$ is as follows.
\begin{align}\label{eq:nash_infty}
&\mathcal{E}(\bm{c}, c_{\tt stick})=\notag \\
&\begin{cases}
    \left\{\left(0,\frac{k}{k+1}\right)\right\} \, \cup\, \{(r_{\mathcal F}, r_{\mathcal B})\,:\, k\leq r_{\mathcal F} \leq 1, r_{\mathcal B}=0\} \\ \hspace{2cm}\text{if $c_{\tt stick}=0$ (Case 1),}\\
    \left\{\left(0,\frac{k}{k+1}\right)\right\} \,\cup\, \{(1-c_{\tt stick}, c_{\tt stick})\} \\ \hspace{2cm}\text{else if $c_{\tt stick}\leq \alpha$ (Case 2),}\\
    \left\{\left(0,\frac{k}{k+1}\right)\right\} \,\cup\, \{(\beta, c_{\tt stick})\} \\ \hspace{2cm}\text{else if $\alpha < c_{\tt stick}\leq \frac{k}{k+1}$ (Case 3),}\\
    \{(0, c_{\tt stick})\} \hspace{2.3mm} \text{otherwise (Case 4)}
\end{cases}
\end{align}

\noindent Here, $\alpha$ and $\beta$ are defined in Section~\ref{sec:infinite}.
\end{theorem}

\smallskip
\noindent
We present the proof of Theorem~\ref{thm:inf_nash} in Appendix~\ref{sec:pf_in}. 
Comparing with Theorem~\ref{thm:eq}, the state $(0,\frac{k}{k+1})$ also becomes another Nash equilibrium when the computational power possessed by players (except for $\Omega_{\tt stick}$) is infinitesimal. 
Note that this state indicates the stable coexistence of $coin_{\tt A}$ and $coin_{\tt B}.$
Indeed, when $\norm{\bm{c}}_2$ is closer to 0, the difference among payoffs of players in \FM, \AM, and \BM would also be closer to 0 at the state $(0,\frac{k}{k+1})$. 
Therefore, under the assumption that players possess infinitesimal power, payoffs of players in \FM, \AM, and \BM are the same at the state $(0,\frac{k}{k+1})$ while the mining difficulties of $coin_{\tt A}$ and $coin_{\tt B}$ are maintained as $\frac{1}{k+1}$ and $\frac{k}{k+1}$, respectively.
Meanwhile, at the remaining equilibria except for the state $(0,\frac{k}{k+1})$, only the $coin_{\tt B}$-factions $\Omega_{\tt stick}$ conduct $coin_{\tt B}$-mining after the $coin_{\tt B}$-mining difficulty increases. 
In particular, if no $coin_{\tt B}$-faction sticking to $coin_{\tt B}$-mining exists, loyal mining power to $coin_{\tt B}$ is 0 in the Nash equilibria.}
Note that, in this case, \FM and \AM would continuously conduct $coin_{\tt A}$-mining, because the mining difficulty of $coin_{\tt B}$ has not decreased after the previous increase in difficulty.
These players would not also change their strategy 
because the mining difficulty of $coin_{\tt B}$ increases to a significantly high value due to the heavy occurrence of fickle mining.

\smallskip
\noindent \textbf{Example.}
Considering the case $c_{\tt stick}=0,$ we give an example where $(r_{\mathcal F}=0.2,r_{\mathcal B}=0), k=0.3,$ and the initial mining difficulty of $coin_{\tt B}$ is 0.4.
The state $(0.2,0)$ is not a Nash equilibrium according to Theorem~\ref{thm:inf_nash}.
Because fickle miners continuously conduct the $coin_{\tt A}$-mining, the mining difficulty of $coin_{\tt A}$ is maintained as 1, and players in \FM and \AM earn the payoff of 1. 
If a player moves into \BM, the player would earn $\frac{0.3}{0.4}$ for a while in the beginning.
However, because the mining difficulty of $coin_{\tt B}$ decreases after \BM finds several blocks, the player who moves to \BM would eventually earn $\frac{0.3}{0.2}$ consistently.
Note that the time duration in which the mining difficulty of $coin_{\tt B}$ is close to 0 is negligible compared to the time duration in which the mining difficulty of $coin_{\tt B}$ is 0.2.
Therefore, the payoff of \BM is $\frac{0.3}{0.2},$ and rational players tend to move to \BM due to the higher payoff.
This means that the state $(0.2,0)$ is not a Nash equilibrium.

\subsection{Dynamics in game $\mathcal{G}(\bm{c}, c_{\tt stick})$} \label{sec:infinite}

In this section, we analyze dynamics in the game $\mathcal{G}(\bm{c}, c_{\tt stick})$ and study how a state can reach an equilibrium. 

\begin{figure}[t]
\centering
\includegraphics[width=0.65\columnwidth]{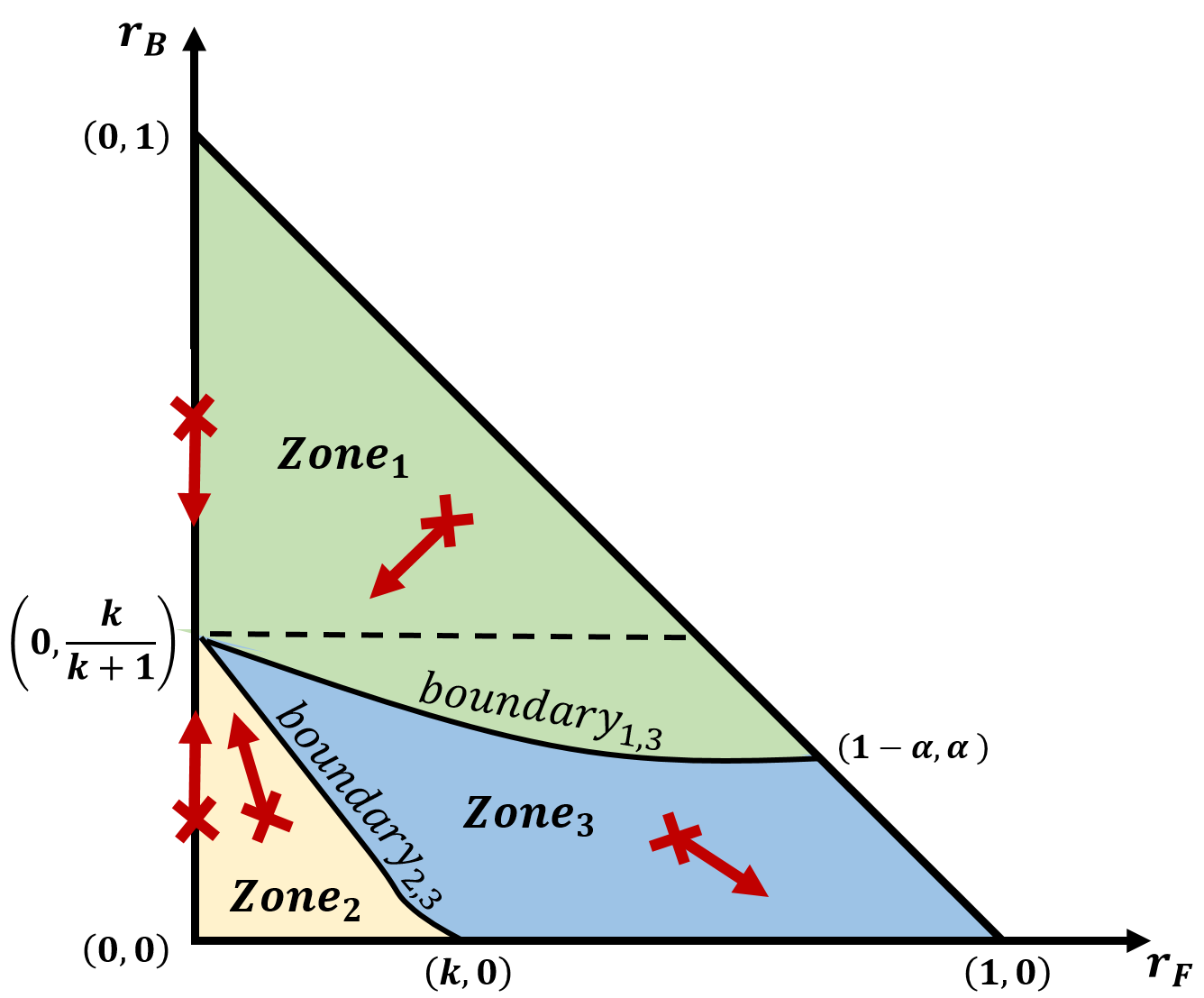}
\caption{Horizontal and vertical axes give the values of $r_{\mathcal F}$ and $r_{\mathcal B}$, respectively, and $(r_{\mathcal F}, r_{\mathcal B})$-coordinates of vertices in zones are marked.
At the vertex of \1 and \3, $\alpha$ is a solution of equation $N_{\tt in}r_{\mathcal B}^3+N_{\tt de}r_{\mathcal B}(1+k)-kN_{\tt de}=0$ for $r_{\mathcal B}$.
All points in \1, \2, and \3 move in directions $(-,-)$, $(-,+)$, and 
$(+,-)$, respectively.}
\label{fig:zones}
\end{figure}

\begin{figure}[t]
\centering
\includegraphics[width=0.75\columnwidth]{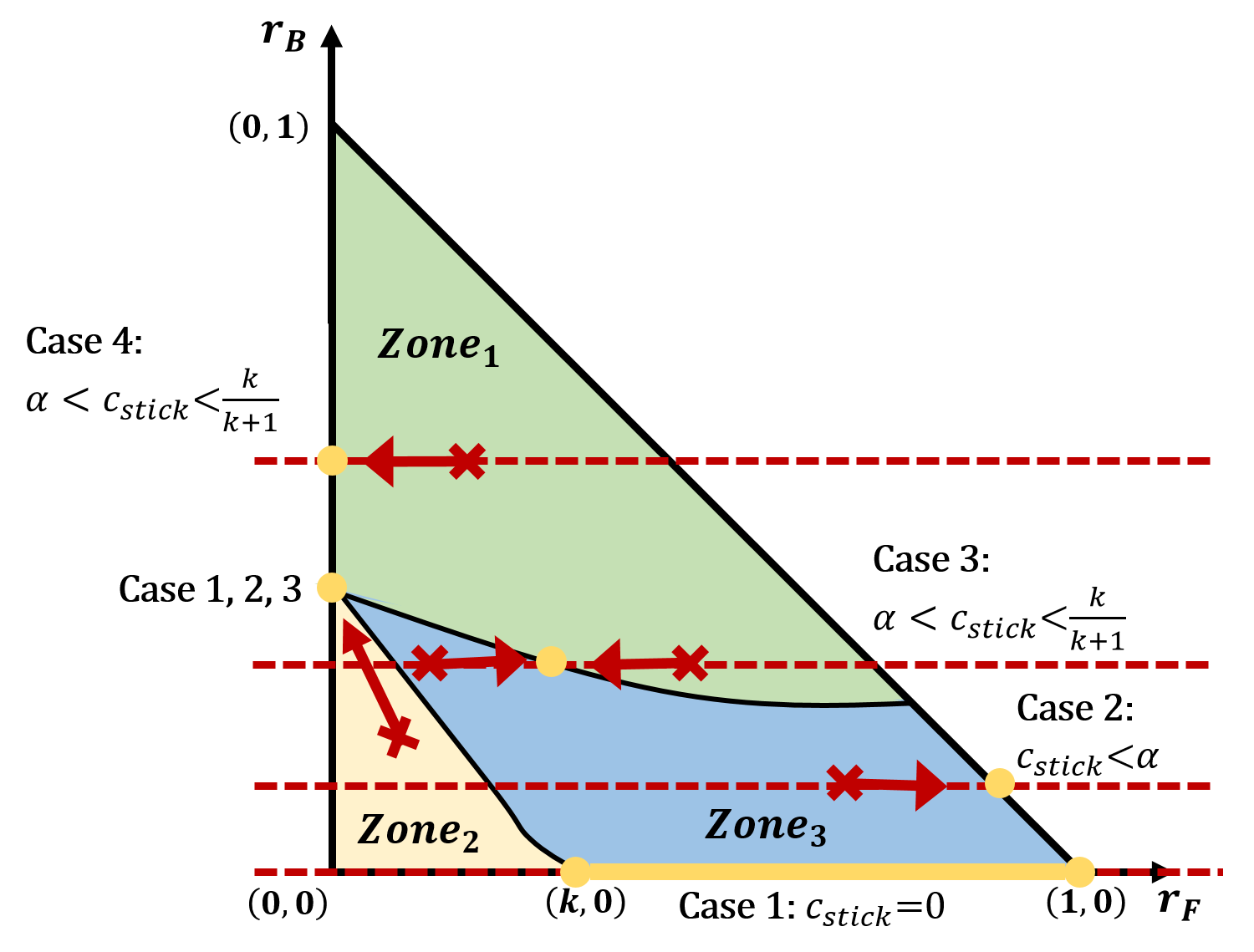}
\caption{Yellow points and line represent equilibria for each case.}
\label{fig:eqs}
\end{figure}

\blue{\smallskip\noindent \textbf{Best response dynamics.}
In game $\mathcal{G}(\bm{c}, c_{\tt stick})$, point $(r_{\mathcal F},r_{\mathcal B})$ reaches either of the two types of Nash equilibria: the stable coexistence of two coins and the lack of loyal miners to $coin_B.$}
Figure~\ref{fig:zones} represents dynamics in game $\mathcal{G}(\bm{c}, c_{\tt stick})$, where horizontal and vertical axes are $r_{\mathcal F}$ and $r_{\mathcal B}$ values, respectively. 
A line, $boundary_{1,3}$, represents 

\begin{small}
\begin{equation}
\begin{aligned}
&\frac{r_{\mathcal B}}{(1-r_{\mathcal F}-r_{\mathcal B})N_{\tt in}r_{\mathcal B}^2+(1-r_{\mathcal B})N_{\tt de}(r_{\mathcal F}+r_{\mathcal B})^2}\\
&=\frac{k}{N_{\tt in}r_{\mathcal B}^2+N_{\tt de}(r_{\mathcal F}+r_{\mathcal B})^2}.\label{eq:line1}
\end{aligned}
\end{equation}
\end{small}

\noindent On the line, the payoffs of \FM (i.e., $U_{\mathcal{F}}(r_{\mathcal F},r_{\mathcal B})$) and \AM (i.e., $U_{\mathcal{A}}(r_{\mathcal F},r_{\mathcal B})$) are the same.
In addition, the line does not intersect with the line $(0\le r_{\mathcal F} \le 1, r_{\mathcal B}=0)$ and has an intersection
$(1-\alpha, \alpha)$ with the line $r_{\mathcal F}+r_{\mathcal B}=1$ for $0\le r_{\mathcal F} \le 1$,
where $\alpha$ is a solution of equation $N_{\tt in}r_{\mathcal B}^3+N_{\tt de}r_{\mathcal B}(1+k)-kN_{\tt de}=0$ for $r_{\mathcal B}$.
The equation $N_{\tt in}r_{\mathcal B}^3+N_{\tt de}r_{\mathcal B}(1+k)-kN_{\tt de}=0$ has only one solution $\alpha$, and 
it is between 0 and $\frac{k}{1+k}$.
Another line, $boundary_{2,3}$, represents

\begin{small}
\begin{equation}
\begin{aligned}
&\frac{(r_{\mathcal F}+r_{\mathcal B})}{(1-r_{\mathcal F}-r_{\mathcal B})N_{\tt in}r_{\mathcal B}^2+(1-r_{\mathcal B})N_{\tt de}(r_{\mathcal F}+r_{\mathcal B})^2}\\
&=\frac{k}{N_{\tt in}r_{\mathcal B}^2+N_{\tt de}(r_{\mathcal F}+r_{\mathcal B})^2}, \label{eq:line2}
\end{aligned}
\end{equation}
\end{small}

\noindent and the payoffs of \FM (i.e., $U_{\mathcal{F}}$) and \BM (i.e., $U_{\mathcal{B}}$) are the same on the line.
The line does not intersect with the line $r_{\mathcal F}+r_{\mathcal B}=1$ for $0\le r_{\mathcal F} \le 1$ and has an intersection $(k,0)$ with the line $(0\le r_{\mathcal F} \le 1, r_{\mathcal B}=0)$.
Moreover, it is most profitable among the three strategies to continually conduct $coin_{\tt A}$-mining (\A) in a zone above $boundary_{1,3}$. 
We let this zone be \1.
In the zone below $boundary_{2,3}$, it is most profitable to continually conduct $coin_{\tt B}$-mining (\B), and the zone is denoted as \2.
In the zone between $boundary_{1,3}$ and $boundary_{2,3}$, fickle mining (\F) is the most profitable, and this zone is denoted as \3.
\textit{Note that the range of zones changes if the coin price changes because boundaries are functions of $k.$}

The moving direction of point $(r_{\mathcal F},r_{\mathcal B})$ is expressed as a red arrow in Figure~\ref{fig:zones}. 
For ease of reading, we express directions in which values $r_{\mathcal F}$ and $r_{\mathcal B}$ increase ($+$) or decrease ($-$) as $(\pm,\pm)$.
For example, $(+,+)$ indicates the direction in which both values, $r_{\mathcal F}$ and $r_{\mathcal B}$, increase.
In \1, \A is the most profitable strategy, and thus every point in \1 moves in the direction $(-,-)$.
In \2, because \B is the most profitable strategy,
every point moves in the direction $(-,+)$.
Finally, in \3, as \F is the most profitable strategy, 
every point in \3 moves in the direction $(+,-)$. 
Figure~\ref{fig:zones} shows the directions in the three zones (\1, \2, and \3).

\blue{\smallskip
\noindent\textbf{2D-Illustration of movement towards equilibria.}
To determine which equilibrium can be reached within each zone, we represent all Nash equilibria in game $\mathcal{G}(\bm{c}, c_{\tt stick})$ depending on a value of $c_{\tt stick}$ as yellow points and line in Figure~\ref{fig:eqs}. 
In the figure, the red dash lines represent $r_{\mathcal B}=c_{\tt stick}$ for each case. As described in Section~\ref{sec:finite}, there are two types of equilibrium points: 1) a lack of loyal miners and 2) stable coexistence of two coins. 
The equilibrium point representing a lack of loyal miners would be located on a red dash line $r_{\mathcal B}=c_{\tt stick}$, and we can see that all cases have this equilibrium. 
For Cases~1, 2, and 3, the second type of equilibrium (i.e., $(0, \frac{k}{k+1})$) representing stable coexistence of two coins is also found.  
A point $(r_{\mathcal F}, r_{\mathcal B})$ moves in the direction depending on its zone. In the meantime, if the point meets the line $r_{\mathcal B}=c_{\tt stick},$ then the point moves toward an equilibrium located on the line $r_{\mathcal B}=c_{\tt stick}$ as shown in Figure~\ref{fig:eqs}. 
In particular, the value of $r_{\mathcal F}$ in the equilibrium on the red dash line representing Case 3 is denoted by $\beta$, where the equilibrium is the intersection point between $boundary_{1,3}$ and the red dash line. 
Note, a point in \2 would not meet a red dash line because the point in \2 moves in the direction $(-,+)$ and can always be above the red dash line.
Therefore, such points in \2 are likely to reach the stable coexistence of $coin_{\tt A}$ and $coin_{\tt B}.$
However, some points (near to $boundary_{2,3}$) in \2 can also move into \3 when more miners of \AM than that of \FM revise their strategies, and then it is possible to reach the equilibrium, representing a lack of loyal miners to $coin_{\tt B}$.}

\section{Application to Bitcoin System}
\label{sec:application}

\begin{figure}[ht]
    \centering
    \includegraphics[width=.8\columnwidth]{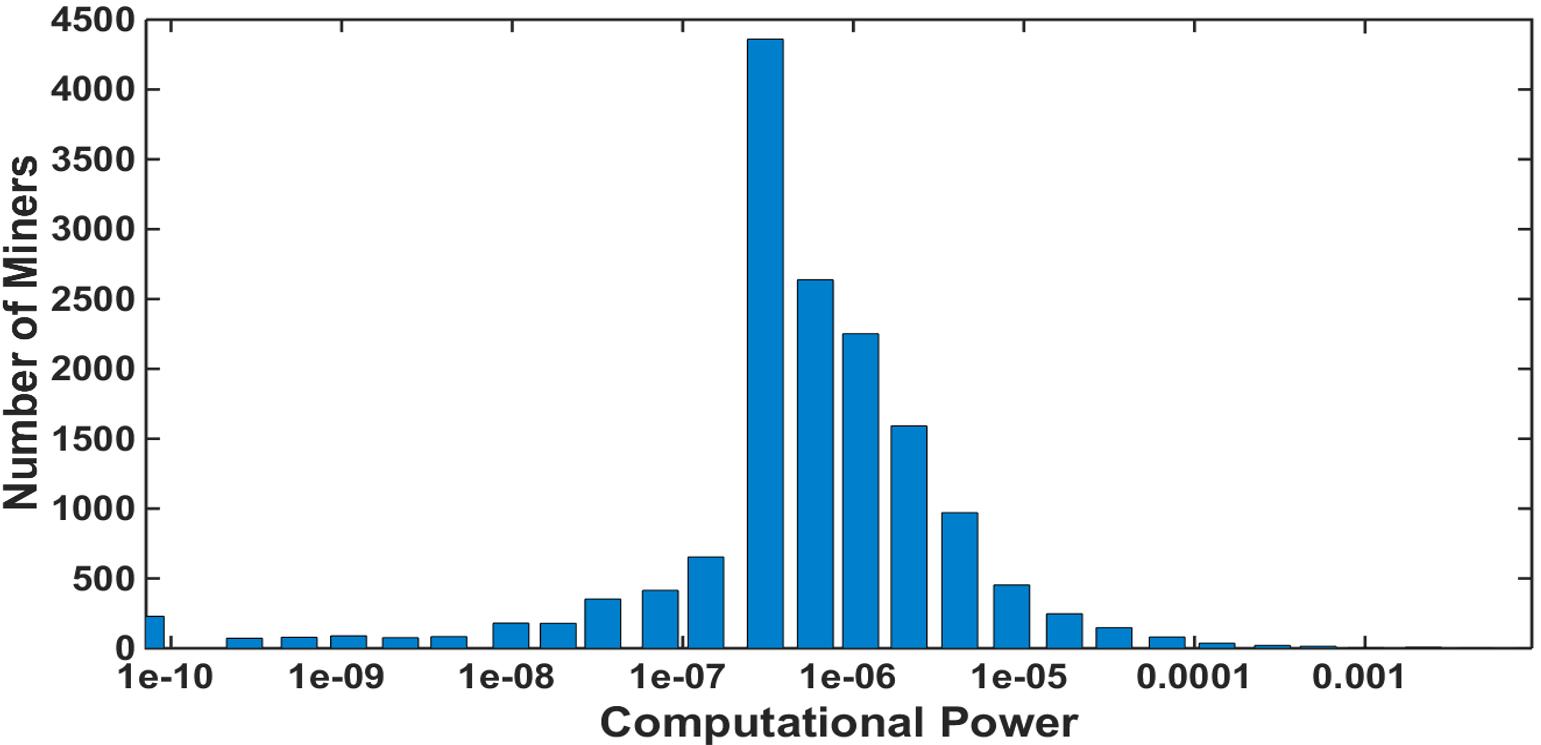}
    \caption{The computational power distribution in Slush.}
    \label{fig:slush}
\end{figure}

In this section, we apply our game model to Bitcoin as a case study. 
\blue{Specifically, we consider game $\mathcal{G} (\bm{c}, c_{\tt stick})$ when players possess sufficiently small mining power.
To see if this assumption is reasonable, we investigate the mining power distribution in the Bitcoin system, referring to the power distribution provided by Slush~\cite{slush}. 
The distribution is depicted in Figure~\ref{fig:slush} where the $x$-axis represents the range of the relative computational power $c_i$ and the $y$-axis represents the number of miners possessing computational power in the corresponding range. 
The figure shows that 1) most miners possess sufficiently small mining power, and 2) even the maximum computational power is less than $10^{-2}.$ Note that BITMAIN's $c_i$ is about $3\cdot 10^{-2}$ as of Dec. 2018.
Moreover, even though mining pools currently possess large computational power, 
the miners in pools can individually decide which coin to mine. 
We also recognize the distribution of computational power is significantly biased toward a few miners, as shown in Figure~\ref{fig:slush}. 
However, this fact does not imply that $\norm{\bm{c}}_2$ is large. 
Referring to the data provided by Slush, $\norm{\bm{c}}_2$ is only about 0.05, where this value is equivalent to that for the case where all miners possess $2.5\times 10^{-3}$ computational power.\footnote{We calculated this assuming that other pools have the computational power distribution similar to Slush.} 
Therefore, most miners (and most mining power) would follow dynamics of game $\mathcal{G} (\bm{c}, c_{\tt stick})$. 
As a result, we can apply game $\mathcal{G} (\bm{c}, c_{\tt stick})$ to the practical systems.}

Now, we describe how game $\mathcal{G} (\bm{c}, c_{\tt stick})$ is applied to the Bitcoin system. 
As described in Section~\ref{sec:preliminary}, Bitcoin was split into BTC and BCH in Aug. 2017.
Thus, we can map BTC and BCH to $coin_{\tt A}$ and $coin_{\tt B}$, respectively.
For the mining difficulty adjustment algorithm of BCH,
we should consider two types of BCH mining difficulty adjustment algorithms: those that BCH have before and after Nov. 13, 2017. 
This is because the mining difficulty adjustment algorithm of BCH changed through a hard fork of BCH (on Nov. 13, 2017).

\smallskip\noindent\textbf{Before Nov. 13, 2017. }
First, we consider the mining difficulty adjustment algorithm of BCH before Nov. 13, 2017. 
In this algorithm, not only the mining difficulty is adjusted for every 2016 block, but also EDA can occur as described in Section~\ref{sec:preliminary}.  
Note that EDA occurs if the mining is significantly difficult in comparison with the current mining power, i.e., EDA is used only for \textit{decreasing} the BCH mining difficulty. 
Therefore, the value of $N_{\tt in}$ is 2016 because the BCH mining difficulty can increase after 2016 blocks are found.
Meanwhile, when the BCH mining difficulty decreases, the value of $N_{\tt de}$ varies depending on $r_{\mathcal F}$ and $r_{\mathcal B}$, ranging between 6 and 2016.
Thus, we can consider the expected number of blocks found until the mining difficulty decreases (i.e, the mean of $N_{\tt de}$ denoted by $E[N_{\tt de}]$) instead of $N_{\tt de}$, and $E[N_{\tt de}]$ as a function of $r_{\mathcal F}$ and $r_{\mathcal B}$ would continuously vary from 6 to 2016.
If $r_{\mathcal F}$ is 0, $E[N_{\tt de}]$ is 2016 because EDA does not occur, and if $r_{\mathcal B}$ is 0, $E[N_{\tt de}]$ is 6.

As a result, the Bitcoin system before Nov. 13, 2017 can be $\mathcal{G} (\bm{c}, c_{\tt stick})$ where $E[N_{\tt de}]$ substitutes for $N_{\tt de}.$
This game $\mathcal{G} (\bm{c}, c_{\tt stick})$ has also Nash equilibria and dynamics as shown in Figure~\ref{fig:zones} 
because $E[N_{\tt de}]$ is a continuous function of $r_{\mathcal F}$ and $r_{\mathcal B}.$

\smallskip\noindent\textbf{After Nov. 13, 2017. }
Next, we consider the Bitcoin system after Nov. 13, 2017.
In this case, the BCH mining difficulty adjustment algorithm is different from that assumed in our game because the mining difficulty is adjusted for every block by considering the generation time of the past 144 blocks as a moving time window. 
Despite that, game $\mathcal{G} (\bm{c}, c_{\tt stick})$ can be applied to this system.
Indeed, in general, our results for game $\mathcal{G} (\bm{c}, c_{\tt stick})$ would appear in the Bitcoin system regardless of the BCH mining difficulty adjustment algorithm, shown below. 

\smallskip
\begin{theorem}\label{thm:bch2}
Consider the game $\mathcal{G} (\bm{c}, c_{\tt stick})$ when $\norm{\bm{c}}_2\approx 0.$ 
Then when the mining difficulty of $coin_{\tt B}$ is adjusted every block or in a short time period, the set $\mathcal{E} (\bm{c}, c_{\tt stick})$ is \eqref{eq:nash_infty} presented in Theorem~\ref{thm:inf_nash}.
In addition, $\mathcal{G} (\bm{c}, c_{\tt stick})$ under this mining difficulty adjustment algorithm of $coin_{\tt B}$ has dynamics such as in Figure~\ref{fig:zones}.
\end{theorem}
\smallskip

Because the current BCH mining difficulty is adjusted every block, Theorem~\ref{thm:bch2} implies that results for game $\mathcal{G} (\bm{c}, c_{\tt stick})$ is also applied to the current Bitcoin system even though the BCH mining difficulty adjustment algorithm changed. 
The proof of Theorem~\ref{thm:bch2} is presented in Appendix~\ref{sec:diff}. 

\begin{figure*}[!htb]
  \centering
  \begin{tabularx}{16cm}{c r}
  \multicolumn{1}{m{0.5cm}}{(a)} & \multicolumn{1}{m{15cm}}{\includegraphics[width=15cm]{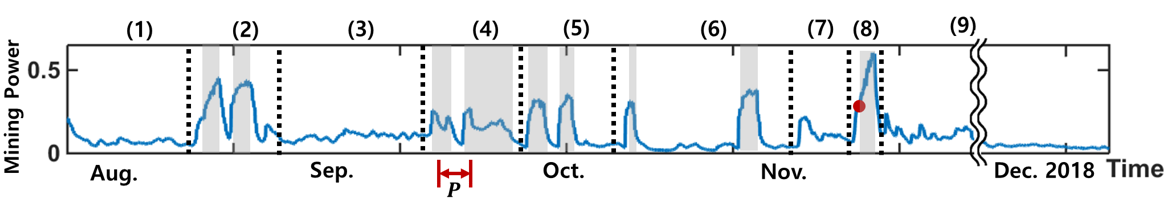}} \\[-7pt]
  \multicolumn{1}{m{0.5cm}}{(b)} &\multicolumn{1}{m{15cm}}{\includegraphics[width=15cm]{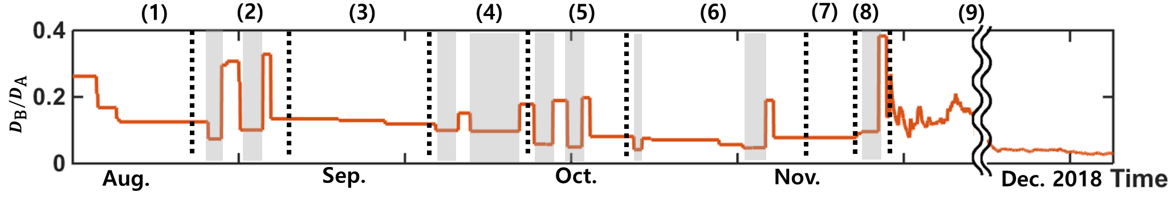}} \\[-7pt]
  \multicolumn{1}{m{0.5cm}}{(c)} &\multicolumn{1}{m{15cm}}{\includegraphics[width=15cm]{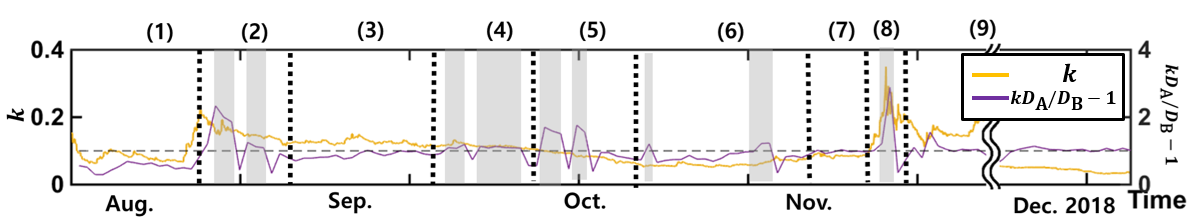}}\\[-6pt]
  \multicolumn{1}{m{0.5cm}}{(d)} &\parbox{15.3cm}{\centering \includegraphics[width=14.7cm]{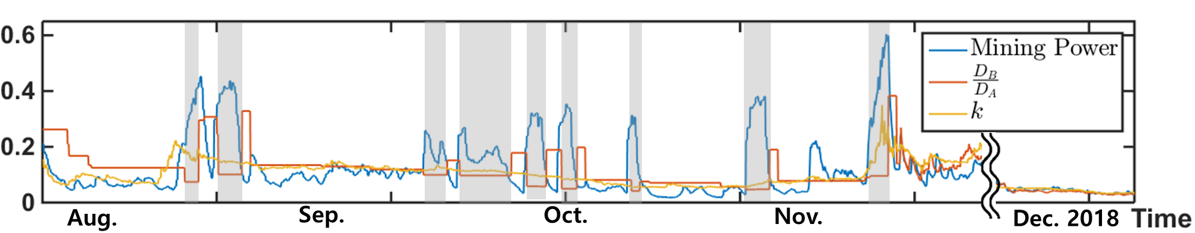}}
  \end{tabularx}
    \vspace{-1mm}
    \caption{The data for the Bitcoin system from early Aug. 2017 to Dec. 2018 is represented. 
    Figure~7a, 7b, and 7c represent (a) relative mining power of BCH to the total mining power, (b) the ratio between mining difficulties of BCH and BTC, (c) the ratio between prices of BCH and BTC, and BCH mining profitability. 
    Figure~7d shows the data for mining power, price, and mining difficulty of BCH. 
    In the gray zones, fickle miners conduct BCH mining. 
The data from Dec. 2017 to Nov. 2018 are omitted because they are similar to the data for Dec. 2018. Each point represents a data captured every hour.}
    \vspace{-2mm}
    \label{fig:bch_data}
\end{figure*}

\begin{figure*}[!htb]
	\centering
    \captionsetup[subfloat]{farskip=0pt,captionskip=-1pt}
    \subfloat[Figure~\ref{fig:bch_data}-(1)]{
      \includegraphics[width=0.25\textwidth]{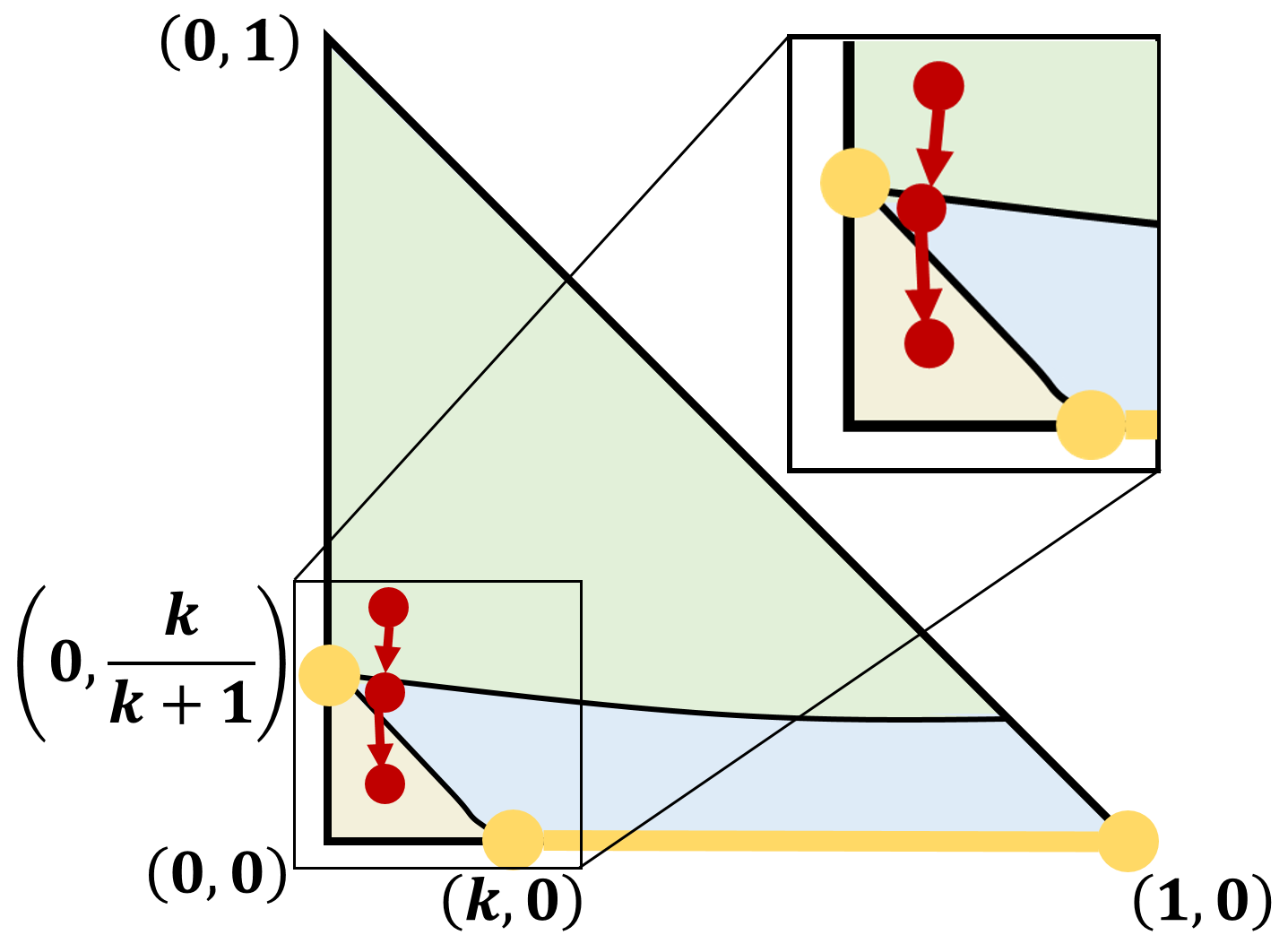}
      \label{fig:btc1}
   }
    \subfloat[Figure~\ref{fig:bch_data}-(2)]{
      \includegraphics[width=0.25\textwidth]{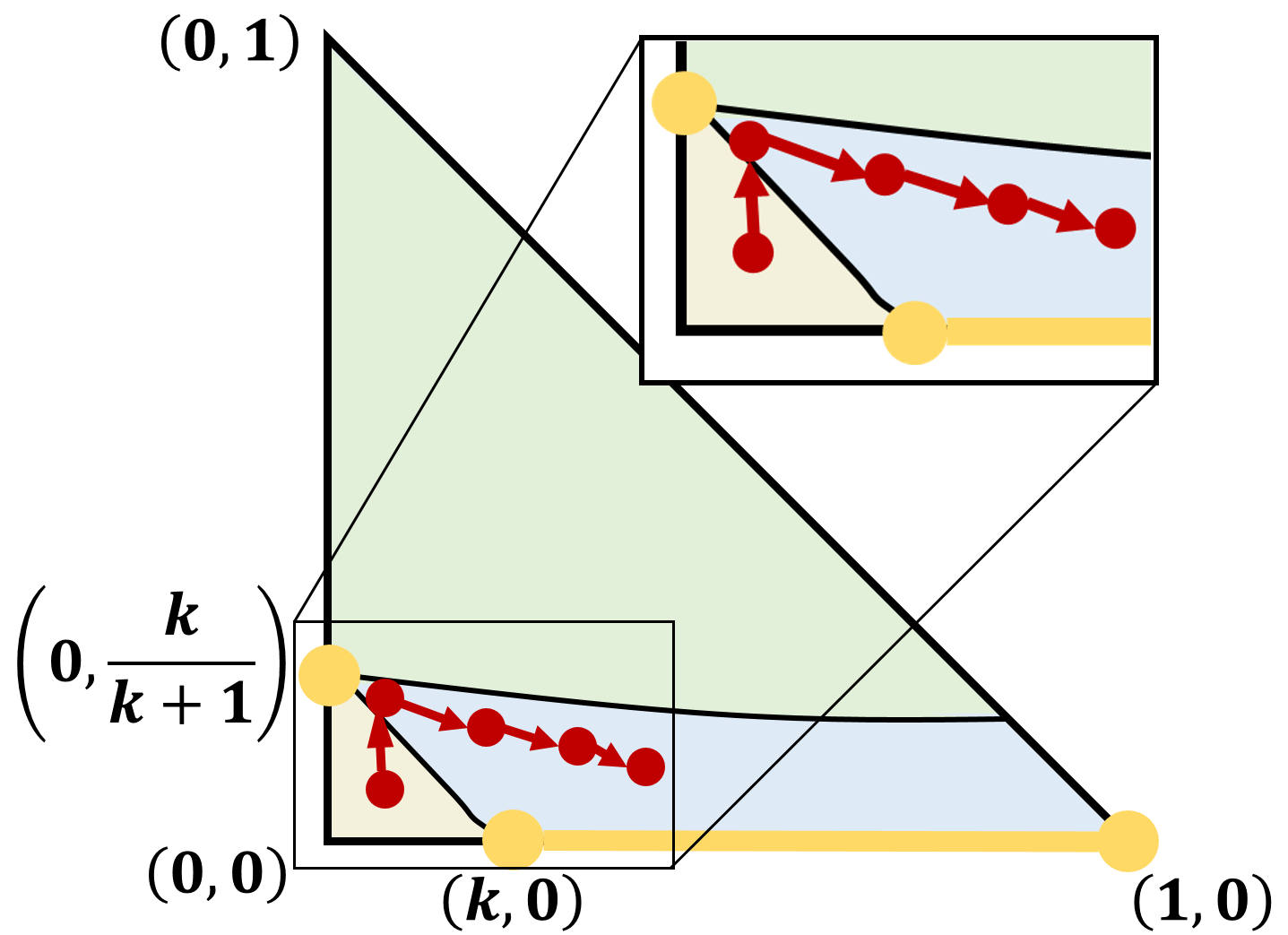}
      \label{fig:btc2}
   }
    \subfloat[Figure~\ref{fig:bch_data}-(3)]{
      \includegraphics[width=0.25\textwidth]{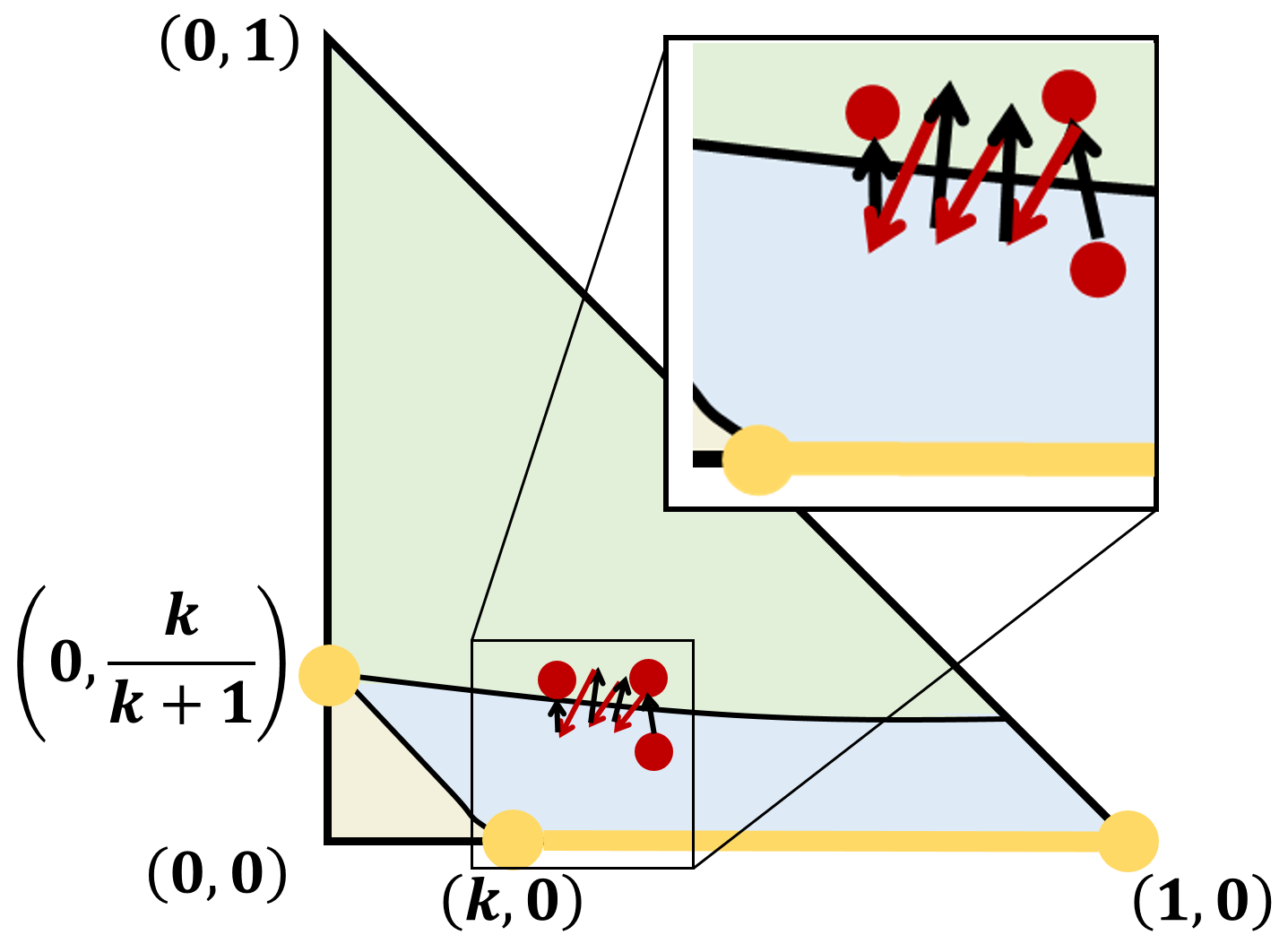}
      \label{fig:btc3}
   }\\
    \subfloat[Figure~\ref{fig:bch_data}-(4)]{
      \includegraphics[width=0.25\textwidth]{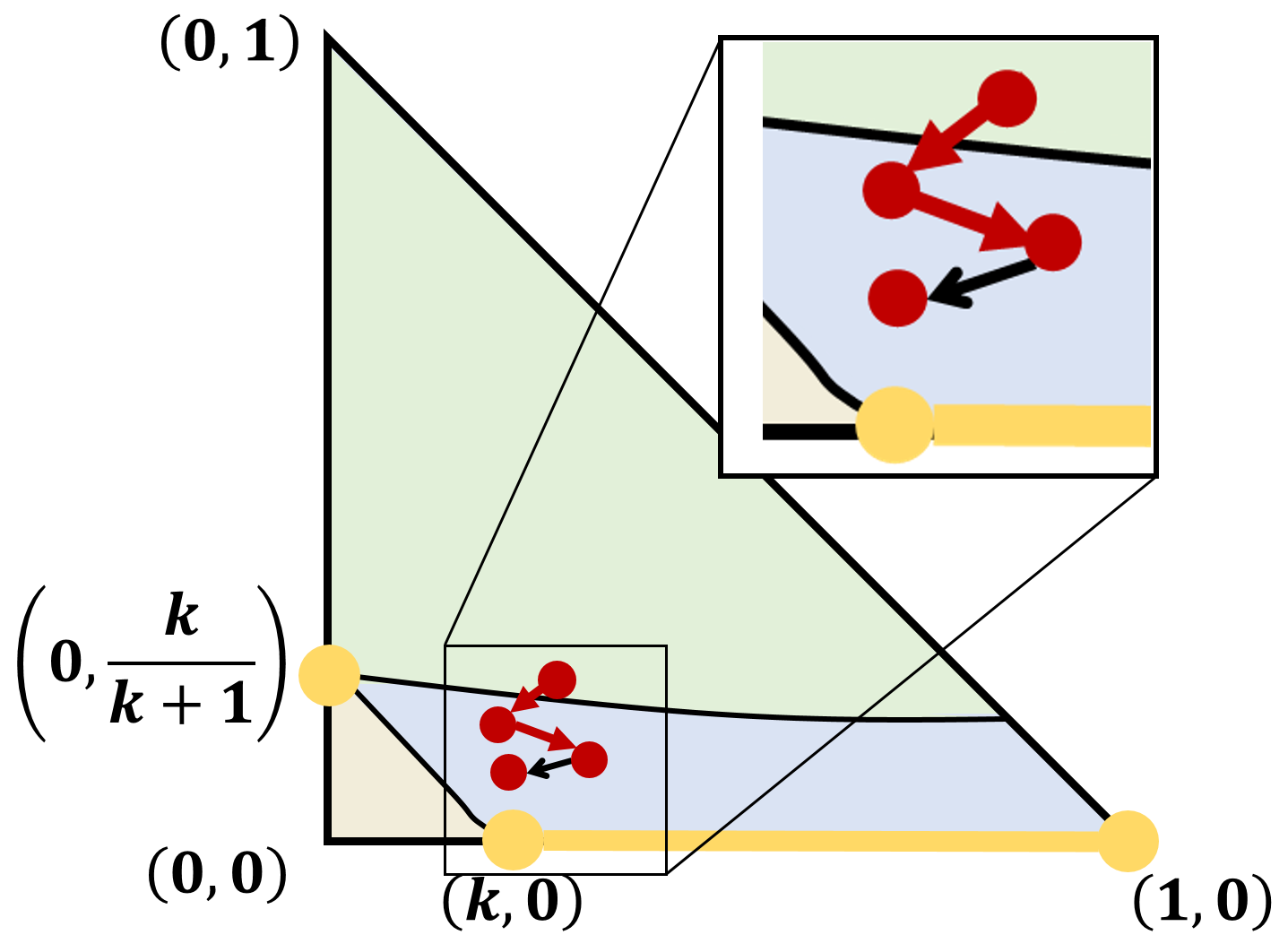}
      \label{fig:btc4}
   }
    \subfloat[Figure~\ref{fig:bch_data}-(5)]{
      \includegraphics[width=0.25\textwidth]{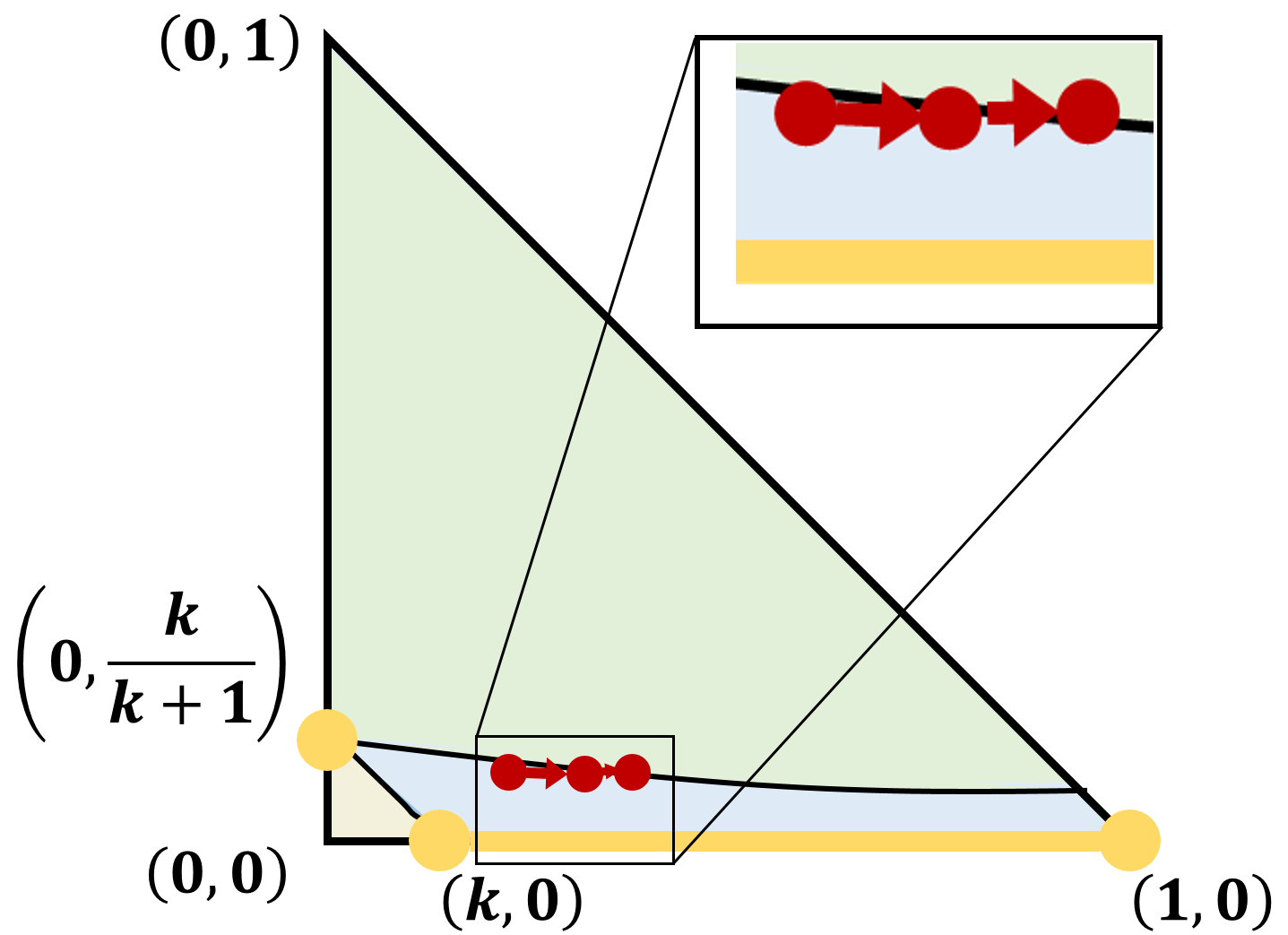}
      \label{fig:btc5}
   }
    \subfloat[Figure~\ref{fig:bch_data}-(6)]{
      \includegraphics[width=0.25\textwidth]{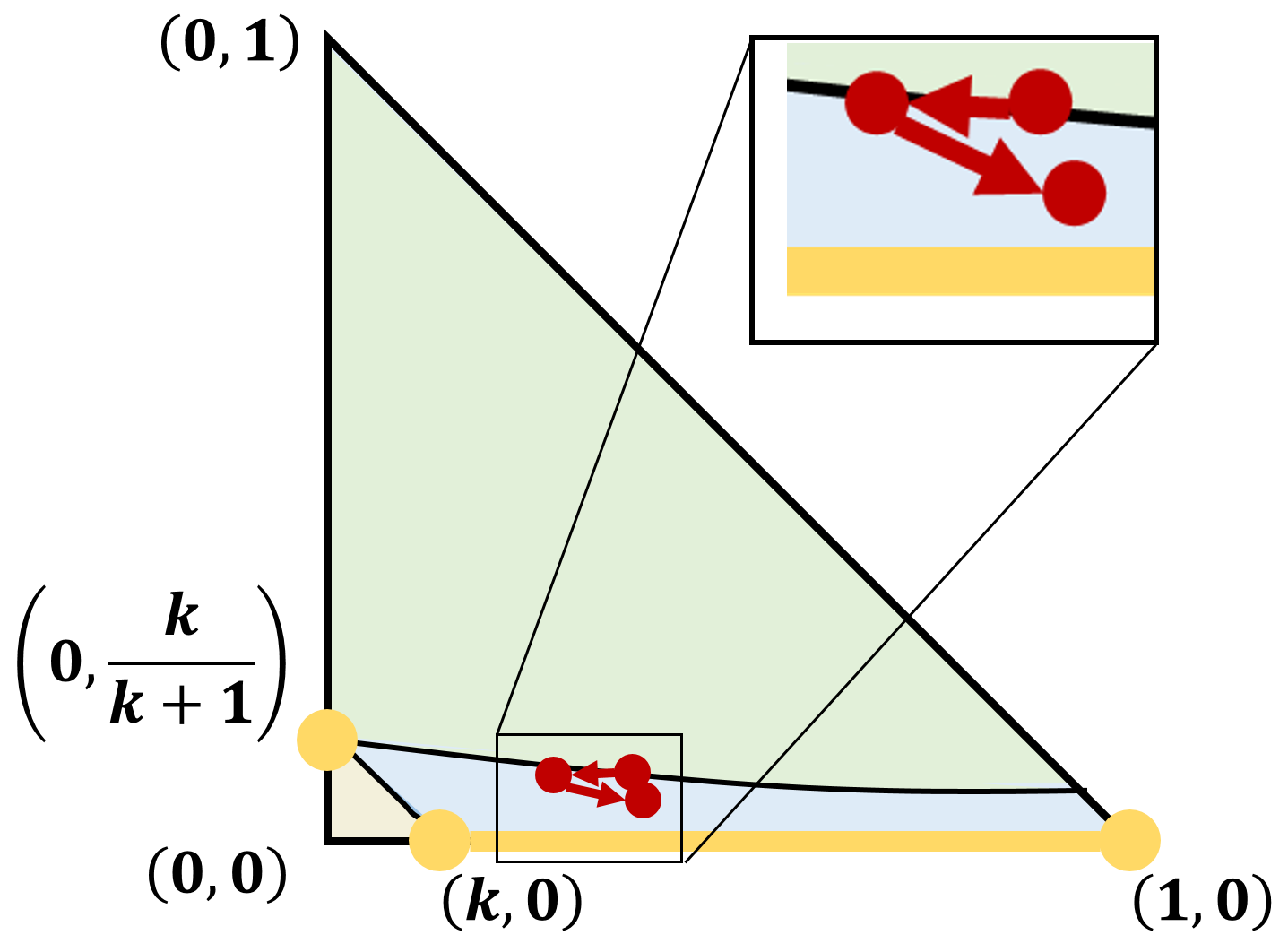}
      \label{fig:btc6}
   }\\
    \subfloat[Figure~\ref{fig:bch_data}-(7)]{
      \includegraphics[width=0.25\textwidth]{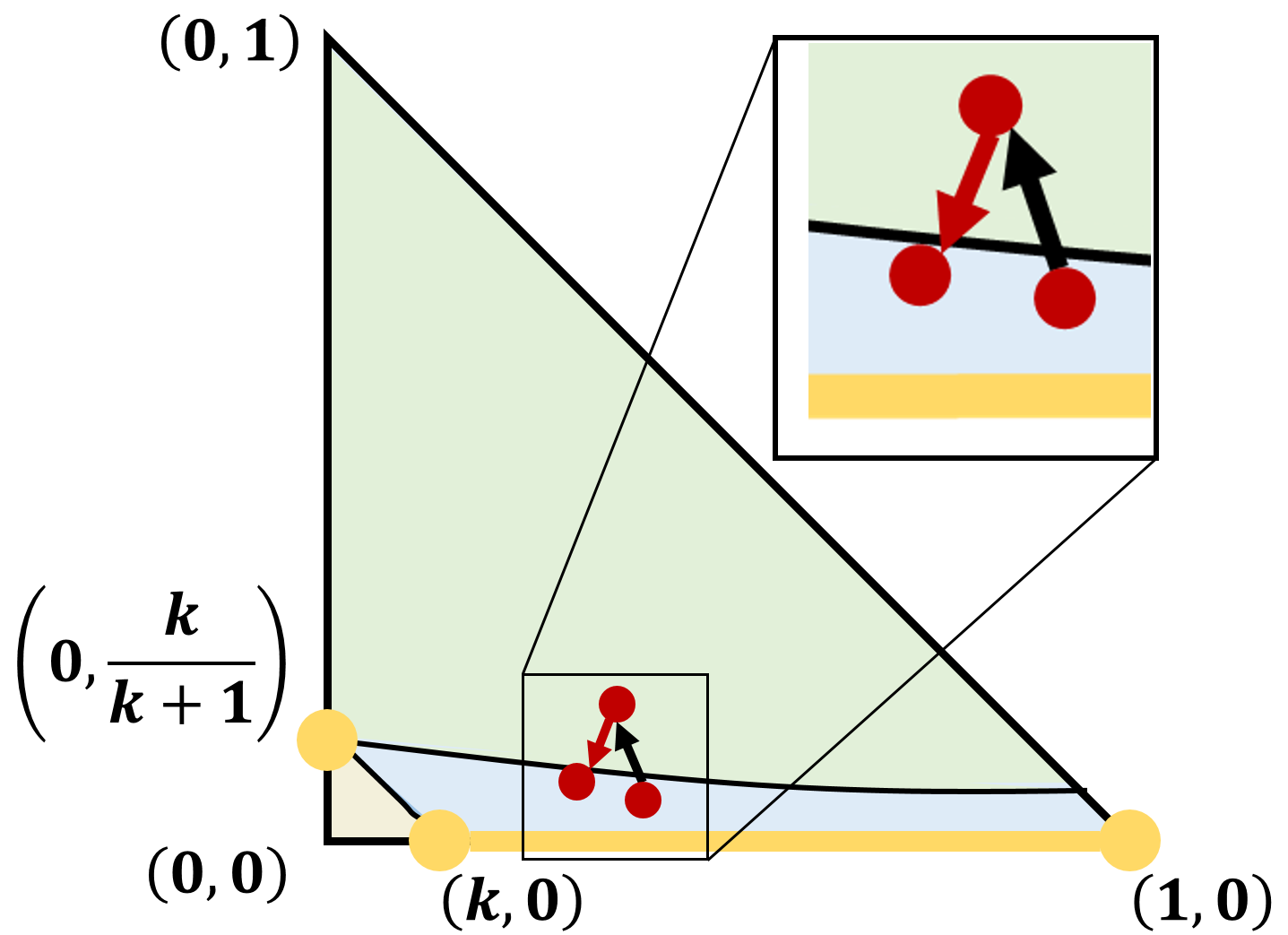}
      \label{fig:post1}
   }
    \subfloat[Figure~\ref{fig:bch_data}-(8)]{
      \includegraphics[width=0.25\textwidth]{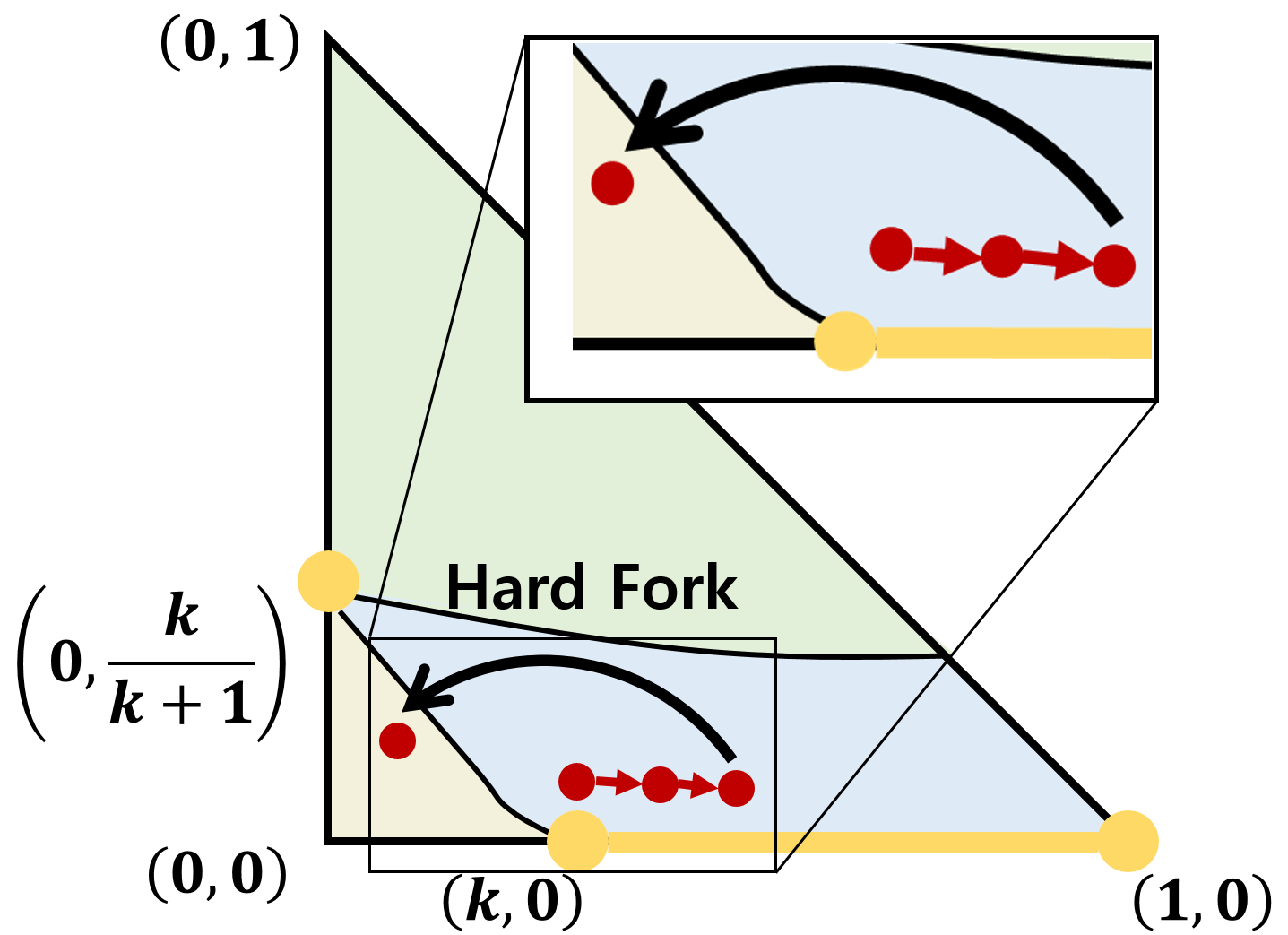}
      \label{fig:post2}
   }
    \subfloat[Figure~\ref{fig:bch_data}-(9)]{
      \includegraphics[width=0.25\textwidth]{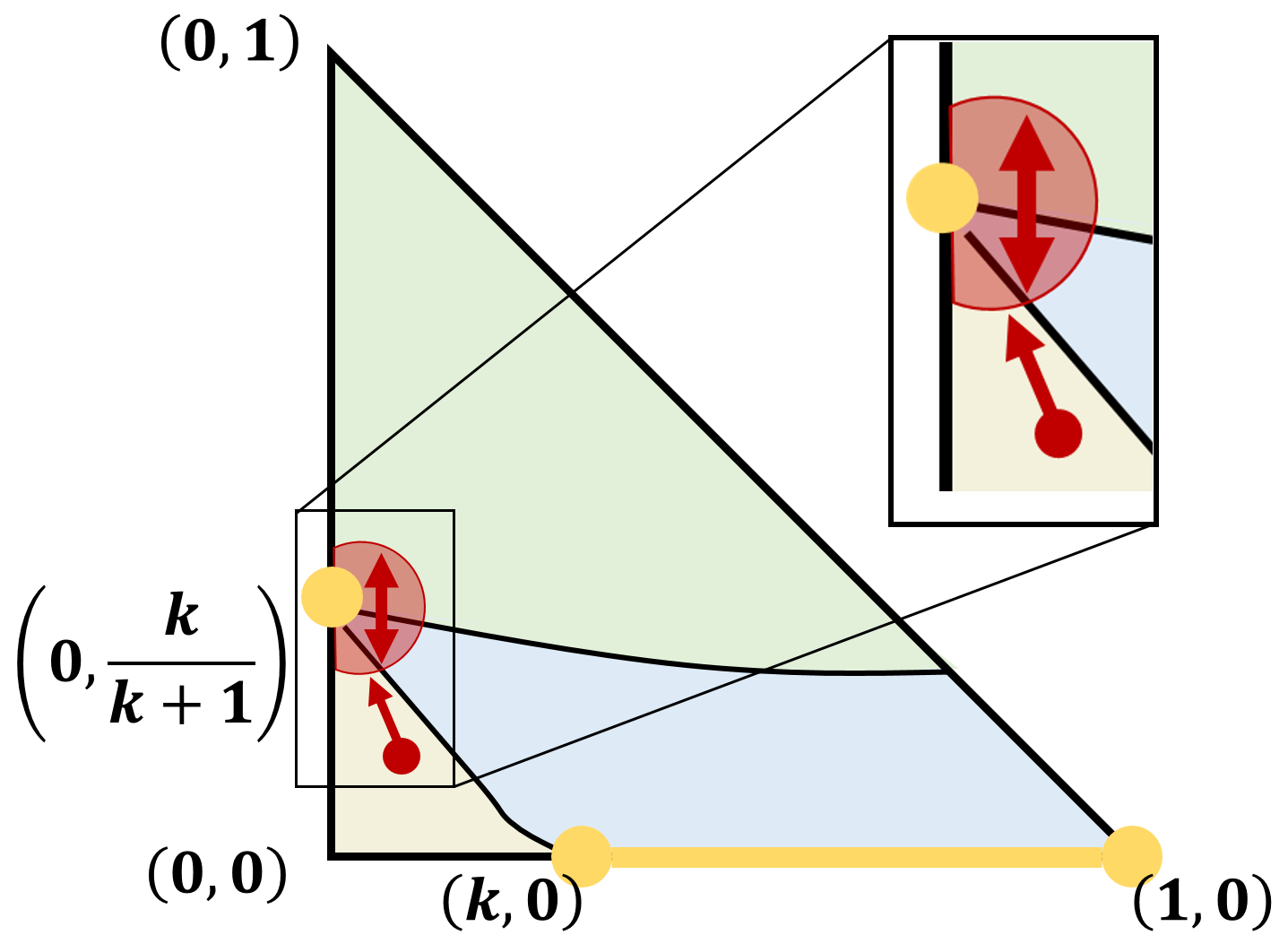}
      \label{fig:post3}
   }
    \caption{Points and movements of Figure~\ref{fig:bch_data}.
Figure~\ref{fig:btc1} $\sim$ \ref{fig:post3} correspond to parts (1)$\sim$(9) in Figure~\ref{fig:bch_data}. 
Red arrows represent movement in agreement with
our model, whereas black arrows represent movement deviating from 
our model. Each upper right square presents enlarged points and directions. \vspace{-4mm}}
    \label{fig:btc}
    \vspace{-2mm}
\end{figure*}

\section{Data analysis}
\label{sec:data}

\subsection{BTC vs. BCH}
\label{subsec:btc}

We analyze the mining power data in the Bitcoin system to identify to which equilibrium the state has been moving.
Moreover, through this data analysis, we can find out empirically 
how much our theoretical model agrees with practical results.
For data analysis of the Bitcoin system, we collected the mining power data of BTC and BCH from the release date of BCH (Aug.\ 1, 2017) until the time of writing (Dec. 10, 2018) from CoinWarz~\cite{coinwarz}.
Figure~7a represents the mining power history of BCH, where the mining power is expressed as a fraction of the total power in BTC and BCH, i.e., 
$$\frac{\text{BCH mining power}}{\text{BTC mining power}+\text{BCH mining power}}.$$
\blue{In addition, we represent the data history of a ratio between difficulties of BCH and BTC (i.e., $\frac{D_{\tt B}}{D_{\tt A}}$) and a relative price of BCH to that for BTC (i.e., $k$) in Figure~7b and 7c, respectively. 
The price of BCH is depicted as a yellow line in Figure 7c (see the left $y$-axis). 
Moreover, Figure 7c represents the relative BCH mining profitability ($\frac{kD_{\tt A}}{D_{\tt B}}-1$) to the BTC mining profitability as a purple line, and the black dashed line represents $\frac{kD_{\tt A}}{D_{\tt B}}-1=0$ (see the right $y$-axis for the two lines).
For this profitability, to increase reliability of data, we collected the daily BCH profitability from CoinDance~\cite{coindance}, and thus a purple point is a data captured every day. 
Note that $\frac{D_{\tt B}}{D_{\tt A}}$ is less than $k$ in the case where the purple line is above the black dashed line.  
Figure~7d simultaneously shows all data histories (except for the BCH mining profitability) presented in Figure~7a$\sim$7c.}
In Figure~\ref{fig:bch_data}, the data from Dec. 2017 to Nov. 2018 are omitted because they are similar to the data for Dec. 2018. Figure~\ref{fig:btc1}$\sim$\ref{fig:post3} correspond to parts (1)$\sim$(9) of Figure~\ref{fig:bch_data}, respectively, where the area of three zones has changed because the relative price $k$ of BCH to that for BTC has fluctuated quite frequently. 

As another case study, we examine the mining power data of Bitcoin ABC and Bitcoin SV from Nov. 1, 2018 to Dec. 20, 2018 to analyze a special situation where $c_{\tt stick}$ suddenly increases due to the ``hash war" caused by a hard fork in the BCH system. We describe this in Section~\ref{subsec:stick}. 

\smallskip
\blue{\noindent\textbf{Methodology.}
We first describe how to determine $r_{\mathcal F}$ and $r_{\mathcal B}$ of each state. 
According to the definition of fickle mining (Definition~\ref{def:fickle}), fickle miners would conduct BCH mining from when $\frac{D_{\tt B}}{D_{\tt A}}$ \textit{changes} to a value less than $k$ to when $\frac{D_{\tt B}}{D_{\tt A}}$ \textit{changes} to a value greater than $k.$ 
This is because $D_{\tt B}$ is always less than $r_{\mathcal F}+r_{\mathcal B}$ and greater than $r_{\mathcal B}$ (see Figure 7d). 
Therefore, Figure~7a represents the value of $r_{\mathcal F}+r_{\mathcal B}$ during the period. 
We indicate the fickle mining periods in gray before the hard fork of BCH (Nov. 13, 2017) in Figure~\ref{fig:bch_data}. 
Figure~7d shows that $\frac{D_{\tt B}}{D_{\tt A}}$ changes to a value less than and greater than $k$ at the start and end of these periods, respectively. 
As a result, in Figure~7a, we can find out the value of $r_{\mathcal F}+r_{\mathcal B}$ for the gray colored periods and the value of $r_{\mathcal B}$ for non-colored periods. 
Here, we can see that the mining power of BCH has fluctuated considerably when the ratio of the BCH mining difficulty to the BTC mining difficulty ($\frac{D_{\tt B}}{D_{\tt A}}$) \textit{changes} to a value less than $k$. 
Moreover, when the coin mining difficulties do not change while BCH mining is more profitable than BTC mining, large \textit{peaks} (i.e., a sudden increase) do not appear. 
This fact is confirmed, referring to the purple line in non-colored zones (e.g., part (3) in Figure 7c).
As a result, we can consider that those fluctuations occur due to fickle miners between BTC and BCH. 

If a miner switches the coin to mine without changes in the coin mining difficulty, this implies that the miner's strategy changes (e.g., from \A to \B). 
From the method described above, we can determine the mining power $r_{\mathcal F}$ used for fickle mining and the mining power $r_{\mathcal B}$ used for BCH-only mining. 
The points and directions are marked roughly in Figure~\ref{fig:btc}. 
The red arrow represents movement in agreement with our analysis, whereas the black arrow represents movement deviating from our analysis.

Next, we explain Figure~\ref{fig:btc} by matching it with each part of Figure~\ref{fig:bch_data}.}

\smallskip\noindent\textbf{The beginning of the game. }
In Figure~\ref{fig:bch_data}-(1), the status point is initially in \1, 
and then it moves to \2 as shown in Figure~\ref{fig:btc1}, as the BCH mining power decreases.

\smallskip\noindent\textbf{Towards the lack of BCH loyal miners. }
In Figure~7a-(2), two peaks occur when the BCH mining difficulty decreases to values less than $k,$ \blue{and these peaks appear in the gray colored periods.}
Therefore, we can know that these peaks occur due to fickle miners.
The first peak indicates that more and more miners started fickle mining (i.e., increase in $r_{\mathcal F}$). 
This is because the upflow of the first peak is less steep than that for other peaks, and the downflow of the first peak is steeper than the upflow of the first peak, 
indicating that $r_{\mathcal F}$ increases from near 0 up to near 0.4. 
Furthermore, one can see that $r_{\mathcal B}$ increased at the beginning of Figure~7a-(2). 
\blue{Remark that Figure~7a shows the value of $r_{\mathcal B}$ in a non-colored zone.}
In addition, the BCH mining power in the valley between two peaks of Figure~7a-(2) is greater than the mining power at the end of Figure~7a-(1).
This fact shows again that $r_{\mathcal B}$ increased at the beginning of Figure~7a-(2). 
After that, because the end of Figure~7a-(2) is less than the valley between the two peaks of Figure~7a-(2), we can know that $r_{\mathcal B}$ decreased while $r_{\mathcal F}$ increased in Figure~7a-(2).
Figure~\ref{fig:btc2} represents these movements described above.

In the beginning of Figure~7a-(3), $r_{\mathcal B}$ slightly increases, and it does not correspond with our model;
we regard this as a momentary phenomenon because of a
decrease in the BCH mining difficulty. 
\blue{Figure~7b shows that the BCH mining difficulty decreased at the beginning of the part (3).
However, even though the BCH mining difficulty decreased, peaks due to fickle mining do not appear because the relative BCH mining difficulty did not decrease to a value less than $k$ as shown in Figure~7d.}
As a result, as can be seen in Figure~\ref{fig:btc3}, the point moves alternatively between \1 and \3. 
One can see that $r_{\mathcal F}$ decreased compared with the mining power in the peaks of Figure~7a-(4) and the peaks in Figure~7a-(2);
this might be because the moving direction in \1 is $(-,-)$.

Next, the peaks in the period $P$ presented in Figure~7a-(4) appeared due to fickle miners because the BTC mining difficulty increased. 
We can check that $\frac{D_{\tt B}}{D_{\tt A}}$ in the period $P$ decreased to a value less than $k$ through Figure~7d. 
Note that the fact that the BTC mining difficulty increased makes the value of $\frac{D_{\tt B}}{D_{\tt A}}$ decrease.
Indeed, the two peaks of the period $P$ show that $r_{\mathcal F}$ decreases and then increases because $r_{\mathcal F}+r_{\mathcal B}$ is represented in the period $P$ of Figure~7a.
This may be explained according to our model as follows: 
the state was near to the boundary between \1 and \3 at the beginning of Figure~\ref{fig:bch_data}-(4), and then the state entered \3 while moving in the direction $(-,-)$ (the moving direction in \1) as in Figure~\ref{fig:btc4}.
Then, the state in \3 moved in the direction $(+,-)$ in agreement with our game, and one can see that the third peak (i.e., the beginning of the second gray colored zone in Figure~7a-(4)) is higher than the second peak. 
After that, $r_{\mathcal F}$ decreases (see the second gray colored zone in Figure~7a-(4)), showing a deviation from our model, which is indicated by the black arrow in Figure~\ref{fig:btc4}.
\blue{Indeed, considering this case as well as Figure~\ref{fig:bch_data}-(3), we observe such noises in the case where $\frac{D_{\tt B}}{D_{\tt A}}$ changes to a value close to $k.$ }

Next, as shown in Figure~\ref{fig:btc5}, 
the point in \3 moves in the direction $(+,-)$ again because peaks in Figure~7a-(5) are higher than that for Figure~7a-(4).
Moreover, in Figure~7c-(4)$\sim$(6), $k$ is roughly decreasing and even drops to about 0.055 in a few cases.
In the meantime, the point passes $boundary_{1,3}.$

Because the state entered \1, $r_{\mathcal F}$ starts to decrease, moving in the direction $(-,-)$ (as shown in Figure~\ref{fig:btc6}).
Therefore, the first peak in Figure~7a-(6) is smaller than the last peak in Figure~7a-(5). 
Then, because the second peak is higher than the first peak in Figure~7a-(6), one can see that the point moved in the direction $(+,-)$ in \3 in agreement with our model, which is, in turn, depicted in Figure~\ref{fig:btc6}.

As can be seen in Figure~\ref{fig:post1}, $r_{\mathcal B}$ first increases in Figure~7a-(7), and the point enters \1; 
this is a deviation from our analysis, which may be explained because the BCH mining is momentarily more profitable than the BTC mining at the time. 
\blue{Here, we can see again the noise in the case where the value of $\frac{D_{\tt B}}{D_{\tt A}}$ is close to $k.$}
However, $r_{\mathcal B}$ decreases again in agreement with our model. 
In addition, one can see that $r_{\mathcal F}$ decreases in the meantime because the starting height of the peak in Figure~7a-(8), which is marked by a red point, is less than that of the final peak in Figure~7a-(6). 
Therefore, the point in \1 moved in the direction $(-,-)$ and entered \3, conforming with our analysis.

Then, in the second week of Nov. 2017, the price of BCH was suddenly pumped ($k \approx 0.4$ in some cases).
Therefore, \2 widens in Figure~\ref{fig:post2}.
Also, the point in \3 continuously moves in the direction $(+,-)$, and $r_{\mathcal F}$ even increases to over 0.5. 
It can be seen that the peak in Figure~\ref{fig:bch_data}-(8) has a right-angle trapezoid with a positive slope, which indicates that $r_{\mathcal F}$ continuously 
increases even though it was \textit{already} high.
\blue{From the history, we observe that the Bitcoin system often reaches the lack of BCH loyal miners.}
However, a breakthrough exists even in this bad situation.
If $k$ continuously increases, \2 widens, and it makes the state enter \2 and reach close to the coexistence equilibrium.
As a result, considering the state of Bitcoin as of Nov. 13, 2017, $k$ had to increase to a minimum of 0.5 in order for the mining power engaging in fickle mining to decrease.

\smallskip\noindent\textbf{Close to coexistence. }
However, at the end of Figure~\ref{fig:bch_data}-(8), another hard fork occurred in BCH for updating the difficulty adjustment algorithm, and this influenced the status as an external factor. 
Consequently, the point jumped into \2 due to this hard fork as shown in Figure~\ref{fig:post2}.
After the hard fork, the point moves in the direction $(-,+)$, reaching close to coexistence. 
This is shown by this fact that fluctuations became stable more and more in the beginning of Figure~7a-(9). 
Note that peaks occur in a short time after the hard fork because the BCH mining difficulty is quickly adjusted.
Even though the state has been close to coexistence, fickle mining is still possible and observed as described in Section~\ref{sec:preliminary}.
In addition, as the price continuously changes, the point sometimes enters \3 where fickle mining increases, alternating up and down in the red semicircle in Figure~\ref{fig:post3}.
In other words, fickle mining will not completely cease.
Therefore, if the Bitcoin state largely deviates from the equilibrium of coexistence due to external factors such as a \textit{sudden} change in prices, then it is still possible to reach the lack of BCH loyal miners.

\smallskip
\blue{\noindent\textbf{Influence of the lack of BCH loyal miners. }
We observe that the Bitcoin system suffered from the lack of BCH loyal miners before Nov. 13, 2017. 
Consequently, the BCH transaction process speed periodically became low, and it even took about four hours to generate one block in some cases. 
Moreover, we can see that BCH was significantly centralized during the period in which the BCH mining difficulty is high. 
For example, when considering blocks generated from Oct. 2 to Oct. 4, only two accounts generated about 70 \% of blocks and there were only five miners who conducted BCH mining. 
We note that, in blockchain systems using a PoW mechanism, high mining power is an essential factor for high security blockchain systems.
In practice, BCH before Nov. 13, 2017 was susceptible to double spending attacks with only 1$\sim$2\% of the total computational power in the Bitcoin system. 
There is also selfish mining~\cite{eyal2014majority}, which makes the attacker unfairly earn the extra reward while others suffer a loss.
Because of a decrease in $r_{\mathcal B},$ these attacks can be executed with relatively small mining power. 
As a result, fickle mining, which heavily occurred before Nov. 13, 2017, weakened the performance, decentralization level, and security of the BCH system.}

\smallskip
\noindent\textbf{Influence of the hard fork of BCH. }
Next, we discuss why Bitcoin moved toward different equilibria before and after Nov. 13, 2017. 
First, in the Bitcoin system before Nov. 13, 2017, $r_{\mathcal F}$ considerably increased as can be seen in Figure~7a-(2).
Meanwhile, after Nov. 13, 2017, $r_{\mathcal F}$ did not considerably increase even though the point passed \3.
This can be attributed to the different difficulty adjustment algorithms before and after Nov. 13, 2017; the mining difficulty of BCH is currently adjusted faster than that before Nov. 13, 2017.
Therefore, currently, to conduct fickle mining, miners must switch between BTC and BCH relatively fast; this would make the current fickle mining in the Bitcoin system annoying.
Then, can we regard the current state of BCH to be safe if the system avoids external factors such as a \textit{sudden} change in prices?
We delay the answer until Section~\ref{sec:attack}. 

\blue{
\subsection{The "hash war" between Bitcoin ABC and Bitcoin SV}
\label{subsec:stick}

According to our model, we also describe the ``hash war" that recently occurred between Bitcoin ABC (ABC) and Bitcoin SV (BSV), which are derived from the original BCH on Nov. 15, 2018. 
In this paper, we call `Bitcoin ABC' ABC rather than BCH to avoid confusion with the original BCH even though Bitcoin ABC is currently regarded as BCH~\cite{hashwar}.
This war was caused by the conflict over a BCH update that adds a new \textit{opcode}, where the BCH factions split into a reformist group and an opposing group.
As a result, this conflict caused the two factions to make their own chain, where the reformist group is the ABC faction led by Roger Ver (the owner of \textit{Bitcoin.com}~\cite{bitcoin.com}) and Jihan Wu (the cofounder of Bitmain and also the owner of BTC.com~\cite{btc.com} and Antpool~\cite{antpool}) and the opposing group is the BSV faction led by Craig Wright and Calvin Ayre (the CEO of Coingeek~\cite{coingeek}). 
This split of the original BCH was achieved by a hard fork on Nov. 15, 2018, and each faction wanted its own chain to be the longest chain in order to unify the divided BCH. 
This fact makes both factions desperately conduct mining of their coins with vast computational power; thus the hash war occurred from Nov. 15, 2018 to Nov. 24, 2018. 
Such behavior of ABC and BSV factions would influence on a general miner who choose its coin among BTC, ABC, and BSV, and we analyze this situation by dividing into two games: 1) a game between BTC and ABC and 2) another game between BTC and BSV. 
In both games, $c_{\tt stick}$ became significantly high during the hash war period, and we can consider this situation as Case 4 ($c_{\tt stick}>\frac{k}{k+1}$).

\begin{figure}[ht]
  \centering
  \includegraphics[width=.9\columnwidth]{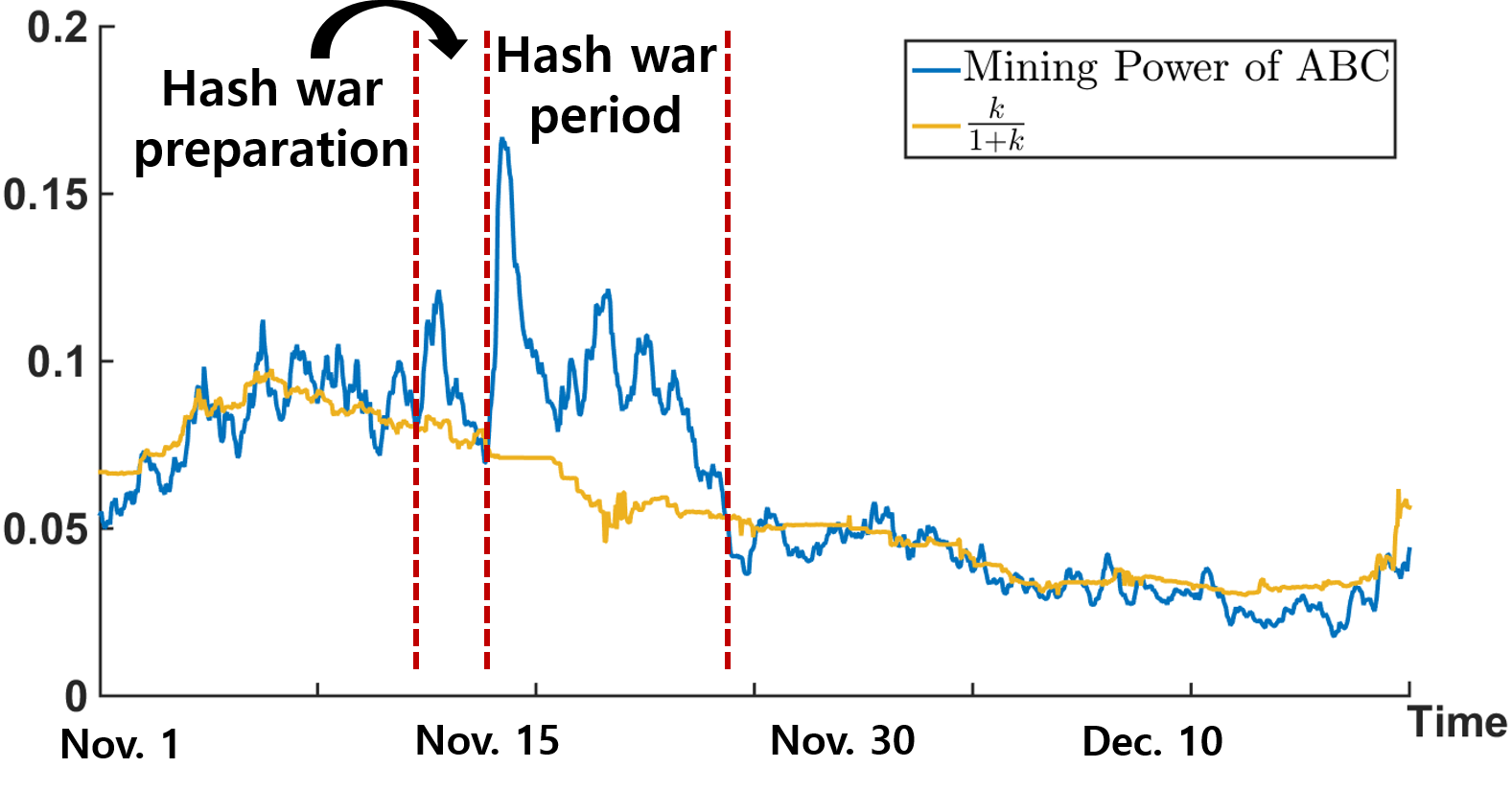}
\caption{The data for ABC from \textbf{Nov. 1, 2018} to Dec. 20 2018 is represented. 
The mining power of ABC is expressed as a relative value to the total power in BTC and ABC, and 
$k$ indicates a relative price of ABC to that for BTC. 
}
    \label{fig:bitcoincash}
\end{figure}

\begin{figure}[ht]
  \centering
  \includegraphics[width=.9\columnwidth]{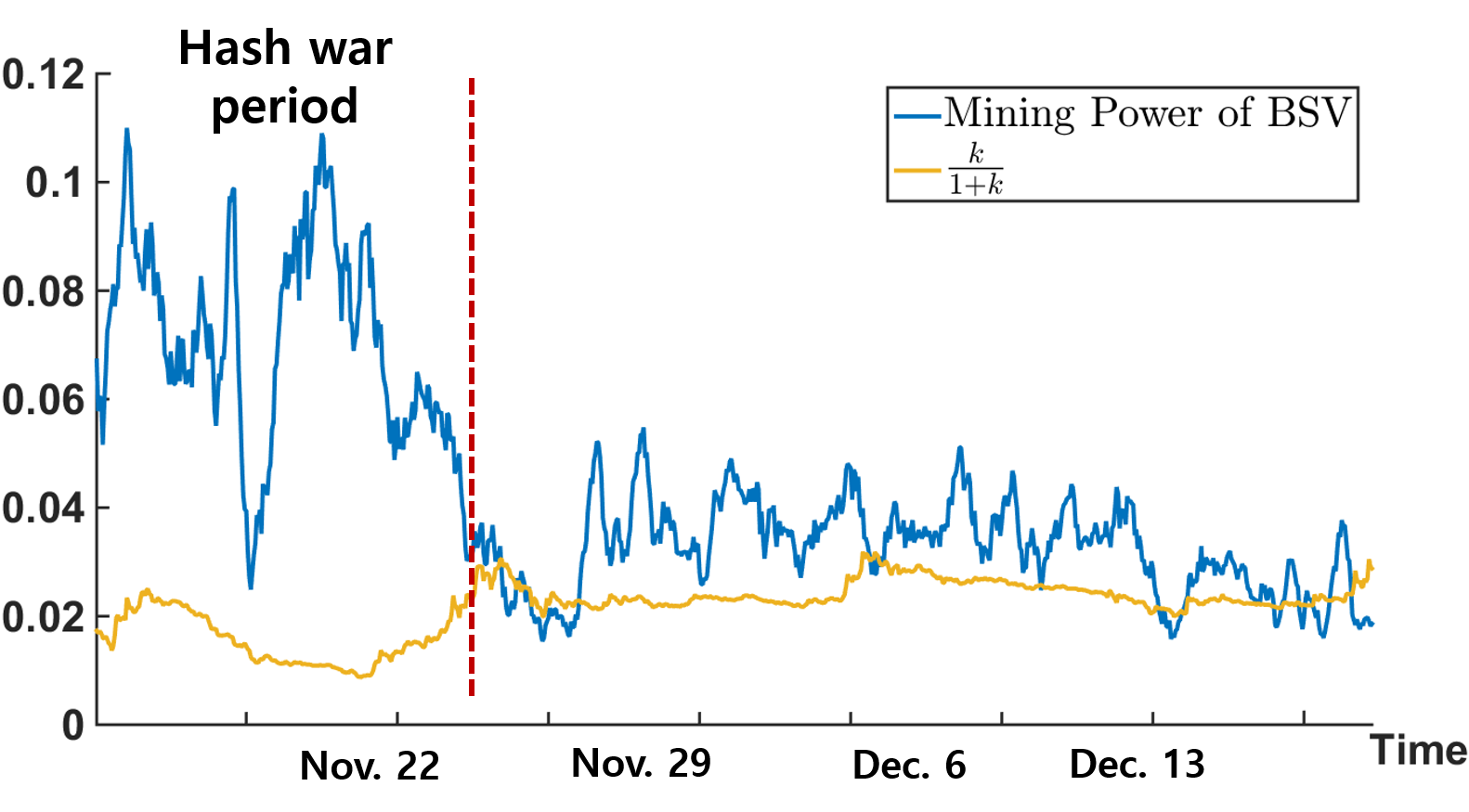}
\caption{The data for BSV from \textbf{Nov. 15, 2018} to Dec. 20 2018 is represented. 
In this figure, mining power of BSV is expressed as a relative value to the total power in BTC and BSV, and 
$k$ indicates a relative price of BSV to that for BTC. 
}
    \label{fig:bitcoinsv}
\end{figure}


\begin{figure}[ht]
  \centering
  \includegraphics[width=.9\columnwidth]{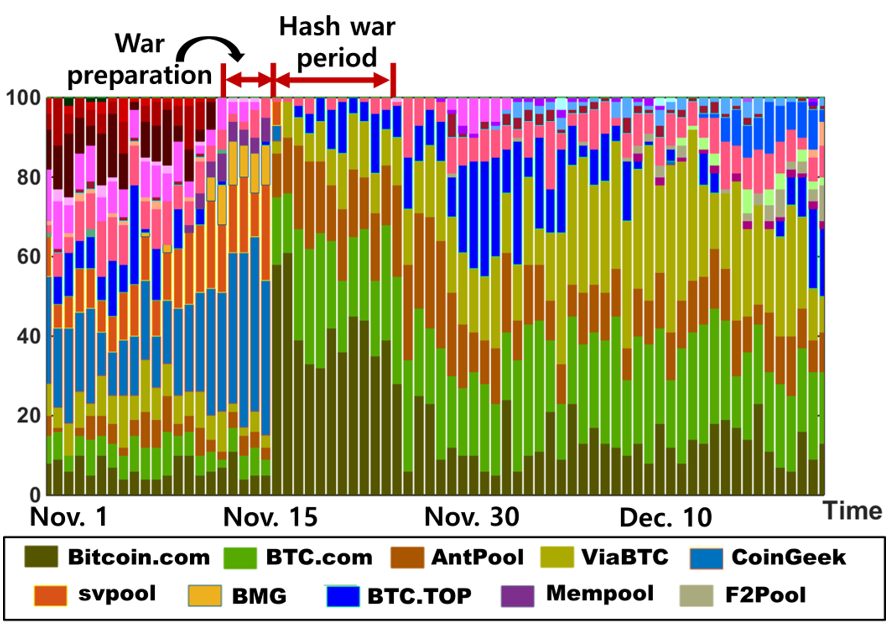}
\caption{The $x$ and $y$-axes represent time from \textbf{Nov. 1, 2018} to Dec. 20, 2018 and the number of ABC blocks generated by each miner in previous 100 blocks, respectively.
The name of a miner corresponding to each color is presented at the bottom of this figure. 
}
    \label{fig:bitcoincash_dist}
\end{figure}

\begin{figure}[ht]
  \centering
  \includegraphics[width=.9\columnwidth]{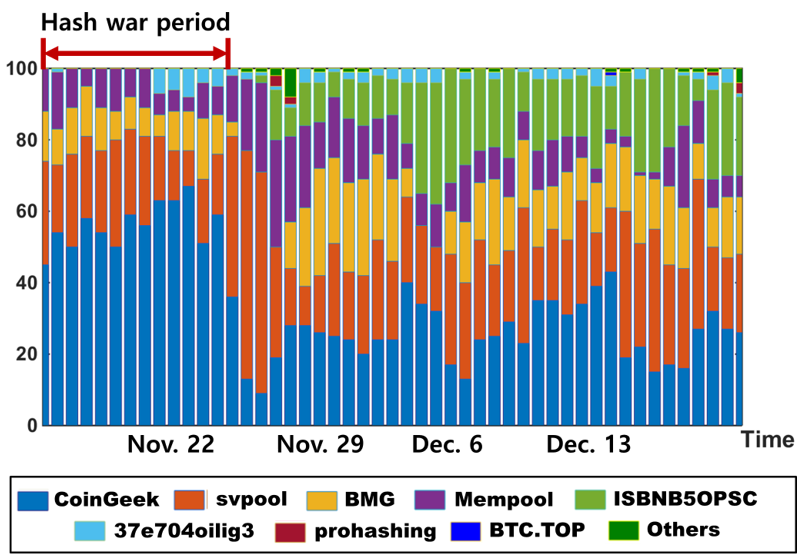}
\caption{The $x$ and $y$-axes represent time from \textbf{Nov. 15, 2018} to Dec. 20, 2018 and the number of BSV blocks generated by each miner in previous 100 blocks, respectively.
The name of a miner corresponding to each color is presented at the bottom of this figure. 
}
    \label{fig:bitcoinsv_dist}
\end{figure}

\begin{figure}[ht]
  \centering
  \includegraphics[width=0.65\columnwidth]{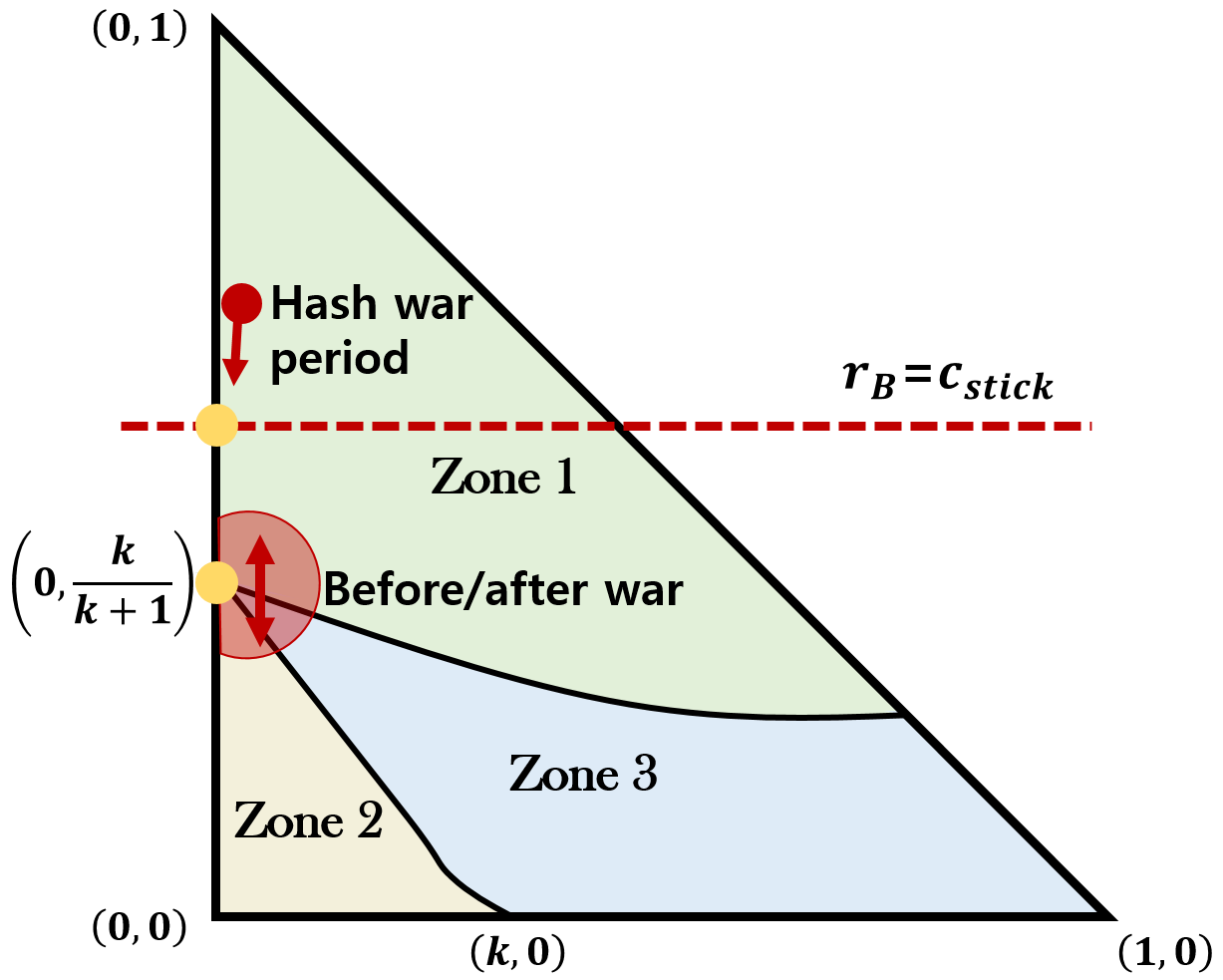}
\caption{This figure describes the movement of state for hash war period and the movement of state before and after war. 
}
    \label{fig:movement_war}
\end{figure}

To analyze a phenomenon that appeared due to the hash war, we collect the data for ABC and BSV. 
Figure~\ref{fig:bitcoincash} and \ref{fig:bitcoinsv} show the ABC data history from Nov. 1, 2018 to Dec. 20, 2018 and the BSV data history from Nov. 15, 2018 to Dec. 20, 2018, respectively. 
Note that BSV was released on Nov. 15, 2018. 
In Figure~\ref{fig:bitcoincash}, the mining power of ABC is presented as a relative value to the total mining power of ABC and BTC, and $\frac{k}{k+1}$ is also presented, where $k$ indicates a relative price of ABC to that for BTC. 
Figure~\ref{fig:bitcoinsv} depicts the data history of BSV like Figure~\ref{fig:bitcoincash}.
These figures show that the state $(r_{\mathcal{F}}, r_{\mathcal{B}})$ in the two games was above the state $(0, \frac{k}{1+k})$ during the hash war period. 

Moreover, to determine the movement of the state for the hash war period, we investigate the history of ABC computational power distribution among miners from Nov. 1, 2018 to Dec. 20, 2018 and that for BSV from Nov. 15, 2018 to Dec. 20, 2018. 
This is because it would be hard to determine the movement of the state through just the mining power history (i.e., Figure~\ref{fig:bitcoincash} and \ref{fig:bitcoinsv}) because $c_{\tt stick}$ significantly changed during this period. 
Figure~\ref{fig:bitcoincash_dist} and \ref{fig:bitcoinsv_dist} represent the changes in the mining power distribution of ABC and BSV over time, respectively. 
To do this, we crawled coinbase transactions and analyzed the number of blocks mined by each miner among previous 100 blocks.
In these figures, each miner corresponds to one color, and the length of one colored bar represents the number of blocks generated by the corresponding miner among 100 blocks. 
Therefore, the number of colors in the entire bar indicates the number of active miners at the corresponding time.
Note that only names of ten miners are presented in Figure~\ref{fig:bitcoincash_dist}.

First, we consider the game between BTC and ABC. 
One can see that the state $(r_{\mathcal F}, r_{\mathcal B})$ jumps to a point above $(\frac{k}{k+1},0)$ for the hash war preparation period (from Nov. 13, 2018 to Nov.15, 2018) through Figure~\ref{fig:bitcoincash}. 
Such an increase in the ABC mining power may be explained because the mining power of BSV factions such as CoinGeek, svpool, BMG pool, and Mempool increased from the hash war preparation~\cite{mempool} as shown in Figure~\ref{fig:bitcoincash_dist}. 
In other words, the increase in the ABC mining power for the hash war preparation is because $c_{\tt stick}$ increased. 
On the other hand, Figure~\ref{fig:bitcoincash_dist} shows that some miners left the ABC system during the war preparation (the colors that appeared at the top of the figure before the war preparation period disappeared from the war preparation period). 
This fact indicates that the state moves toward the line $r_{\mathcal B}=c_{\tt stick}$ in the case that $c_{\tt stick}$ is large. 
Note that the reason why the ABC mining power decreases at the end of the hash war preparation period (i.e., the start of the hash war) is that BSV factions move to the BSV system.

Next, for the hash war period, the ABC mining power increased because the ABC factions such as Bitcoin.com increased their mining power (i.e., $c_{\tt stick}$ increased)~\cite{hashwar}. 
However, there were only a few loyal ABC miners during this period. For example, at the start of the hash war, only five miners exist: Bitcoin.com, BTC.com, AntPool, ViaBTC, and BTC.TOP. 
Note that all of them are the ABC factions (ViaBTC and BTC.TOP announced that they support ABC~\cite{viabtc_support, btc.top_support}). 
As a result, we can see that this state is close to the state $r_{\mathcal B}=c_{\tt stick},$ which represents a lack of BCH loyal miners.  
This state makes the ABC system severely centralized. 
In particular, one miner (Bitcoin.com) possessed about 60 \% of the total computational power in some cases, which indicates the breakage of censorship resistance. 
Meanwhile, after the hash war (i.e., when $c_{\tt stick}$ is less than $\frac{k}{k+1}$), one can see that more other miners gradually enter the ABC system (see the increase in the number of colors after the hash war in Figure~\ref{fig:bitcoincash_dist}). 
In addition, Figure~\ref{fig:bitcoincash} shows that the state is close to $\frac{k}{k+1}$ after the hash war. 
As a result, the state moves as shown in Figure~\ref{fig:movement_war}. 

Second, we describe the game between BTC and BSV through Figure~\ref{fig:bitcoinsv} and \ref{fig:bitcoinsv_dist}. 
As shown in Figure~\ref{fig:bitcoinsv}, the state is above $(0,\frac{k}{k+1})$ for the hash war period because $c_{\tt stick}$ is significantly high. 
This fact is also presented in Figure~\ref{fig:bitcoinsv_dist}. 
Note that CoinGeek, svpool, BMG, and Mempool are BSV factions. 
Therefore, the state was close to $r_{\mathcal B}=c_{\tt stick}$ at the time. 
Similar to ABC, BSV also suffered from the severe centralization due to a lack of loyal miners. 
However, the other miners have entered the BSV system after the hash war, and the state became close to $(0,\frac{k}{k+1})$. 
Therefore, Figure~\ref{fig:movement_war} represents the state movement, and this result empirically confirms our theoretical analysis. 

Here, note that when the state is located above $\frac{k}{k+1},$ $\Omega_{\tt stick}$ suffers a loss. 
This fact makes the state $c_{\tt stick}>\frac{k}{k+1}$ would not last for a long time.  
Therefore, the hash war was also not able to continue for a long time, and the hash war ended with BSV's surrender~\cite{surrender}.}

\section{Broader Implications}
\label{sec:attack}

In this section, we describe broader implications of our game model.
More precisely, we first describe the risk of automatic mining, and then explain how one coin can exploit this risk to {\emph{intentionally} steal the loyal miners from other less valued coins} with negligible efforts and resources. 

\subsection{A potential risk of automatic mining}
\label{sec:auto}

As described above, the current state of Bitcoin is close to coexistence between BTC and BCH 
because faster BCH mining difficulty adjustment makes \textit{manual} fickle mining inconvenient. 
We introduce another possible mining scheme called \emph{automatic mining}, which can be less affected by faster mining difficulty adjustment. Automatic mining is designed for miners to automatically switch the coin to mine to the likely most profitable one of the compatible coins by analyzing their mining difficulty and coin prices in real time unlike fickle mining. 
Here, note that all automatic miners almost \textit{simultaneously} change their coin when not only mining difficulty but also coin prices changes. 
Indeed, automatic mining can be considered to be automatically choosing the most profitable one among three strategies, \F, \A, and $\mathcal{B}$ in real time.
Automatic mining has been executed in the Bitcoin system~\cite{automatic} and has already become popular in the altcoin system~\cite{multipool}.
Indeed, mining power increases and decreases by more than a factor of four in most altcoins several times a day~\cite{altcoin_auto}.
We describe a simple implementation of automatic mining below.

\begin{figure}[ht]
    \centering
    \includegraphics[width=.9\columnwidth]{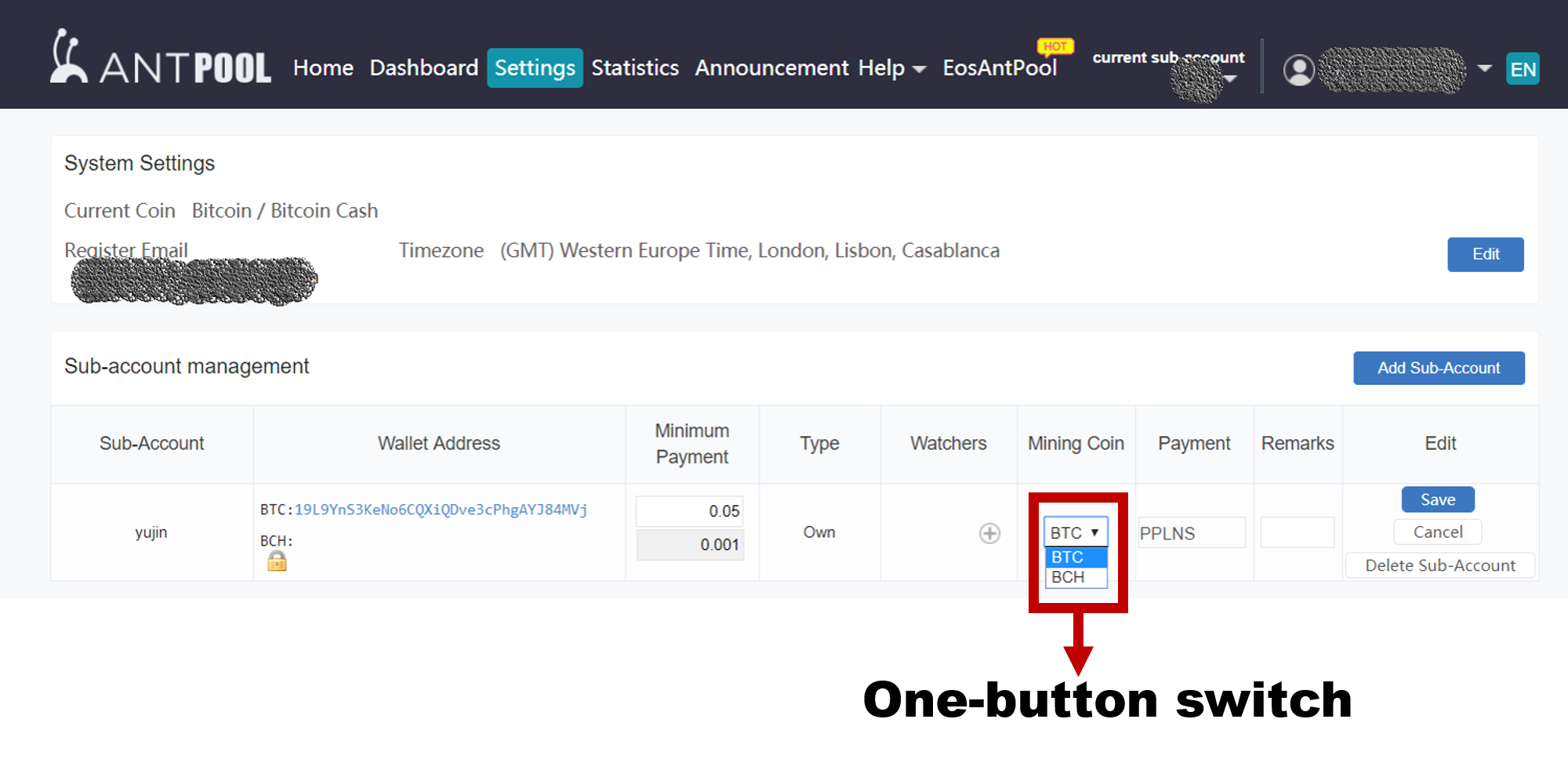}
    \caption{One-button switching mining in Antpool}
    \label{fig:one-button}
\end{figure}

Currently, many mining pools, including BTC.com, Antpool, and ViaBTC, support interactive user interfaces for switching the coin to mine by just clicking one button.
Figure~\ref{fig:one-button} represents the one-button switching mining feature provided by Antpool.
This feature makes automatic mining easier without technical difficulties in implementing this approach.
For example, a miner can conduct automatic mining in Antpool as follows. 
\begin{enumerate}
    \item First, the miner saves an HTTP header with its cookies to maintain the login session.
    \item To determine which coin is more profitable, the miner calculates the mining profitability of BTC and BCH.
    In real-world settings, this can be simply implemented by using real-time coin prices~\cite{BTCprice, BCHprice} and the coin mining difficulty.
    \item If BTC mining is more profitable than BCH mining, the miner sends an HTTP request, which includes the saved HTTP header and data for switching to BTC mining. 
    Otherwise, the miner sends an HTTP request to conduct BCH mining.
    \item The above steps are repeated.
\end{enumerate}
As shown in the code~\cite{automatic_mining}, this automatic mining can be executed within about 50 lines in Python.

Large-scale automatic mining makes the state of the coin system enter \3. 
As a simple example, we can consider an extreme case wherein the entire computational power is involved in automatic mining. 
In this case, {any initial state except for $(0,\frac{k}{k+1})$ immediately reaches the equilibrium $r_{\mathcal B}=c_{\tt stick}$} as soon as all miners start automatic mining.
This is because all automatic miners should simultaneously choose the same coin and would eventually mine $coin_{\tt A}$ when the mining difficulty of $coin_{\tt B}$ increases. 

Then, we have the following question: What ratio of automatic mining power is needed to {reach the lack of $coin_{\tt B}$-loyal miners?}
As shown in Figure~\ref{fig:zones}, 
the state $(r_{\mathcal F},r_{\mathcal B})$ cannot be in \2 when $r_{\mathcal F}$ is not less than $k.$ 
Therefore, $(r_{\mathcal F},r_{\mathcal B})$ where $r_{\mathcal F}\geq k$ would move in the decreasing direction of $r_{\mathcal B}$.
Further, even manual miners who do not conduct automatic mining would prefer $coin_{\tt A}$ rather than $coin_{\tt B}$ at states in \3 where $r_{\mathcal F}\geq k$ because $coin_{\tt A}$-only mining is more profitable than $coin_{\tt B}$-only mining at the states; loyal miners of $coin_{\tt B}$ should generate blocks with high difficulty. 
Therefore, when a fraction $k$ of the total mining power is involved in the automatic fickle mining, the state moves towards a lack of $coin_{\tt B}$-loyal miners. 
As of Dec. 2018, because $k$ in the Bitcoin system is about 0.05, if 5\% of the total mining power in the Bitcoin system 
is involved in automatic mining, the automatic miners would conduct (automatic) fickle mining and the state would enters \3.
Note that if automatic miners of which the total mining power is 5\% conduct $coin_{\tt A}$-only (or $coin_{\tt B}$-only) mining, the state would enter \2 (or \1). 
This is contradiction because the automatic miners should choose the most profitable strategy. 
{As a result, when only 5\% of the total mining power is involved in the automatic mining, the number of BCH loyal miners decreases and the BCH system is finally becoming more centralized.}



\subsection{Injuring rivalry coins}
\label{subsec:rivalry}
In Section~\ref{sec:application}, we explained how our game $\mathcal G (\bm{c},c_{\tt stick})$ can be applied to the Bitcoin system regardless of the BCH mining difficulty adjustment algorithm. 
To generalize our game model, we here consider two types of possible mining difficulty adjustment algorithms: The first type of algorithm is to adjust the mining difficulty in a long time period (e.g., two weeks) while the second type of algorithm is to adjust the mining difficulty every block or in a short time period in order to promptly respond to the changes in the mining power. 
In the real-world, both types of these mining difficulty adjustment algorithms are mostly used. For example, BTC and Litecoin are the cryptocurrency systems using the first type, 
while many altcoins including BCH, Ethereum (ETH), and Ethereum Classic (ETC) are currently using the second type. 

We can generalize our game model to any coin system satisfying the following conditions.
\begin{enumerate}
    \item Two existing coins share the same mining hardware.
    \item The more valued coin $coin_{\tt A}$ between those coins has the first type of mining difficulty adjustment algorithm. 
\end{enumerate}

We note that there is no restriction on the mining difficulty adjustment algorithm for the less valued $coin_{\tt B}$ in our game model $\mathcal G_\infty$. 
When $coin_B$ has the first type of mining difficulty adjustment algorithm, our model can be applied according to Section~\ref{sec:model}. Note that we modeled our game in Section~\ref{sec:model}, assuming that $coin_B$ has the first type of mining difficulty adjustment algorithm. 
In addition, in Section~\ref{sec:application}, we described why our game can be applied to when $coin_B$ has the second type of mining difficulty adjustment algorithm. 
Therefore, regardless of $coin_{\tt B}$ mining difficulty adjustment algorithm, in the coin system 
satisfying the above two conditions, the $coin_{\tt B}$-loyal miners would leave if at least $k$ fraction of the total mining power is involved in automatic mining. 

 

Next, we explain {how the more valued coin can steal loyal miners from the other less valued rivalry coin. 
If $coin_{\tt A}$ utilizes the first type of mining difficulty adjustment algorithm, the number of $coin_{\tt B}$-loyal miners would naturally decrease due to the automatic mining.} 
Again note that this situation periodically weakens the health of the $coin_{\tt B}$ system in terms of security and decentralization. On one hand, if $coin_{\tt A}$ has a mining difficulty adjustment algorithm different from the first type 
(i.e., different from that in Assumption~\ref{ass:c}), our game model may not be applied. 
For example, when considering the Ethereum system consisting of ETH and ETC, ETH corresponding to $coin_{\tt A}$ has a different difficulty adjustment algorithm from that which we assumed in our game. 
In this case, even if $r_{\mathcal B}=0,$ the complete downfall of $coin_{\tt B}$ (e.g., ETC) may not occur and the mining power of $coin_{\tt A}$ and $coin_{\tt B}$ would fluctuate heavily. 
Therefore, to follow our game and so steal the loyal miners from $coin_{\tt B}$, $coin_{\tt A}$ should change its mining difficulty adjustment algorithm through a hard fork. 
We can see that some cryptocurrency systems (e.g., BCH, ETH, and ETC) have often performed hard forks to change their mining difficulty adjustment algorithms~\cite{bch_hardfork, eth_hardfork, etc_hardfork}. 
This indicates that cryptocurrency systems can practically update their mining difficulty adjustment algorithms if needed.

In conclusion, if the mining difficulty adjustment algorithm for $coin_{\tt A}$ is changed to the first type of mining difficulty adjustment algorithms, {a lack of loyal miners for $coin_{\tt B}$ might be reached due to automatic mining.} 


\section{Discussion}
\label{sec:discuss}

In this section, we first discuss \blue{how $coin_{\tt B}$ can maintain its loyal miners} 
and consider environmental factors that may 
affect our game analysis results.


\blue{\subsection{Maintenance of $coin_{\tt B}$-loyal miners}}


\blue{As described in Section~\ref{subsec:rivalry}, $coin_{\tt B}$ cannot prevent the rivalry coin from stealing loyal miners by changing its difficulty adjustment algorithm alone.}
Surely, the most straightforward way to avoid the risk is to not use the mining hardware compatible with $coin_{\tt A}$. That is, a proprietary mining algorithm, requiring customized mining hardware which is not compatible with $coin_{\tt A}$, should be introduced for $coin_{\tt B}$. However, this solution is not applicable in practice for small and medium-sized mining operators because it is expensive to develop customized mining hardware (e.g., ASICs). In fact, because many altcoins use a mining algorithm that can be implemented in CPU or GPU, automatic mining endangers their mining power, weakening their security. 

The second way is to use auxiliary proof-of-work (or merged mining), which makes a miner conduct mining more than two coins at the same time~\cite{merged_mining}. Therefore, our first assumption in Section~\ref{sec:model} is not satisfied by merged mining, and our game results would not be applied. 
This is also regarded as a potential solution to 51\% attacks because it significantly increases mining power of altcoins~\cite{merged_mining2}. 
However, despite of such definite advantages, most projects do not adopt merged mining because of following reasons: It is complex to implement merged mining, and miners should do additional work~\cite{merged_mining2}.

The another way is to increase the price of $coin_{\tt B}$ through price manipulation. 
However, as far as we know, the problem of maintaining the increased coin price through price manipulation is not well-studied. 
Moreover, we can consider a way to increase the relative incentive of $coin_{\tt B}$ mining to $coin_{\tt A}$ mining, where it can be achieved by increasing the block reward or decreasing the average time of block generation. 
\blue{Even though this method may help prevent the rivalry coin from stealing loyal miners, it would cause other side effects such as inflation or the increase in fork rate~\cite{carlsten2016instability,gervais2016security}.}



Lastly, $coin_{\tt B}$ can change its consensus protocol, the PoW mechanism, to another protocol.
However, \blue{this process would not be supported by existing miners in $coin_{\tt B}$.}
For example, Ethereum is planning to switch from a proof-of-work mechanism to a proof-of-stake mechanism for several years. 
However, note that if the consensus protocol is just changed through a hard fork, \blue{the existing miners may leave because they can lose their own merits (e.g., powerful hardware capability) for mining $coin_{\tt B}$.}

\subsection{Environmental factors}
\label{sec:practical}

In practice, miners' behavior can deviate from our model because of the following environmental factors. 

\smallskip\noindent\textbf{Not all miners are rational.}
First, miners are not always rational or wise. 
Even if fickle mining or $coin_{\tt A}$ mining is more profitable than $coin_{\tt B}$ mining, 
some miners may be reluctant to engage in fickle mining or $coin_{\tt A}$ mining because they may not recognize the profitability in doing so.
\blue{However, our data analysis confirms that most miners are rational. 
In addition, if miners use the automatic mining function, they would always follow the most profitable strategy.}

\smallskip
\blue{\noindent\textbf{Some miners consider the long-term price of coins.}
Because price prediction is significantly difficult~\cite{longterm}, we believe that most miners behave depending on the short-term price of a coin rather than the long-term price. 
For example, who could have predicted the hash war between ABC and BSV in advance? 
Therefore, as can be seen from the history of the Bitcoin system, most miners behave depending on short-term profits. To model more realistic and general situations, our model considered both rational miners who are interested in short-term profits and $coin_B$ factions ($\Omega_{\tt stick}$) which are interested in long-term profits.}

\blue{\smallskip
\noindent\textbf{Some miners prefer the stable coexistence of coins.}
Some miners may want the stable coexistence of coins for coin market stability,  
and they may try to reach the equilibrium representing the coexistence of coins regardless of their profits.
If the fraction of such miners is large, a state would move to the equilibrium $(0,\frac{k}{k+1})$ regardless of its zones.}
Based on historical observations of the Bitcoin system, however, the fraction of these miners seems unlikely to be high in the real-world. 

\smallskip
\noindent\textbf{Other selfish mining.}
In this study, we considered only fickle mining, which is a type of 
rational mining.
However, miners engaging in various form of selfish mining~\cite{eyal2014majority,eyal2015miner,luu2015power,kwon2017selfish} might cause a deviation from our analysis.

\section{Conclusion}
\label{sec:conclude}

\blue{In this study, we modeled and analyzed the game between two coins for fickle mining, and our results imply that fickle mining can lead to a lack of loyal miners in the less valued coin system. 
We confirm that this lack of loyal miners can weaken the overall health of coin systems by analyzing real-world history.}
In addition, our analysis is extended to the analysis of automatic mining, which shows a potentially severe risk of automatic mining. 
As of Dec. 2018, BCH's loyal miners would leave if more than about 5\% of the total mining power in BTC and BCH is involved in automatic mining. 
\blue{Moreover, we explained how one coin can steal the loyal miners from other less valued rivalry coins in the highly competitive coin market by generalizing our game model. 
We believe that this is one of the serious threats for a cryptocurrency system using a PoW mechanism.}




\section*{Acknowledgment}
\addcontentsline{toc}{section}{Acknowledgment}
We are very grateful to the anonymous reviewers and Andrew Miller, the contact point for major revision of this paper.

\bibliographystyle{IEEEtran}
\bibliography{references}

\begin{appendices}

\section{Proof of Lemma~\ref{lem:dynamics} and Theorem~\ref{thm:eq}} \label{sec:pf_eq}

\blue{In order for player $i$ to not change its strategy at $(r_{\mathcal F},r_{\mathcal B})$, the below inequalities should be satisfied.

\begin{align}
&\begin{cases}
&U_{\mathcal{F}}(r_{\mathcal F},r_{\mathcal B})\geq U_{\mathcal{A}}(r_{\mathcal F}-c_i,r_{\mathcal B}),\\
&U_{\mathcal{F}}(r_{\mathcal F},r_{\mathcal B})\geq U_{\mathcal{B}}(r_{\mathcal F}-c_i,r_{\mathcal {B}}+c_i)\label{eq:fm}
\end{cases}\\
&\begin{cases}
&U_{\mathcal{A}}(r_{\mathcal F},r_{\mathcal B})\geq U_{\mathcal{F}}(r_{\mathcal F}+c_i,r_{\mathcal B}), \\
&U_{\mathcal{A}}(r_{\mathcal F},r_{\mathcal B})\geq U_{\mathcal{B}}(x,r_{\mathcal {B}}+c_i)\label{eq:am}
\end{cases}\\
&\begin{cases}
&U_{\mathcal{B}}(r_{\mathcal F},r_{\mathcal B})\geq U_{\mathcal{F}}(r_{\mathcal F}+c_i,r_{\mathcal B}-c_i),\\
&U_{\mathcal{B}}(r_{\mathcal F},r_{\mathcal B})\geq U_{\mathcal{A}}(x,r_{\mathcal B}-c_i)\label{eq:bm}
\end{cases}
\end{align}
\eqref{eq:fm} represents that a fickle miner's payoff decreases when the fickle miner moves to \AM (i.e., $U_{\mathcal{F}}(r_{\mathcal F},r_{\mathcal B})\geq U_{\mathcal{A}}(r_{\mathcal F}-c_i,r_{\mathcal B})$) or 
when it moves to \BM (i.e., $U_{\mathcal{F}}(r_{\mathcal F},r_{\mathcal B})\geq U_{\mathcal{B}}(r_{\mathcal F}-c_i,r_{\mathcal {B}}+c_i)$). }
Similarly,  \eqref{eq:am} and \eqref{eq:bm} represent that players in \AM and \BM cannot increase
their payoff by changing their strategy, respectively.

\blue{To prove Lemma~\ref{lem:dynamics}, we first consider the case that $c_{\tt stick}=0,$ and have the following steps.
\begin{enumerate}
    \item First, we find all states characterized as $(r_{\mathcal F},0)$ in which player $i$ does not change its strategy.  
    \item Second, we show that there is no state characterized as $(0,r_{\mathcal B})$ in which player $i$ does not change its strategy.
    \item Finally, there exists $\varepsilon>0$ such that, for any player $i$ with $c_i<\varepsilon,$ a player $i$ can change its strategy at state $(r_{\mathcal F},r_{\mathcal B})$ where $r_{\mathcal B}$ is positive.
\end{enumerate}}

\noindent\textbf{First step: }
We find all states characterized as $(r_{\mathcal F},0)$ in which player $i$ does not change its strategy, where we denote such a state by $S$. 
In order for $(0,0)$ to be $S$, it is sufficient that \eqref{eq:am} is satisfied. 
Meanwhile, when $r_{\mathcal F}$ is greater than 0 and less than 1, in order for $(r_{\mathcal F},0)$ to be $S$, 
not only \eqref{eq:am} but also \eqref{eq:fm} should be satisfied.
If $r_{\mathcal F}$ is 1, only \eqref{eq:fm} should be satisfied.

First, we consider the condition for $(0,0)$ to be $S$.
The payoff $U_{\mathcal{A}}(0,0)$ of players in \AM is 1, and the payoff $U_{\mathcal{F}}(c_i,0)$ of a player who changes its strategy from \A to \F is also 1. 
Because $U_{\mathcal{A}}(0,0)\geq U_{\mathcal{F}}(c_i,0),$ it is sufficient to show $U_{\mathcal{A}}(0,0)\geq U_{\mathcal{B}}(0,c_i)$ in order for $(0,0)$ to be $S$.
The payoff $U_{\mathcal{B}}(0,c_i)$ is $\frac{k}{c_i}$, and thus, 
the state $(0,0)$ cannot be a $S$ for $c_i$ less than $k.$

Next, we consider $(r_{\mathcal F},0)$ where $r_{\mathcal F}$ is greater than 0. 
In \eqref{eq:fm}, $U_{\mathcal{F}}(x,0)$ and $U_{\mathcal{A}}(r_{\mathcal F}-c_i,0)$ are 1. 
Moreover, in \eqref{eq:fm}, $U_{\mathcal{B}}(r_{\mathcal F}-c_i,c_i)\leq U_{\mathcal{F}}(x,0)$ can be arranged as follows. 

\begin{align}
&\frac{{kN_{\tt in}c_i}+kN_{\tt de}r_{\mathcal F}}{N_{\tt in}c_i^2+N_{\tt de}r_{\mathcal F}^2}\leq 1\\
\Leftrightarrow &\,\, 0 \leq N_{\tt de}r_{\mathcal F}^2-kN_{\tt de}r_{\mathcal F}-\frac{N_{\tt in}}{n}(k-c_i) \notag\\
\Leftrightarrow &\,\, r_{\mathcal F} \leq \frac{k}{2}-\frac{\sqrt{N_{\tt de}^2k^2+4N_{\tt de}N_{\tt in}(kc_i-c_i^2)}}{2N_{\tt de}} \text{  or  }\label{eq:con1} \\
&\,\,\frac{k}{2}+\frac{\sqrt{N_{\tt de}^2k^2+4N_{\tt de}N_{\tt in}(c_ik-c_i^2)}}{2N_{\tt de}} \leq r_{\mathcal F}\label{eq:con2}
\end{align}
If $c_i$ is less than $k$, \eqref{eq:con1} cannot be satisfied because the right-hand side is negative.
Also, if 
\begin{align*}
    &c_i \leq \frac{kN_{\tt in}-\sqrt{k^2 N_{\tt in}^2-4N_{\tt de}N_{\tt in}(1-k)}}{2N_{\tt in}}\leq \text{  or  } \\ 
    &k^2 N_{\tt in}^2-4N_{\tt de}N_{\tt in}(1-k) \leq 0,
\end{align*}
the left-hand side of \eqref{eq:con2} is less than or equal to 1, and $(1,0)$ is $S$. 

By \eqref{eq:am}, $U_{\mathcal{B}}(r_{\mathcal F},c_i)$ should be less than or equal to 1 in order that $(r_{\mathcal F},0)$ where $r_{\mathcal F}$ is greater than 0 and less than 1 is $S$.
Referring to \eqref{eq:con2}, the following is satisfied:
\begin{equation*}
\resizebox{\hsize}{!}{$
\frac{k}{2}+\frac{\sqrt{N_{\tt de}^2k^2+4N_{\tt de}N_{\tt in}(c_ik-c_i^2)}}{2N_{\tt de}} \leq r_{\mathcal F}+c_i
\Rightarrow U_{\mathcal{B}}(r_{\mathcal {F}},c_i) \leq 1.$}
\end{equation*}
Therefore, when 
$$r_{\mathcal F}\leq \frac{k}{2}+\frac{\sqrt{N_{\tt de}^2k^2+4N_{\tt de}N_{\tt in}(c_ik-c_i^2)}}{2N_{\tt de}},$$ 
both \eqref{eq:fm} and \eqref{eq:am} are satisfied. 
As a result, when $$c_i\leq\frac{kN_{\tt in}-\sqrt{k^2 N_{\tt in}^2-4N_{\tt de}N_{\tt in}(1-k)}}{2N_{\tt in}},$$
the all points $(\frac{k}{2}+\frac{\sqrt{N_{\tt de}^2k^2+4N_{\tt de}N_{\tt in}(c_ik-c_i^2)}}{2N_{\tt de}}\leq r_{\mathcal F}\leq 1,0)$ are $S$. 

\smallskip
\noindent\textbf{Second step: }
As the second step, we show that game $\mathcal{G} (\bm{c},c_{\tt stick})$ does not have state $(0,r_{\mathcal B})$ where $r_{\mathcal B}$ is positive and player $i$ does not change its strategy. 
In order for $(0,r_{\mathcal B})$ where $r_{\mathcal B}$ is greater than 0 and less than 1 to be $S$, both \eqref{eq:am} and \eqref{eq:bm} should be satisfied for player $i$.
First, we consider the inequality $U_{\mathcal{F}}(c_i,r_{\mathcal B}) \leq U_{\mathcal{A}}(0,r_{\mathcal B})$.
This inequality is expressed as follows: 

\begin{align}
&U_{\mathcal{F}}(c_i,r_{\mathcal B}) \leq U_{\mathcal{A}}(0,r_{\mathcal B})\notag\\
&\Leftrightarrow 
\frac{N_{de}(c_i+r_{\mathcal{B}})^2}{(1-c_i-r_{\mathcal{B}})N_{\tt in}r_{\mathcal B}^2+(1-r_{\mathcal B})N_{\tt de}(r_{\mathcal{B}}+c_i)^2}+\notag\\
&\hspace{5mm}\frac{kN_{\tt in}r_{\mathcal{B}}}{N_{\tt in}r_{\mathcal B}^2+N_{de}(r_{\mathcal{B}}+c_i)^2}\leq\frac{1}{1-r_{\mathcal B}}\notag\\
&\Leftrightarrow k(1-r_{\mathcal B})\left((1-c_i-y)N_{\tt in}r_{\mathcal B}^2+(1-r_{\mathcal B})N_{de}(r_{\mathcal {B}}+c_i)^2\right)\notag\\
&\hspace{5mm}\leq r_{\mathcal{B}}\left(1-c_i-r_{\mathcal{B}}\right)\left(N_{\tt in}r_{\mathcal B}^2+N_{de}(r_{\mathcal {B}}+c_i)^2\right)\notag\\
&\Leftrightarrow kN_{\tt de}c_i(1-r_{\mathcal B})(r_{\mathcal{B}}+c_i)^2
\leq ((1+k)r_{\mathcal B}-k)\times\notag\\
&\hspace{5mm}\left(1-c_i-r_{\mathcal B}\right)\left(N_{\tt in}r_{\mathcal B}^2+N_{de}(r_{\mathcal{B}}+c_i)^2\right)
\label{eq:pf1}
\end{align}
The another inequality $U_{\mathcal{F}}(c_i,r_{\mathcal B}-c_i) \leq U_{\mathcal{B}}(0,r_{\mathcal B})$ 
can be expressed as follows:

\begin{align}
&U_{\mathcal{F}}(c_i,r_{\mathcal B}-c_i) \leq U_{\mathcal{B}}(0,r_{\mathcal B})\notag\\
&\Leftrightarrow \frac{N_{\tt de}r_{\mathcal B}^2}{(1-r_{\mathcal B})N_{\tt in}(r_{\mathcal B}-c_i)^2+(1-r_{\mathcal B}+c_i)N_{\tt de}r_{\mathcal B}^2}+\notag\\
&\hspace{5mm}\frac{kN_{\tt in}(r_{\mathcal B}-c_i)}{N_{\tt in}(r_{\mathcal B}-c_i)^2+N_{\tt de}r_{\mathcal B}^2}\leq\frac{k}{r_{\mathcal {B}}}\notag\\
&\Leftrightarrow \left(N_{\tt in}(r_{\mathcal B}-c_i)^2+N_{\tt de}r_{\mathcal B}^2\right)\label{eq:pf2}\\
&\hspace{5mm}\times \left(N_{\tt de}r_{\mathcal B}^3-k(1-r_{\mathcal B})(N_{\tt de}r_{\mathcal B}^2-N_{\tt in}c_i(r_{\mathcal B}-c_i))\right)\notag\\
&\hspace{5mm}\leq k\left(N_{\tt de}r_{\mathcal B}^2-N_{\tt in}c_i(r_{\mathcal B}-c_i)\right)N_{\tt de}r_{\mathcal B}^2c_i\notag
\end{align}
\eqref{eq:pf2} is greater than or equal to $N_{\tt de}r_{\mathcal B}^2$.
Therefore, the following inequality 

\begin{align}
&N_{\tt de}r_{\mathcal B}^3-k(1-r_{\mathcal B})(N_{\tt de}r_{\mathcal B}^2-N_{\tt in}c_i(r_{\mathcal B}-c_i))\leq \notag\\
&kc_i
\left(N_{\tt de}r_{\mathcal B}^2-N_{\tt in}c_i(r_{\mathcal B}-c_i)\right)\notag\\
&\Leftrightarrow  N_{\tt de}r_{\mathcal B}^3-k(1+c_i-r_{\mathcal B})(N_{\tt de}r_{\mathcal B}^2-N_{\tt in}c_i(r_{\mathcal B}-c_i))\leq 0
\label{eq:pf3}
\end{align}
should be satisfied.
We denote the left-hand side of \eqref{eq:pf3} by a function $f(c_i)$ of $c_i$.
Moreover, if \eqref{eq:pf4} is satisfied, \eqref{eq:pf1} cannot
be certainly satisfied as follows:

\begin{small}
\begin{align}
&((1+k)r_{\mathcal B}-k)N_{\tt in}<N_{de}(k(1+c_i)-(1+k)r_{\mathcal B})\label{eq:pf4}\\
\Rightarrow&((1+k)r_{\mathcal B}-k)N_{\tt in}r_{\mathcal B}^2<N_{de}(k(1+c_i)-(1+k)r_{\mathcal B})(r_{\mathcal {B}}+c_i)^2\notag\\
\Leftrightarrow & ((1+k)r_{\mathcal B}-k)(N_{\tt in}r_{\mathcal B}^2+N_{de}(r_{\mathcal {B}}+c_i)^2)<kN_{\tt de}c_i(r_{\mathcal {B}}+c_i)^2\notag\\
\Rightarrow &((1+k)r_{\mathcal B}-k)\left(1-c_i-r_{\mathcal B}\right)\left(N_{\tt in}r_{\mathcal B}^2+N_{de}(r_{\mathcal {B}}+c_i)^2\right)\notag\\
&<kN_{\tt de}c_i(1-r_{\mathcal B})(r_{\mathcal {B}}+c_i)^2\notag
\end{align}
\end{small}
Thus, if \eqref{eq:pf4} is satisfied for all $r_{\mathcal B}$ that satisfies 
\eqref{eq:pf3}, 
there would not exist $S=(0,r_{\mathcal B})$ where $r_{\mathcal B}$ is greater than 0 and less than 1 because any state 
$(0,r_{\mathcal B})$ does not satisfy both \eqref{eq:am} and \eqref{eq:bm}.

We find a condition of $c_i$ such that there is no $S=(0,r_{\mathcal B})$ where $r_{\mathcal B}$ is greater than 0 and less than 1.
In other words, we find a range of $c_i$ such that \eqref{eq:pf4} is satisfied for all $r_{\mathcal B}$ that satisfies \eqref{eq:pf3}.
Eq.~\eqref{eq:pf4} is equivalent to the following inequality 
$$r_{\mathcal B}<\frac{k}{1+k}\left(1+\frac{N_{\tt de}c_i}{N_{\tt de}+N_{\tt in}}\right).$$
When $r_{\mathcal B}$ is $\frac{k}{1+k}\left(1+\frac{N_{\tt de}c_i}{N_{\tt de}+N_{\tt in}}\right)$, 
$f(c_i)$ is a quadratic equation of $c_i$, which has a negative coefficient of $c_i^2$. 
Therefore, we can easily find a number $l$ such that, for all $c_i<l$, 
$f(c_i)$ is positive when $r_{\mathcal B}$ is $\frac{k}{1+k}\left(1+\frac{N_{\tt de}c_i}{N_{\tt de}+N_{\tt in}}\right)$.
Then, we find the derivative $\frac{\partial f(c_i)}{\partial r_{\mathcal {B}}}$, and it is expressed as
\begin{equation}
3N_{de}(1+k)r_{\mathcal B}^2-2k\left(N_{de}(1+c_i)+N_{\tt in}c_i\right)r_{\mathcal {B}}+kN_{\tt in}c_i+2kN_{\tt in}c_i^2.\label{eq:pf5}
\end{equation}
In order that the derivative is non-negative when $r_{\mathcal B}$ is not less than $\frac{k}{1+k}\left(1+\frac{N_{\tt de}c_i}{N_{\tt de}+N_{\tt in}}\right),$ a solution for $r_{\mathcal B}$ of \eqref{eq:pf5} should not exist, or 
all solutions for $r_{\mathcal B}$ should be less than $\frac{k}{1+k}\left(1+\frac{N_{\tt de}c_i}{N_{\tt de}+N_{\tt in}}\right).$
If solutions exist, they are positive. 
Therefore, when the sum of solutions, $\frac{2k\left(N_{de}(1+c_i)+N_{\tt in}c_i
\right)}{3N_{de}(1+k)},$ is less than $\frac{k}{1+k}\left(1+\frac{N_{\tt de}c_i}{N_{\tt de}+N_{\tt in}}\right)$,
the solutions are less than $\frac{k}{1+k}\left(1+\frac{N_{\tt de}c_i}{N_{\tt de}+N_{\tt in}}\right).$
In other words, when $\frac{1}{c_i}$ is greater than $2+\frac{2N_{\tt in}}{N_{\tt de}}-\frac{3N_{\tt de}}{N_{\tt de}+N_{\tt in}}$, 
the solutions are less than $\frac{k}{1+k}\left(1+\frac{N_{\tt de}c_i}{N_{\tt de}+N_{\tt in}}\right).$
As a result, if $\frac{1}{c_i}>\max\{\frac{1}{l},2+\frac{2N_{\tt in}}{N_{\tt de}}-\frac{3t}{N_{\tt de}+N_{\tt in}}\},$ 
$f(c_i)$ is positive for $r_{\mathcal F} \geq \frac{k}{1+k}\left(1+\frac{N_{\tt de}c_i}{N_{\tt de}+N_{\tt in}}\right).$
This means that, for small $c_i,$ \eqref{eq:pf1} cannot be satisfied for all $r_{\mathcal B}$ that satisfies \eqref{eq:pf3}, 
and there is no $S=(0,r_{\mathcal B})$ where $r_{\mathcal B}$ is greater than 0 and less than 1.

For a state $(0,1),$ \eqref{eq:bm} should be satisfied to be $S$.
However, the state $(0,1)$ does not satisfy \eqref{eq:bm} except for when $\frac{1}{k}\geq c_i.$
Note that $k$ is not greater than 1.
Therefore, $(0,1)$ cannot be $S$. 

\smallskip\noindent\textbf{Third step: }
To do the third step, we consider the game when a player possesses sufficiently small power. 
When $r_{\mathcal B}$ is positive, inequality $\lim_{c_i\rightarrow 0}U_{\mathcal{F}}(r_{\mathcal F}+c_i,r_{\mathcal B}) \leq 
U_{\mathcal{A}}(r_{\mathcal F},r_{\mathcal B})$ is as follows.
\begin{align}
&\lim_{c_i\rightarrow 0}U_{\mathcal{F}}(r_{\mathcal F}+c_i,r_{\mathcal B}) \leq U_{\mathcal{A}}
(r_{\mathcal F},r_{\mathcal B})\notag\\
\Leftrightarrow &\,\, U_{\mathcal{F}}(r_{\mathcal F},r_{\mathcal B})\leq U_{\mathcal{A}}(r_{\mathcal F},r_{\mathcal B})\notag\\
\Leftrightarrow &\,\,\frac{k}{N_{\tt in}r_{\mathcal B}^2+N_{\tt de}(r_{\mathcal F}+r_{\mathcal B})^2}\\
&\leq \frac{r_{\mathcal {B}}}{(1-r_{\mathcal F}-r_{\mathcal B})N_{\tt in}r_{\mathcal B}^2+(1-r_{\mathcal B})N_{de}(r_{\mathcal F}+r_{\mathcal B})^2}\\
\Leftrightarrow &\,\,k((1-r_{\mathcal F}-r_{\mathcal B})N_{\tt in}r_{\mathcal B}^2+(1-r_{\mathcal B})N_{\tt de}(r_{\mathcal F}+r_{\mathcal B})^2)\notag\\
&\leq r_{\mathcal B}(N_{\tt in}r_{\mathcal B}^2+N_{de}(r_{\mathcal F}+r_{\mathcal B})^2)
\label{eq:pf1_c}
\end{align}
Also, inequality $\lim_{c_i\rightarrow 0}
U_{\mathcal{F}}(r_{\mathcal F}+c_i,r_{\mathcal B}-c_i) \leq U_{\mathcal{B}}(r_{\mathcal F},r_{\mathcal B})$ is as follows. 
%
\begin{align}
\lim_{c_i\rightarrow 0}&
U_{\mathcal{F}}(r_{\mathcal F}+c_i,r_{\mathcal B}-c_i) \leq U_{\mathcal{B}}(r_{\mathcal F},r_{\mathcal B})\notag\\
\Leftrightarrow &\,\, U_{\mathcal{F}}(r_{\mathcal F},r_{\mathcal B})\leq U_{\mathcal{B}}(r_{\mathcal F},r_{\mathcal B})\notag\\
\Leftrightarrow & \,\,(r_{\mathcal F}+r_{\mathcal B})(N_{\tt in}r_{\mathcal B}^2+N_{\tt de}(r_{\mathcal F}+r_{\mathcal B})^2)\notag\\
&\hspace{-8mm}\leq k((1-r_{\mathcal F}-r_{\mathcal B})N_{\tt in}r_{\mathcal B}^2+(1-r_{\mathcal B})N_{de}(r_{\mathcal F}+r_{\mathcal B})^2)
\label{eq:pf2_c}
\end{align}
The solution which satisfies both \eqref{eq:pf1_c} and \eqref{eq:pf2_c}
is only $(0,\frac{k}{1+k})$.
When $r_{\mathcal B}$ is greater than $\frac{k}{1+k}$, only \eqref{eq:pf1_c} is satisfied
for $(0,r_{\mathcal B})$.
Meanwhile, if $r_{\mathcal B}$ is less than $\frac{k}{1+k}$, only \eqref{eq:pf2_c} is 
satisfied for $(0,r_{\mathcal B})$.
Therefore, the range of $(r_{\mathcal F},r_{\mathcal B})$, which satisfies \eqref{eq:pf1_c}, is always above that for \eqref{eq:pf2_c} except for $(0,\frac{k}{k+1})$ and $r_{\mathcal B}=0$ (see Figure~\ref{fig:zones}).
It means that there exists a value $\varepsilon$ such that, for all $c_i<\varepsilon$ and given a positive real number $\delta$,
the line where $U_{\mathcal{F}}(r_{\mathcal F}+c_i,r_{\mathcal B}) =
U_{\mathcal{A}}(r_{\mathcal F},r_{\mathcal B})$ is always above the line where $U_{\mathcal{F}}(r_{\mathcal F}+c_i,r_{\mathcal B}-c_i) = U_{\mathcal{B}}(r_{\mathcal F},r_{\mathcal B})$ when $r_{\mathcal F}$, $r_{\mathcal B}$, and $1-r_{\mathcal F}-r_{\mathcal B}$ are in $[\delta, 1]$, $[c_i,1]$, and $[0,1]$, respectively.
For ease of reading, we denote by $boundary_{\mathcal A}$ the line where $U_{\mathcal{F}}(r_{\mathcal F}+c_i,r_{\mathcal B}) =
U_{\mathcal{A}}(r_{\mathcal F},r_{\mathcal B})$ when $r_{\mathcal F}$, $r_{\mathcal B}$, and $1-r_{\mathcal F}-r_{\mathcal B}$ are in $[0,1]$, $[c_i,1]$, and $[0,1]$, respectively. 
Also, we denote by $boundary_{\mathcal B}$ the line where $U_{\mathcal{F}}(r_{\mathcal F}+c_i,r_{\mathcal B}-c_i) = U_{\mathcal{B}}(r_{\mathcal F},r_{\mathcal B})$ when $r_{\mathcal F}$, $r_{\mathcal B}$, and $1-r_{\mathcal F}-r_{\mathcal B}$ are in $[0,1]$, $[c_i,1]$, and $[0,1]$, respectively. 

Moreover, the derivative $\frac{\partial r_{\mathcal {B}}}{\partial r_{\mathcal {F}}}|_{r_{\mathcal F}=0}$ on $boundary_{\mathcal A}$ is greater than that the derivative $\frac{\partial r_{\mathcal {B}}}{\partial r_{\mathcal {F}}}|_{r_{\mathcal F}=0}$ on $boundary_{\mathcal B}.$
Because $\frac{\partial r_{\mathcal {B}}}{\partial r_{\mathcal {F}}}$ is a continuous function, there exists a positive real number $\delta'$ such that, for all $x\in [0,\delta']$, 
the derivative $\frac{\partial r_{\mathcal {B}}}{\partial r_{\mathcal {F}}}|_{r_{\mathcal F}=x}$ on $boundary_{\mathcal A}$ is greater than the derivative $\frac{\partial r_{\mathcal {B}}}{\partial r_{\mathcal {F}}}|_{r_{\mathcal F}=x}$ on $boundary_{\mathcal B}.$
Then, there exists a number $\varepsilon^\prime$ such that, for all $c_i<\varepsilon^\prime$ and $x\in [0,\delta']$, 
the derivative $\frac{\partial r_{\mathcal {B}}}{\partial r_{\mathcal {F}}}|_{r_{\mathcal F}=x}$ on $boundary_{\mathcal A}$ is greater than that the derivative $\frac{\partial r_{\mathcal {B}}}{\partial r_{\mathcal {F}}}|_{r_{\mathcal F}=x}$ on $boundary_{\mathcal B}.$
Also, as described above, there exists a number $\varepsilon$ such that, for all $c_i<\varepsilon$, $boundary_{\mathcal A}$ is above $boundary_{\mathcal B}$ when $r_{\mathcal F}$,
$r_{\mathcal B}$, and $1-r_{\mathcal F}-r_{\mathcal B}$ are in $[\delta', 1]$, $[c_i,1]$, and $[0,1]$, respectively.

In the second step, we showed that $(0,r_{\mathcal B})$ cannot be $S$ where player $i$ does not change its strategy, when $\frac{1}{c_i}>\max\{\frac{1}{l}, 2+\frac{2N_{\tt in}}{N_{\tt de}}-\frac{3t}{N_{\tt de}+N_{\tt in}}\}.$
Therefore, 
$$\forall \frac{1}{c_i}>\max\{\frac{1}{l}, 2+\frac{2N_{\tt in}}{N_{\tt de}}-\frac{3t}{N_{\tt de}+N_{\tt in}}, \frac{1}{\varepsilon^\prime}\} \text{  and  } \forall r_{\mathcal{F}}\in[0,\delta'],$$ 
a range for \eqref{eq:am} is always above that for \eqref{eq:bm} without any intersection.
As a result, there exist $\varepsilon^{\prime\prime}$ as $$\varepsilon^{\prime\prime}=\max\{\frac{1}{l}, 2+\frac{2N_{\tt in}}{N_{\tt de}}-\frac{3t}{N_{\tt de}+N_{\tt in}}, \frac{1}{\varepsilon}, \frac{1}{\varepsilon^\prime}\}$$
such that, for all $c_i<\varepsilon^{\prime\prime},$ $(r_{\mathcal F},r_{\mathcal B})$ where $r_{\mathcal B}$ is positive is not $S$ in the game $\mathcal{G}(\bm{c},c_{\tt stick}).$

By the above three steps, if $c_{\tt stick}=0$, there exists $\varepsilon^{\prime\prime}$ such that, for all $c_i<\varepsilon^{\prime\prime},$ $S$ is characterized as presented in Lemma~\ref{lem:dynamics}. 
If $c_{\tt stick}>0,$ from this result, we can easily see that the value of $r_{\mathcal B}$ of $S$ is equal to $c_{\tt stick}.$ 
To characterize $S$ in this case, it is sufficient to have the second and third steps described above. 

Moreover, by Lemma~\ref{lem:dynamics}, the Nash equilibria in game $\mathcal{G}(\bm{c},c_{\tt stick})$ are characterized as presented in Theorem~\ref{thm:eq}. 
This completes the proof.

\section{Proof of Theorem~\ref{thm:inf_nash}} \label{sec:pf_in}

In this section, we show that all Nash equilibria in the game $\mathcal{G}(\bm{c}, c_{\tt stick})$ when players possess sufficiently small mining power. 
We first consider when $c_{\tt stick}$ is 0. 
In order for a state $(r_{\mathcal F},r_{\mathcal B})$ to be a Nash equilibrium in the game $\mathcal{G}(\bm{c}, c_{\tt stick})$, the following equation should be satisfied:
\begin{equation*}
    \sum_{s\in \mathcal{S}_{\tt max}}r_s=1 \, \text{ when }\,
    \mathcal{S}_{\tt max}=\argmax_{s\in \{\mathcal{F}, \mathcal{A}, \mathcal{B}\}} U_s(r_{\mathcal F},r_{\mathcal B})
\end{equation*}
The above equation means that all players belong to the most profitable group among \FM, \AM, and \BM.
In other words, in order for a point $(r_{\mathcal F},r_{\mathcal B})$ to be an equilibrium, either 1) $U_\mathcal{F}, U_\mathcal{A},$ and $U_\mathcal{B}$ have the same value at the point, or 
2) all miners should be in the most profitable group at the point. 
If both of them are not satisfied, some players would change their strategy to the most profitable one.

First, we consider that three payoffs are the same.
The case that payoffs of \FM and \AM are the same is equal to
\begin{equation}
\begin{aligned}
r_{\mathcal B}=0 \,\,&\text{or}\,\, \frac{k}{N_{\tt in}r_{\mathcal B}^2+N_{\tt de}(r_{\mathcal F}+r_{\mathcal B})^2}=\\
&\frac{r_{\mathcal {B}}}{(1-r_{\mathcal F}-r_{\mathcal B})N_{\tt in}r_{\mathcal B}^2+(1-r_{\mathcal B})N_{\tt de}(r_{\mathcal F}+r_{\mathcal B})^2}.
\end{aligned}
\label{eq:simeq1}
\end{equation}
The case that payoffs of \FM and \BM are the same equal to
\begin{equation}
\begin{aligned}
r_{\mathcal F}+r_{\mathcal B}=0 \,\,&\text{or}\,\,\frac{k}{N_{\tt in}r_{\mathcal B}^2+N_{\tt de}(r_{\mathcal F}+r_{\mathcal B})^2}=\\
&\frac{r_{\mathcal F}+r_{\mathcal B}}{(1-r_{\mathcal F}-r_{\mathcal B})N_{\tt in}r_{\mathcal B}^2+(1-r_{\mathcal B})N_{\tt de}(r_{\mathcal F}+r_{\mathcal B})^2}.
\end{aligned}
\label{eq:simeq2}
\end{equation}
By finding a solution satisfying both \eqref{eq:simeq1} and \eqref{eq:simeq2}, we can derive that three payoffs have the same value at the points $(r_{\mathcal F}=0, r_{\mathcal B}=\frac{k}{k+1})$ and $(r_{\mathcal F}=k, r_{\mathcal B}=0)$.
Therefore, these two points are equilibria.

Second, we consider three cases, when all miners belong to only two groups: 
1) \FM and \AM have the same mining profit density when $r_{\mathcal B}$ is 0, 2) \AM and \BM have the same mining profit density when $r_{\mathcal F}$ is 0, and 3) \FM and \BM have the same mining profit density when $r_{\mathcal F}+r_{\mathcal B}$ is 1. 
In the first case, in order for the case to be equilibria, \FM and \AM are profitable than \BM. 
Therefore, when $r_{\mathcal F}$ is not less than $k$, the case can be an equilibrium.
In other words, $(k\leq r_{\mathcal F} \leq 1, r_{\mathcal B}=0)$ is a Nash equilibrium.
In the second case, given that $r_{\mathcal F}$ is 0, $r_{\mathcal B}$ should be $\frac{k}{k+1}$ in order that \AM and \BM have the same payoff.
We already showed that the point $(0,\frac{k}{k+1})$ is an equilibrium.
The final case is impossible except for when $k$ is 1. 
If $k$ is 1, only point $(1,0)$ belongs to the final case. 
Also, we already showed above that the point $(1,0)$ is an equilibrium.

Finally, we consider three cases, when all players belong to just one group: 1) all players are in \FM, 2) \AM, and 3) \BM. 
As we demonstrated above, the first case $(r_{\mathcal F}=1, r_{\mathcal B}=0)$ is an equilibrium.
The second case represents $(r_{\mathcal F}=0, r_{\mathcal B}=0)$. 
In the second case, \AM has the mining profit density 1. 
However, players in \AM would shift to other groups because payoffs of \FM and \BM diverge to infinity.
Therefore, this case cannot be an equilibrium.
The third case presents $(r_{\mathcal F}=0, r_{\mathcal B}=1)$. 
In this case, \BM has the payoff $k$, and payoffs for other strategies diverge to infinity.
Therefore, players in \BM shift to others, and this is not an equilibrium.
As a result, all equilibria in the game $\mathcal{G}(\bm{c}, c_{\tt stick}=0)$ are $(r_{\mathcal F}=0, r_{\mathcal B}=\frac{k}{k+1})$ and $(k\leq r_{\mathcal F} \leq 1, r_{\mathcal B}=0)$. 

In the same manner, we can determine all Nash equilibria in game $\mathcal{G}(\bm{c}, c_{\tt stick}=0)$ when $c_{\tt stick}>0.$ 
Then the equilibria are \eqref{eq:nash_infty}. 

\section{Proof of Theorem~\ref{thm:bch2}}\label{sec:diff}

\blue{In this section, we first consider $c_{\tt stick}=0.$ }
At the state $(0,\frac{k}{1+k})$, the two payoffs of $coin_{\tt A}$ mining and $coin_{\tt B}$ mining are the same as $1+k$. 
Also, the payoff of \FM has the same value of $1+k,$ because the mining difficulty of both $coin_{\tt A}$ and $coin_{\tt B}$ would not change and thus \FM does not change the coin to mine. 
Therefore, rational miners do not revise their strategies at  $(0,\frac{k}{1+k})$, and the state $(0,\frac{k}{1+k})$ is a Nash equilibrium.
Indeed, in order for the payoffs of \FM, \AM, and \BM to be the same, \FM should not change the coin to be mined like in the state $(0,\frac{k}{1+k})$. 
If not, the $coin_{\tt B}$ mining difficulty would periodically change because of fickle miners.
In this case, we first assume that the payoffs of \AM and \BM are the same. 
Then the $coin_{\tt B}$ mining is more profitable than the $coin_{\tt A}$ mining when the $coin_{\tt B}$ mining difficulty is low. 
Conversely, when the $coin_{\tt B}$ mining difficulty is high, the $coin_{\tt A}$ mining is more profitable than the $coin_{\tt B}$ mining.
Therefore, \FM would earn more profit than that for \AM and \BM because they conduct the $coin_{\tt B}$ mining only when its difficulty is low.
This fact implies that, in a state where \FM changes its preferred coin, the payoffs of \FM, \AM, and \BM cannot be the same.
By using this property, one can easily find that there exist only the states $(0,\frac{k}{1+k})$ and $(k,0)$ where the three payoffs are the same.

In the states $(k <r_{\mathcal F} \leq 1, r_{\mathcal B}=0)$, the mining difficulty of $coin_{\tt A}$ is eventually maintained as 1 while the mining difficulty of $coin_{\tt B}$ is maintained as more than $k.$ 
Thus, the payoffs of \FM and \AM are 1 at the states.
To find the payoff of \BM in the states, we consider states $(k< r_{\mathcal F} \leq 1, r_{\mathcal B}=\delta)$ for sufficiently small $\delta$. 
Note that $U_\mathcal{B}(r_{\mathcal F},0)$ is defined as $\lim_{\delta\rightarrow 0}U_\mathcal{B}(r_{\mathcal F},\delta).$
In $(k< r_{\mathcal F} \leq 1, r_{\mathcal B}=\delta),$ the mining difficulty of $coin_{\tt B}$ would have value $d \in (k,r_{\mathcal F}]$ most of the time, because fickle mining that heavily occurs increases the mining difficulty of $coin_{\tt B}$ by a high value and it takes a significantly long time for \BM with the mining power $\delta$ to find blocks with the high mining difficulty $d.$
Therefore, the payoff $U_\mathcal{B}(k<r_{\mathcal F}\leq 1,0)$ of \BM as $\frac{k}{d}$ is less than $1$.
This means that rational miners do not change their strategies to \B at the states, 
and states $(k\leq r_{\mathcal F} \leq 1, r_{\mathcal B}=0)$ representing the downfall of $coin_{\tt B}$ are Nash equilibria.
Meanwhile, in states $(r_{\mathcal F}<k , r_{\mathcal B}=0)$, rational miners would move to \BM because \BM's payoff is greater than 1 while \FM and \AM's payoffs are $1.$

In addition, like in Figure~\ref{fig:zones}, $boundary_{1,3}$ is always above
$boundary_{2,3}$ except the point $(0,\frac{k}{k+1})$ in the triangle area.
Note that $boundary_{1,3}$ refers to a line on which the payoffs of \FM and \AM are the same while $r_{\mathcal B}>0$, and $boundary_{2,3}$ refers to a line on which the payoffs of \FM and \BM are the same.
When $r_{\mathcal B}=0$, the payoffs of \FM and \AM are always the same because \FM would mine only $coin_A$ eventually.
If we assume that $boundary_{1,3}$ and $boundary_{2,3}$ have another intersection point, not $(0,\frac{k}{k+1})$, in the triangle area, the payoffs of \FM, \AM, and \BM would be the same for at least three points. 
This is a contradiction because the three payoffs are the same at only two points, $(0,\frac{k}{1+k})$ and $(k,0)$.
Moreover, $boundary_{2,3}$ intersects with the line $r_{\mathcal B}=0$ at the point $(k,0)$ because payoffs of \FM, \AM, and \BM are the same at the point. 
Meanwhile, $boundary_{1,3}$ does not intersect with the line $r_{\mathcal B}=0.$ 
This is because \FM is trivially most profitable at $(k\leq r_{\mathcal F} \leq 1, r_{\mathcal B}=\delta)$ for sufficiently small $\delta$, and the difference between payoffs of \FM and \AM (i.e., $U_{\mathcal F}-U_{\mathcal A}$) is a decreasing function of $r_{\mathcal B}$ when $r_{\mathcal F}$ is given. 
Indeed, when $r_{\mathcal F}$ is given, the greater $r_{\mathcal B}$, the smaller $r_{\mathcal A}$ is and the lower $coin_A$ mining difficulty is. 
Therefore, when $r_{\mathcal B}$ increases, the profit, which fickle miners earn by mining $coin_{\tt A}$, increases, and this means that, for given $r_{\mathcal F}$, $U_{\mathcal F}-U_{\mathcal A}$ is a decreasing function of $r_{\mathcal B}$. 
Similarly, for given $r_{\mathcal F}$, $U_{\mathcal F}-U_{\mathcal B}$ is an increasing function of $r_{\mathcal B}.$ 
Because of these facts, $boundary_{2,3}$ does not intersect with the line $r_{\mathcal F}+r_{\mathcal B}=0$ while $boundary_{1,3}$ intersects at one point of the line $r_{\mathcal F}+r_{\mathcal B}=0.$
As a result, even when the mining difficulty of $coin_{\tt B}$ is adjusted in a short time period, the game $\mathcal{G} (\bm{c}, c_{\tt stick}=0)$ has Nash equilibria and dynamics such as in Figure~\ref{fig:zones}. 
This fact also makes the game $\mathcal{G}(\bm{c},c_{\tt stick}>0)$ have dynamics presented in Figure~\ref{fig:zones}. 

\end{appendices}

\end{document}